\documentclass[twocolumn]{aastex62}
\pdfoutput=1
\usepackage{amsmath}
\usepackage{graphicx}
\usepackage[toc,page]{appendix}

\shorttitle{UV-H$_2$ and UV-CO in Protoplanetary Disks}
\shortauthors{Arulanantham et al.}

\begin{document}

\title{UV Fluorescence Traces Gas and Ly$\alpha$ Evolution in Protoplanetary Disks}

\author{Nicole Arulanantham}
\affil{Space Telescope Science Institute, 3700 San Martin Drive, Baltimore, MD 21218, USA}
\author{Kevin France}
\affil{Laboratory for Atmospheric and Space Physics, University of Colorado, 392 UCB, Boulder, CO 80303, USA}
\author{Keri Hoadley}
\affil{Department of Physics, Mathematics, and Astronomy, California Institute of Technology, 1200 East California Blvd., Pasadena, CA 91125, USA}
\affil{Department of Physics and Astronomy, University of Iowa, Van Allen Hall, Iowa City, IA 52242}
\author{P.C. Schneider}
\affil{Hamburger Sternwarte, Gojenbergsweg 112, 21029 Hamburg, Germany}
\author{Catherine C. Espaillat}
\affil{Institute for Astrophysical Research, Department of Astronomy, Boston University, 725 Commonwealth Avenue, Boston, MA 02215, USA}
\author{H.M. G\"{u}nther}
\affil{MIT, Kavli Institute for Astrophysics and Space Research, 77 Massachusetts Ave., Cambridge, MA 02139, USA}
\author{Gregory J. Herczeg}
\affil{Kavli Institute for Astronomy and Astrophysics, Peking University, Yiheyuan 5, Haidian Qu, 100871, Beijing, China}
\author{Alexander Brown}
\affil{Center for Astrophysics and Space Astronomy, University of Colorado, 389 UCB, Boulder, Colorado 80309, USA}

\begin{abstract}

Ultraviolet spectra of protoplanetary disks trace distributions of warm gas at radii where rocky planets form. We combine \textit{HST}-COS observations of H$_2$ and CO emission from 12 classical T Tauri stars to more extensively map inner disk surface layers, where gas temperature distributions allow radially stratified fluorescence from the two species. We calculate empirical emitting radii for each species under the assumption that the line widths are entirely set by Keplerian broadening, demonstrating that the CO fluorescence originates further from the stars $\left(r \sim 20 \, \rm{AU} \right)$ than the H$_2$ $\left(r \sim 0.8 \, \rm{AU} \right)$. This is supported by 2-D radiative transfer models, which show that the peak and outer radii of the CO flux distributions generally extend further into the outer disk than the H$_2$. These results also indicate that additional sources of Ly$\alpha$ photons remain unaccounted for, requiring more complex models to fully reproduce the molecular gas emission. As a first step, we confirm that the morphologies of the UV-CO bands and Ly$\alpha$ radiation fields are significantly correlated and discover that both trace the degree of dust disk evolution. The UV tracers appear to follow the same sequence of disk evolution as forbidden line emission from jets and winds, as the observed Ly$\alpha$ profiles transition between dominant red wing and dominant blue wing shapes when the high-velocity optical emission disappears. Our results suggest a scenario where UV radiation fields, disk winds and jets, and molecular gas evolve in harmony with the dust disks throughout their lifetimes.
\end{abstract}


\section{Introduction}

The advent of the Atacama Large Millimeter/submillimeter Array (ALMA) has revolutionized studies of circumstellar disks, providing images of planet-forming material at unprecedented spatial resolution and sensitivity. By mapping the dust distributions down to $\sim$5 AU from the central star, sub-mm observations have revealed both small and large-scale dust substructures in most disks (see e.g., \citealt{Andrews2018, Huang2018_TWHya, Long2018}). This includes all observed disks in the Taurus star-forming region with effective emission radii larger than 50 au, implying that disks without detected substructures are simply too compact to resolve the features without even higher resolution observations \citep{Long2019}. The ALMA images are consistent with models of disk evolution, which predict asymmetric dust traps \citep{vanderMarel2013}, rings \citep{HLTau2015, Perez2018, Guzman2018, Huang2018_annular}, and spiral arms \citep{Huang2018_spirals} as a result of physical mechanisms like planet-disk interactions, dust-trapping, and X-ray photoevaporation (see e.g., \citealt{Johansen2004, Bae2016, Gorti2009}). However, sensitivity requirements for detecting individual gas species prevent observations of molecular gas at the spatial resolution required to trace key carbon-, oxygen-, and nitrogen-bearing species in regions where rocky planets are expected to form ($r < 5$ AU). Alternative tracers of warm inner disk gas are required to map the material close to the central star. 

Ultraviolet and infrared emission from warm CO (UV-CO, IR-CO; $T \sim 300-1500$ K) and ultraviolet emission from hot H$_2$ (UV-H$_2$; $T > 1500$ K) are readily detected within the inner disk, in regions that are difficult to resolve at sub-mm wavelengths ($r < 20$ AU; see e.g., \citealt{Herczeg2002, Najita2003, Salyk2011_IRCO, Brown2013, France2011_CO, France2012}). The UV-H$_2$ features originate in a thin surface layer of the gas disk that is heated to $T > 1500$ K by UV radiation, allowing H$_2$ to be excited into higher vibrational levels (see e.g., \citealt{Nomura2005, Nomura2007, Adamkovics2016}). These vibrationally excited molecules are then ``pumped'' from the ground electronic state $\left(X^1 \Sigma_{g}^+ \right)$ into the first and second excited electronic states $\left(B^1 \Sigma_{u}^+, \, C^1 \Pi_{u}^{+} \right)$ by Ly$\alpha$ photons, producing fluorescent emission lines as the molecules return to the ground state \citep{Herczeg2002, Herczeg2004}. The features are spectrally resolved in data from \textit{HST}-STIS and \textit{HST}-COS, allowing observers to extract both the empirical radial locations of emitting gas \citep{France2012} and the radial distributions of flux from the spectra \citep{Hoadley2015, Arulanantham2018}. 

The UV-CO emission lines are produced by the same Ly$\alpha$-pumping mechanism as UV-H$_2$ fluorescence \citep{France2011_CO}. However, models of the temperature and column density of emitting gas demonstrate that CO fluorescence originates from a cooler population of molecules $\left( T < 600 \, \text{ K} \right)$ \citep{Schindhelm2012_CO}, likely located at larger radii than hotter UV-H$_2$. In order to extend our current picture of inner disk gas surface layers to more distant regions, we build on those findings in this work by comparing both UV-H$_2$ and UV-CO emission lines from disks around T Tauri stars. When both species are included, we tentatively identify a multi-layered sequence of disk evolution that is consistent with similar trends observed at optical and IR wavelengths. Our analysis is structured as follows:
\begin{enumerate}
\item We first estimate the purely empirical radial locations of the emitting gas, assuming an ideal model where the features originate in a Keplerian disk.
\item We use the molecular gas emission to reconstruct the Ly$\alpha$ radiation fields that reach the disk surfaces and pump H$_2$ and CO into excited electronic states.
\item We then feed the reconstructed Ly$\alpha$ profiles into 2-D radiative transfer models that map the full radial distributions of flux from both species.
\item However, we discover significant trends in deviations between the models and data, which demonstrate that the Ly$\alpha$ radiation reaching the molecular gas is sensitive to the overall state of disk evolution.
\item We discuss these results in the context of the disk dust distributions, presenting statistically significant correlations between Ly$\alpha$ morphologies, UV-H$_2$ and UV-CO fluxes, infrared spectral indices, and the presence of high-velocity [O I] $\lambda$6300 \AA \, emission.
\end{enumerate}
These results lay the groundwork for more complex models of Ly$\alpha$ irradiation in protoplanetary disks, which will be the subject of future work.

\section{Targets and Observations} 

Our sample consists of 12 young stars with circumstellar disks that were observed with the Cosmic Origins Spectrograph onboard the \textit{Hubble Space Telescope} (\textit{HST}-COS; \citealt{Green2012}). Two of the targets were observed twice (AA Tau and RECX-15), and we include both sets of data for a total sample size of 14 spectra. We utilize the spectra acquired with the FUV observing modes, which were reduced with the CALCOS pipeline before being aligned and co-added into a final product for analysis \citep{Danforth2010}. The point source resolution of the FUV modes on \textit{HST}-COS is $\Delta v \sim 15$ km s$^{-1}$ \citep{Osterman2011}, which sets the radial velocity accuracy of the spectral features we measure and the minimum spectrally resolved line widths.

Although there are archival \textit{HST}-COS spectra for many more T Tauri stars, our sample is limited to the objects with distinct CO emission out to $J'' \sim 20$. Basic stellar and disk properties for each target are shown in Table \ref{targ_list}, and details of the original observing programs are provided in Table \ref{obs_props}. Both the G130M and G160M modes were used for most targets, providing wavelength coverage from $\sim$1100-1700 \AA. Various features in the spectra have been presented in previous work, including accretion rate diagnostics \citep{Ardila2013}, fluorescent emission lines from excited electronic states of hot H$_2$ \citep{France2012, Hoadley2015}, $^{12}$CO and $^{13}$CO $A-X$ absorption bands \citep{McJunkin2013}, Ly$\alpha$ emission \citep{Schindhelm2012} and H$_2$ absorption along the Ly$\alpha$ profile \citep{Hoadley2017}, the FUV radiation field \citep{Yang2012, France2014}, and connections to UV photochemistry \citep{Arulanantham2020}.

\begin{deluxetable*}{ccccccc}
\tablecaption{Sample of CTTS with UV-H$_2$ and UV-CO Emission \label{targ_list}
}
\tablewidth{0 pt}
\tabletypesize{\scriptsize}
\tablehead{ \colhead{Target} & \colhead{$d$} & \colhead{$A_V$} & \colhead{$M_{\ast}$} & \colhead{$i_d$} & \colhead{$n_{13-31}$$^{16}$} & \colhead{References \tablenotemark{a}} \\
 & \colhead{[pc]} & \colhead{[mag]} & \colhead{$\left[ M_{\odot} \right]$} & \colhead{$\left[^{\circ} \right]$} & \colhead{} & \colhead{} \\
}
\startdata
AA Tau (2011) & 136.7 & 0.34 & 0.8 & 59.1 & -0.51 & 1, 3, 3, 4 \\
AA Tau (2013) & 136.7 & 0.34 & 0.8 & 59.1 & -0.51 & 1, 3, 3, 4 \\
CS Cha & 175.4 & 0.16 & 1.05 & 24.2 & 2.89 & 1, 3, 3, 13 \\
CW Tau & 130.7 & \nodata & 0.69 & 59 & -0.65 & 1, -, 14, 5 \\
DF Tau & 124.5 & 0.54 & 0.19 & 24 & -1.09 & 1, 3, 3, 6 \\
DM Tau & 144.5 & 0.48 & 0.5 & 35 & 1.3 & 1, 3, 3, 7 \\
LkCa 15 & 158.2 & 0.31 & 0.85 & 49 & 0.62 & 1, 3, 3, 7 \\
RECX-11 & 98.3 & 0.03 & 0.8 & 70 & -0.8 & 1, 3, 3, 8 \\
RECX-15 (2010) & 91.6 & 0.02 & 0.4 & 60 & -0.2 & 1, 3, 3, 8 \\
RECX-15 (2013) & 91.6 & 0.02 & 0.4 & 60 & -0.2 & 1, 3, 3, 8 \\
RY Lupi & 159.1 & 0.10 & 1.71 & 68 & -0.26 & 2, 3, 3, 10 \\
T Cha & 109.6 & \nodata & 1.5 & 67 & 1.67 & 2, -, 12, 12 \\
UX Tau A & 139.4 & 0.51 & 1.3 & 35 & 1.83 & 1, 3, 3, 11 \\
V4046 Sgr & 72.3 & 0 & $0.86+0.69$\tablenotemark{b} & 33 & 0.32 & 1, 3, 17, 15
\enddata
\tablenotetext{a}{(1) \citealt{BJ2018}; (2) \citealt{GaiaDR2}; (3) \citealt{France2017}; (4) \citealt{Loomis2017}; (5) \citealt{Bacciotti2018}; (6) \citealt{Shakhovskoj2006}; (7) \citealt{Andrews2011}; (8) \citealt{Lawson2004}; (9) \citealt{Cabrit2006}; (10) \citealt{vanderMarel2018}; (11) \citealt{Lawson1996}; (12) \citealt{Huelamo2015}; (13) \citealt{Ginski2018}; (14) \citealt{Guilloteau2014}; (15) \citealt{Rodriguez2010}; (16) \citealt{Furlan2009}; (17) \citealt{Rosenfeld2012}}  
\tablenotetext{b}{V4046 Sgr is a binary system, so the values listed here correspond to the two individual stellar masses \citep{Rosenfeld2012}.}
\end{deluxetable*}

\begin{deluxetable*}{ccc}
\tablecaption{\emph{HST}-COS Observations \label{obs_props}
}
\tablewidth{0 pt}
\tabletypesize{\scriptsize}
\tablehead{ \colhead{Target} & \colhead{PI:Program ID\tablenotemark{a}} & \colhead{Observing Modes/Central Wavelengths} 
}
\startdata
AA Tau (2011) & G. Herczeg:11616 & G130M/1291,1327; G160M/1577, 1600, 1623 \\
AA Tau (2013) & K. France:12876 & G130M/1222; G160M/1577, 1589, 1611, 1623 \\
CS Cha & G. Herczeg:11616 & G130M/1291, 1327; G160M/1577, 1600, 1623 \\
CW Tau & K. France:15070 & G130M/1222; G160M/1577, 1589 \\
DF Tau & J. Green:11533 & G130M/1291, 1300, 1309, 1318; G160M/1589, 1600, 1611, 1623 \\
DM Tau & G. Herczeg:11616 & G130M/1291, 1327; G160M/1577, 1600, 1623 \\
LkCa 15 & G. Herczeg:11616 & G130M/1291, 1327; G160M/1577, 1600, 1623 \\
RECX-11 & G. Herczeg:11616 & G130M/1291, 1327; G160M/1577, 1600, 1623 \\
RECX-15 (2010) & G. Herczeg:11616 & G130M/1291, 1327; G160M/1577, 1600, 1623 \\
RECX-15 (2013) & K. France:12876 & G130M/1222; G160M/1577, 1589, 1611, 1623 \\
RY Lupi & C. F. Manara/P. C. Schneider:14469 & G130M/1291, 1318; G160M/1577, 1611 \\
T Cha & A. Brown:15128 & G130M/1291; G160M/1577, 1600 \\
UX Tau A & G. Herczeg:11616 & G130M/1291, 1327; G160M/1577, 1600, 1623 \\
V4046 Sgr & J. Green:11533 & G130M/1291, 1300, 1309, 1318; G160M/1589, 1600, 1611, 1623 \\
\enddata
\tablenotetext{a}{All the \textit{HST}-COS data used in this paper can be found in MAST: \dataset[https://doi.org/10.17909/t9-pxwh-py47]{https://doi.org/10.17909/t9-pxwh-py47}.}
\end{deluxetable*}

The \textit{HST}-COS spectra of all systems in our sample include strong fluorescent H$_2$ emission lines, with most also showing high signal-to-noise fluorescence from the $^{12}$CO $\left(v'-v'' \right) = \left(14-3 \right)$ and $\left(v'-v'' \right) = \left(14-4 \right)$ bands. Two disks with large dust cavities (LkCa15 and RY Lupi; \citealt{Francis2020}) show much weaker UV-CO features than the rest of the group. We include them here to study combinations of disk geometry and Ly$\alpha$ irradiation that can produce strong UV-H$_2$ but less prominent UV-CO. Furthermore, five other disks in our sample have been observed with moderate to high resolution at sub-mm wavelengths: AA Tau \citep{Loomis2017}, CW Tau \citep{Pietu2014, Bacciotti2018}, DM Tau \citep{Kudo2018}, CS Cha, LkCa 15, RY Lupi \citep{Francis2020}, T Cha \citep{Hendler2018}, and V4046 Sgr \citep{Kastner2018, RuizRodriguez2019}. These sub-mm gas and dust images provide additional context for interpreting spectroscopic data from the disks we survey in this work.    

Previous long-slit observations of some T Tauri stars with \textit{HST}-STIS show spatially extended UV-H$_2$ emission pumped by locally generated Ly$\alpha$ photons, in addition to the on-source emission originating from the inner disk. The features trace cavity walls (T Tau; \citealt{Saucedo2003, Walter2003}), shock-excited outflows (RU Lupi, DG Tau; \citealt{Herczeg2005, Herczeg2006, Schneider2013}), and photoevaporative winds (GM Aur; \citealt{Hornbeck2016}). We note that since \textit{HST}-COS is a slitless instrument, any similar spatially extended emission along its dispersion axis would be mapped to distinguishable velocity shifts in the 1-D spectra (see e.g., \citealt{France2012}). However, this effect would be hidden if the emission was extended along the cross-dispersion direction and did not exceed the 1.25'' aperture radius (see e.g., \citealt{France2011_1987A}). 

\section{Methods}

\subsection{UV-H$_2$}

High signal-to-noise emission lines from UV-fluorescent H$_2$ are detected in all targets in our sample. The features originate in flared surface layers of the gas disk, where temperatures are at least $T > 1500$ K \citep{Adamkovics2014, Adamkovics2016}. Vibrationally excited H$_2$ $\left(v \geq 2 \right)$ in these surface layers is pumped into excited electronic states by Ly$\alpha$ photons, producing a cascade of fluorescent emission lines as the molecules decay back to the ground state. A set of features with the same vibration and rotation $\left([v', J' \right])$ upper level in the excited electronic state is referred to as a progression. 

UV-H$_2$ fluxes from the $\left[v', J' \right] = \left[1, 4 \right]$ and $\left[v', J' \right] = \left[1, 7 \right]$ progressions (see Table \ref{H2_lab_props}) were measured using an interactive Python tool (SELFiE; \citealt{Arulanantham2018}). These features typically have higher signal-to-noise than emission lines from other progressions, since molecules are pumped into the upper levels of the transitions by photons with wavelengths near the peak of the Ly$\alpha$ emission line $\left(1215.73, 1216.07 \, \rm{\AA} \right)$. This allows larger fractions of H$_2$ to be pumped into the $\left[v', J' \right] = \left[1, 4 \right]$ and $\left[v', J' \right] = \left[1, 7 \right]$ states (see e.g., \citealt{Herczeg2004}). 

\begin{deluxetable*}{cccc}
\tablecaption{Measured UV-H$_2$ Emission Lines \label{H2_lab_props}
}
\tablewidth{0 pt}
\tabletypesize{\scriptsize}
\tablehead{\colhead{Progression} & \colhead{Pumping Wavelength} & \colhead{Line ID\tablenotemark{a}} & \colhead{Rest Wavelength} \\
\colhead{$\left[ v', J' \right]$} & \colhead{[\AA]} & \colhead{$\left(v' - v'' \right)$sgn($\Delta J$)($J''$)} & \colhead{[\AA]} \\
}
\startdata
$\left[ 1, 7 \right]$ & 1215.73 & (1-6)P(8) & 1467.08 \\
 & & (1-7)R(6) & 1500.45 \\
 & & (1-7)P(8) & 1524.65 \\
 & & (1-8)R(6) & 1556.87 \\
$\left[1,4 \right]$ & 1216.07 & (1-6)R(3) & 1431.01 \\
 & & (1-6)P(5) & 1446.12 \\
 & & (1-7)R(3) & 1489.57 \\
 & & (1-7)P(5) & 1504.76 \\
\enddata
\tablenotetext{a}{sgn$\left(\Delta J \right) = P$ if $\Delta J = -1$ or $R$ if $\Delta J = +1$}
\end{deluxetable*}

The interactive line fitting routine approximates each emission line as a Gaussian profile and linear continuum that are convolved with the wavelength-dependent line-spread function (LSF) for \emph{HST}-COS. Integrated line fluxes, FWHMs, and central wavelengths are then extracted from the best-fit Gaussian properties. Assuming the widths of the observed emission lines are dominated by Keplerian rotational broadening, an empirical radius of UV-H$_2$ emission can be calculated as 
\begin{equation}
\left< R_{H_2} \right> = G M_{\ast} \left( \frac{2 \sin i_d}{FWHM} \right)^2
\end{equation}    
\citep{Salyk2011_IRCO, France2012, Brown2013}. A total of 11 emission lines from the $\left[1, 4 \right]$ and $\left[1, 7 \right]$ progressions were fit with the Gaussian model, and an average FWHM was measured from the best-fit parameters of those features. The average FWHM was then used to estimate the spatial location of the gas with Equation 1.  

\subsection{UV-CO}

The same Ly$\alpha$-pumping mechanism is responsible for UV-CO emission as well \citep{France2011_CO}. Unlike UV-H$_2$, UV-CO is pumped from rotational states in the $v = 0$ level of the ground electronic state, which allows cooler gas to fluoresce. The UV-CO bands observed with \textit{HST}-COS are produced by molecules cascading from $v' = 14$ in the $A^1 \Pi$ electronic state \citep{Schindhelm2012_CO}. 

\begin{figure*}[t!]
	\begin{minipage}{0.48\textwidth}
	\centering
	\includegraphics[width=\linewidth]{DFTau_H2CO_example.pdf}
	\end{minipage}
	\begin{minipage}{0.48\textwidth}
	\centering
	\includegraphics[width=\linewidth]{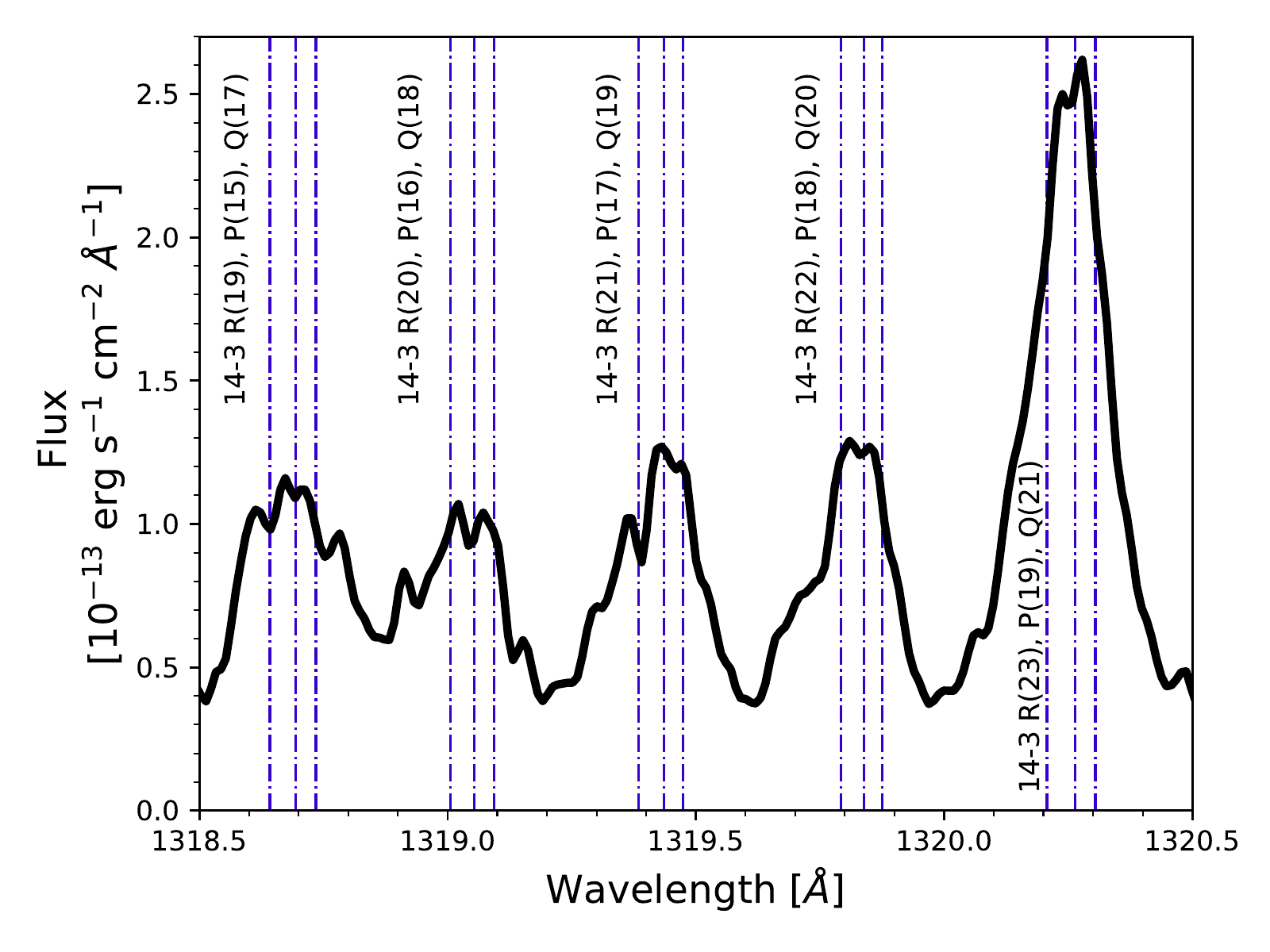}
	\end{minipage}
\caption{\textit{Left:} \textit{HST}-COS spectrum of DF Tau, including a single UV-H$_2$ emission line from the $\left[v', J' \right] = [1,4]$ progression and the full $\left(v'-v'' \right) = (14-3)$ band of UV-CO. The wavelength of the UV-H$_2$ transition is marked with a red dashed line, and lab wavelengths of individual UV-CO transitions are denoted with blue dashed-dotted lines. \textit{Right:} Zoomed in view of the high $J''$ region from the same spectrum. Each blend of three transitions can be decomposed into individual Gaussian emission line components, to empirically estimate the radii from where the UV-CO originates (see Equation 1).}
\label{line_example}
\end{figure*}

In contrast to the spectrally resolved UV-H$_2$ lines, the close spacings of individual rotational states within the $\left(v'-v'' \right) = (14-4)$ and $\left(v'-v'' \right) = (14-3)$ UV-CO bands $\left( \Delta v < 15 \, \rm{km \, s}^{-1} \right)$ make them impossible to resolve with \emph{HST}-COS. However, the high $J''$ states $\left(J'' > 10 \right)$ have energy level spacings that are wide enough to partially separate the rotational structure (see Figure \ref{line_example}). This means that the $J'' > 10$ states appear as four to five distinct emission profiles in both UV-CO bands, with each feature consisting of three blended lines. Although the high $J''$ features are still not spectrally resolved, the separation between each set is wider than the pile-up of states at the bandhead. 

Like the [1,4] and [1,7] progressions of UV-H$_2$, CO molecules are pumped into the upper levels of high $J''$ states via photons at wavelengths where the Ly$\alpha$ line profile peaks in flux $\left( 1215.7 < \lambda < 1217 \, \rm{\AA} \right)$. This produces strong emission lines, even when the temperature is too low to thermally populate the $J'' > 10$ energy levels (T $\leq 500$ K; \citealt{France2011_CO}). The high $J''$ states are so prominent in some systems that the signal-to-noise is greater than that in the bandhead $\left( J'' = 0-10 \right)$. Since the high $J''$ states are also more widely separated, we focus this portion of the analysis on those features. However, our approach is still limited by the spectral resolution of the data, making the spatial constraints less robust than the empirical emitting regions derived from the well-resolved UV-H$_2$ features.

\begin{figure*}[t!]
	\begin{minipage}{0.49\textwidth}
	\centering
	\includegraphics[width=\linewidth]{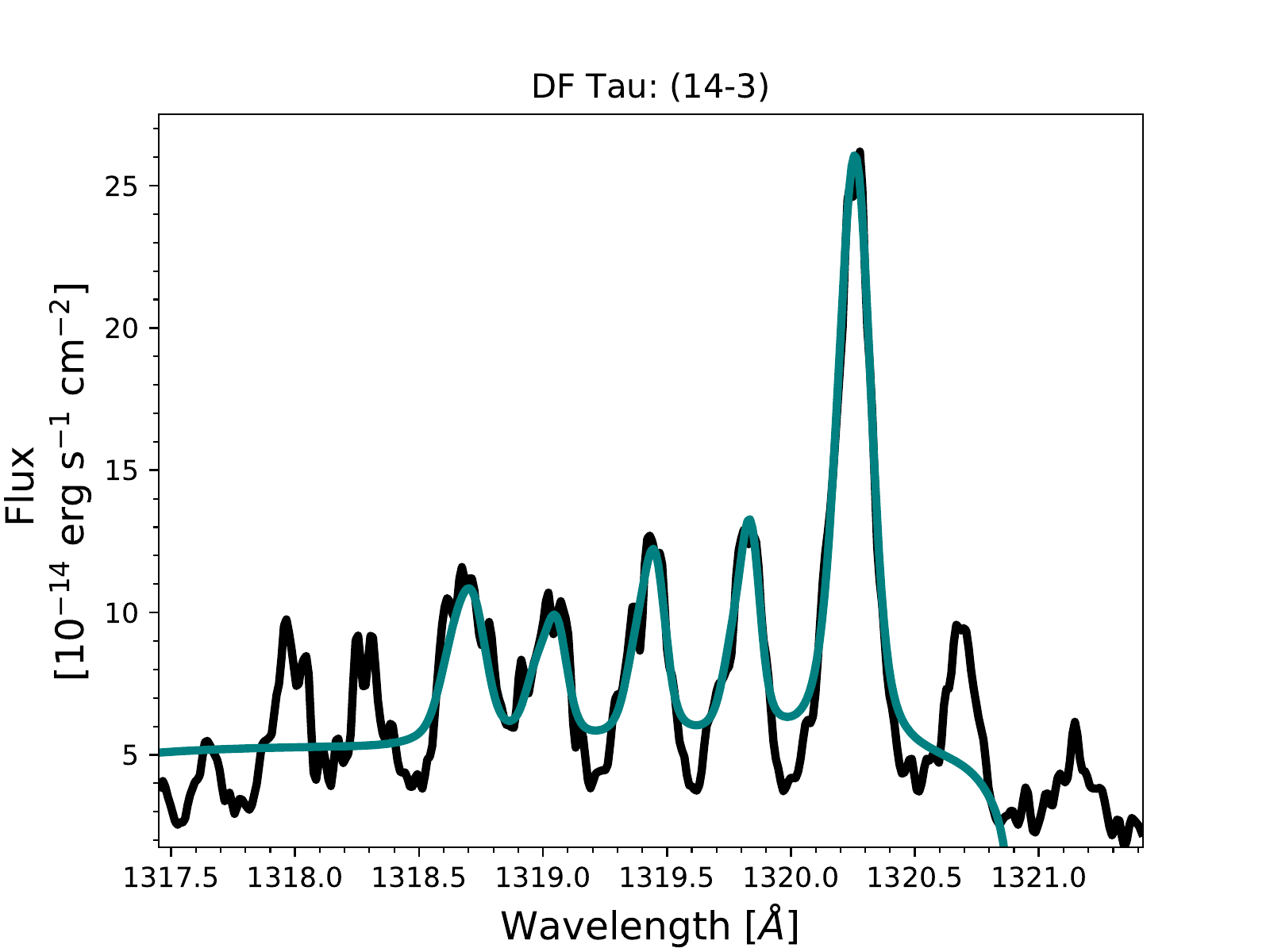}
	\end{minipage}
	\begin{minipage}{0.49\textwidth}
	\centering
	\includegraphics[width=\linewidth]{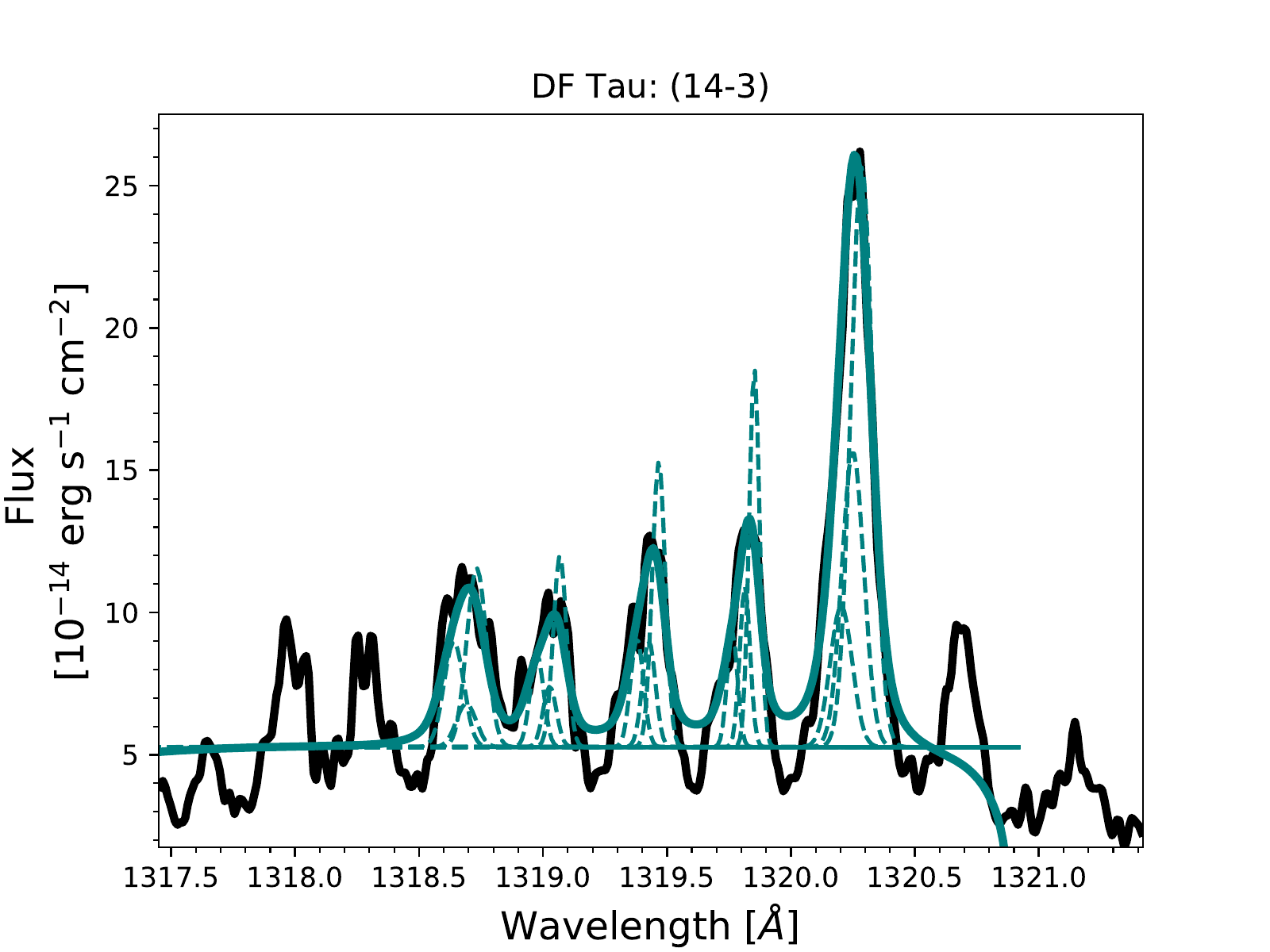}
	\end{minipage}
\caption{\textit{Left:} Partially resolved ``fingers" of UV-CO emission (black curve) were fit with the same interactive fitting routine used on the UV-H$_2$ emission lines. Each finger was fit with a single Gaussian and a linear continuum, with variable amplitude, central wavelength, FWHM, and continuum slope and intercept (teal curve). \textit{Right:} The observed features were further decomposed using multi-Gaussian models, where the widths of individual transitions within a single finger were required to be identical and the width of the blended lines equal to the FWHM from the single Gaussian model. The radial locations of emitting gas were then calculated from the average decomposed widths of all measured fingers. Observed features which are not included in the Gaussian models shown in the figures were not able to be further divided into individual emission lines. The continuum was assumed to be linear for uniformity across the sample of disks, and uncertainties in the slope and intercept are folded into the errors on the empirical radii.}
\end{figure*}

The blended UV-CO features were decomposed by fitting a multi-Gaussian model to the spectra. First, the same interactive line fitting routine used to measure the UV-H$_2$ emission lines was used to fit each partially resolved blend of three lines (called ``fingers" for simplicity) with a single Gaussian model, convolved with the \textit{HST}-COS LSF and superimposed on a linear continuum. The central wavelength, Gaussian line width, and amplitude were left as free parameters for each finger. We included as many fingers as possible, excluding only the features that later did not have high enough S/N to be fit with the multi-Gaussian models. The width of each single Gaussian was then used to constrain the widths of the three narrower emission lines composing the corresponding finger, allowing us to obtain multiple ``measurements" of the unresolved features. This procedure to fit the fingers is equivalent to fitting multiple UV-H$_2$ emission lines from a single progression.  

Next, each finger was further divided using the multi-Gaussian approach. Each model consisted of three Gaussian emission lines, which were first superimposed and then convolved with the \emph{HST}-COS LSF. Individual Gaussians were given the same line width, as expected for emission originating from the same Keplerian radius. The amplitude ratios between the three blended features were fixed based on the Ly$\alpha$ fluxes at the appropriate pumping wavelengths and the branching ratios describing the likelihood of each pathway out of the Ly$\alpha$-pumped upper level. Best-fit models were obtained by varying the amplitude and central wavelength of the transition in the blended triplet with the largest likelihood of transitioning out of its upper level and the single Gaussian line width for all three features. For each target, the radial location of emitting gas was then calculated from the average decomposed line width from all measured fingers. We note that the linear continuum is fine for most systems in our sample, although it is not a perfect approximation. With this in mind, uncertainties in the continuum slopes and intercepts are also folded into the errors on the empirical radii.

We emphasize that the high $J''$ UV-CO emission lines used to derive the empirical radii are not spectrally resolved. While Gaussian decompositions can reproduce the observed features, \citet{Weber2020} point out that such models are still unable to distinguish contributions to the total flux from different emitting regions (e.g., multiple wind scale heights). They note that any emission components originating in winds, rather than the Keplerian disk, will project some velocity signature onto the observed spectra which cannot be untangled with Gaussians alone. Rather than comparing the UV-H$_2$ and UV-CO empirical radii for each individual system, we instead focus on general trends, which provide a starting point for building more complex models.

\section{Results}

\begin{figure*}
\centering
\includegraphics[width=1.0\linewidth]
{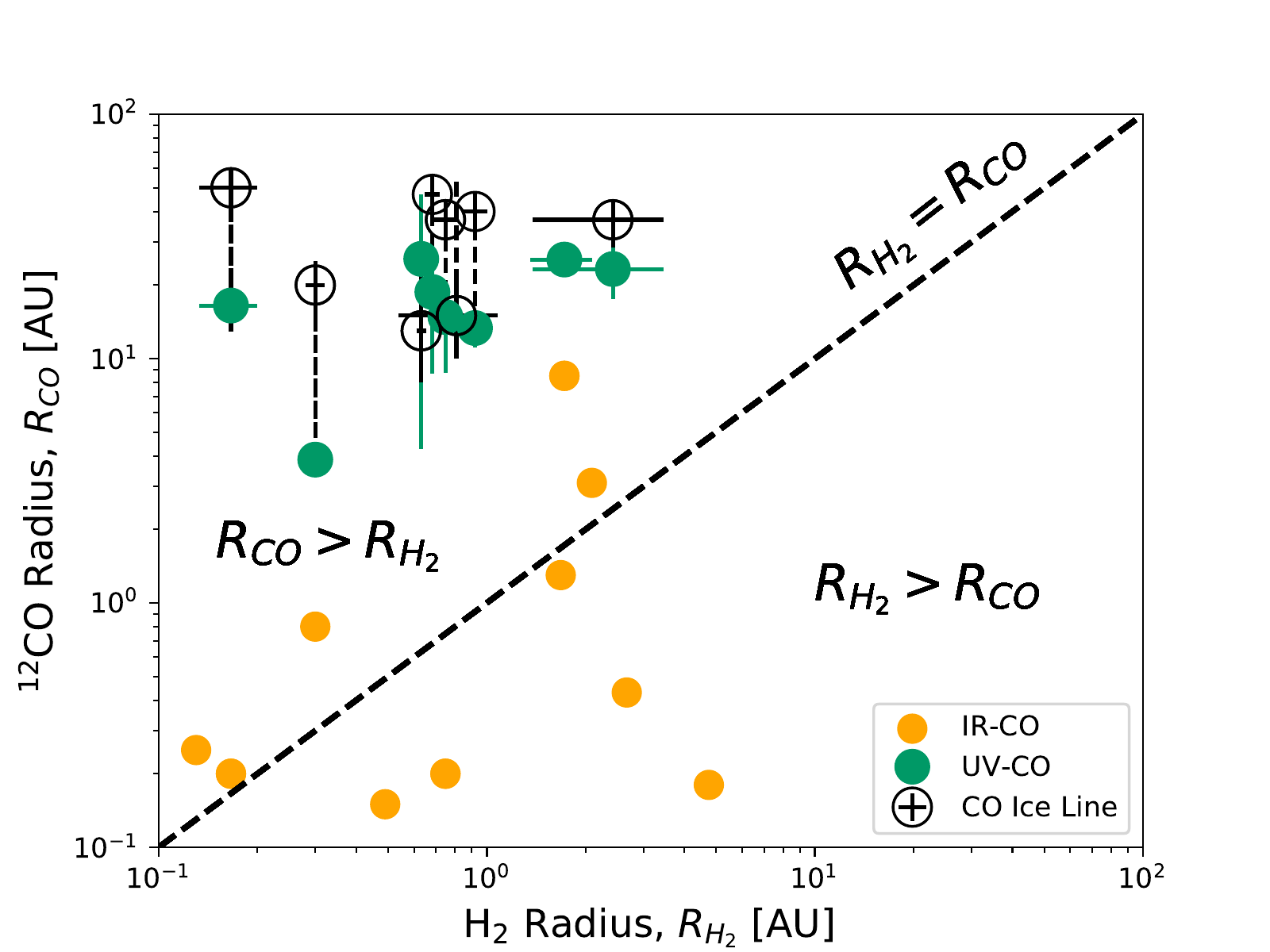}
\caption{Average gas emission radii for $^{12}$CO and H$_2$, which are also presented in Table \ref{H2_measured_props}. We find that UV-CO (teal circles; calculated in this work) originates at more distant radii than UV-H$_2$ in all systems with strong emission from states with $J'' > 10$. In all cases, this emission comes from inside or very close to the CO ice line (open circles; estimated from \citealt{Long2018}). Dashed black lines connect the UV-CO and CO ice line radii for the same target. By contrast, IR-CO emission (yellow circles; \citealt{Salyk2011CO, Brown2013}) appears to trace a population of gas at radial locations closer to the UV-H$_2$ \citep{France2012}.}
\label{Gaussian_Radii_Comp}
\end{figure*}

Consistent with previous studies \citep{Herczeg2004, France2012}, we find that the empirical UV-H$_2$ emitting radius is inside $r < 2$ AU for all targets (see Table \ref{H2_measured_props}). Figure \ref{Gaussian_Radii_Comp} compares the radial locations of UV-CO and UV-H$_2$, demonstrating that the UV-CO originates much further from the central star(s) than the UV-H$_2$ in all seven spectra with strong $J'' > 10$ emission. The UV-CO also appears to trace a different population of gas than the IR-CO emission \citep{Salyk2011CO, Brown2013}, which is produced by material that is roughly radially co-located with the UV-H$_2$ \citep{France2012}. Although the UV-CO radii are more uncertain than the UV-H$_2$ locations, even the blended sets of high $J''$ lines are narrower than the individual UV-H$_2$ features, confirming that the emitting gas is further away. For example, the strongest UV-CO blend in the DF Tau spectrum has $FWHM = 19.8$ km s$^{-1}$, compared to the average UV-H$_2$ line width of $FWHM = 63$ km s$^{-1}$. This further supports our picture of radially stratified fluorescent gas. 

We note that spatially resolved \text{HST}-STIS observations of UV-H$_2$ emission lines from DF Tau show blueshifted components at velocity shifts up to $\sim -40$ km s$^{-1}$ \citep{Herczeg2006}. This excess emission originates in a wind, which \textit{HST}-COS can only distinguish from the on-source disk emission when the source is either oriented along the dispersion direction or the wind extends beyond the 1.25'' aperture radius (see e.g., \citealt{France2011_1987A}). If unresolved wind signatures are present in UV-H$_2$ emission lines from other targets in this sample, the emitting region boundaries derived here would be biased toward smaller radii. However, \citet{Hoadley2015} found no significant asymmetries between the H$_2$ line shapes observed with \textit{HST}-COS from any targets that overlap with this work. \citet{Pontoppidan2011} also note that sub-Keplerian IR-CO emission from slow, uncollimated winds is likely always present but can be treated as an extension of the disk emission, since the gas maintains its Keplerian rotation as it travels up the wind. In the case of the UV-H$_2$ and UV-CO features from targets in our sample, the components will only be resolvable with multi-object UV spectroscopy on a future flagship mission (see e.g., \citealt{France2017_LUVOIR, Decadal2019}) or with UV spectroastrometry (see e.g., \citealt{Pontoppidan2008}).

Figure \ref{Gaussian_Radii_Comp} also includes an estimated CO ice line location for each system, derived from the relationship between stellar luminosity, disk radius, and condensation temperature presented in \citeauthor{Long2018} (\citeyear{Long2018}; $r \propto \sqrt{L_{\ast} / T_{\rm{conden}}^4}$). We find that the empirical UV-CO emission radius is closer to the star than the derived CO ice line location in all but two systems, where the UV-CO emission comes from just beyond the ice line (DM Tau and RECX-15). However, the regions of disks where temperatures are cool enough for CO freeze-out are likely 2-D surfaces instead of radial cutoffs (see e.g. \citealt{Bosman2018, Qi2019}). This thermal structure allows gas-phase CO to survive in surface layers at radii where solids have formed closer to the disk midplane. 

\begin{deluxetable*}{cccccccccc}
\tablecaption{Measured UV-H$_2$ and UV-CO FWHMs and Average Emitting Radii \label{H2_measured_props}
}
\tablewidth{0 pt}
\tabletypesize{\scriptsize}
\tablehead{ \colhead{Target} & \colhead{$\rm{FWHM}_{\rm{H_2}, [1,4]}$} & \colhead{$\left<R_{[1,4]} \right>$} & \colhead{$\rm{FWHM}_{\rm{H_2}, [1,7]}$} & \colhead{$\left<R_{[1,7]} \right>$} & \colhead{$\rm{FWHM}_{\rm{CO}, (14-4)}$\tablenotemark{a}} & \colhead{$\left<R_{CO} \right>$\tablenotemark{b}} \\
 & \colhead{[km s$^{-1}$]} & \colhead{[AU]} & \colhead{[km s$^{-1}$]} & \colhead{[AU]} & \colhead{[km s$^{-1}$]} & \colhead{[AU]} \\
}
\startdata
AA Tau (2011) & $61 \pm 9$ & $0.7 \pm 0.2$ & $60 \pm 3$ & $0.75 \pm 0.07$ & $13 \pm 3$ & $15 \pm 6$ \\
AA Tau (2013) & $43 \pm 6$ & $1.5 \pm 0.4$ & $33 \pm 7$ & $2 \pm 1$ & $11 \pm 1$ & $23 \pm 6$ \\
CS Cha & $30 \pm 11$ & $3 \pm 2$ & $20 \pm 6$ & $7 \pm 4$ & \nodata & \nodata \\
CW Tau & $59 \pm 3$ & $0.51 \pm 0.06$ & $53 \pm 9$ & $0.6 \pm 0.2$ & \nodata & \nodata \\
DF Tau & $63 \pm 3$ & $0.17 \pm 0.02$ & $64 \pm 6$ & $0.17 \pm 0.03$ & $6.4 \pm 0.6$ & $16 \pm 3$ \\
DM Tau & $32 \pm 5$ & $0.6 \pm 0.2$ & $27 \pm 4$ & $0.8 \pm 0.3$ & $3.3 \pm 0.8$ & $53 \pm 26$ \\
LkCa 15 & $59 \pm 9$ & $0.5 \pm 0.1$ & $50 \pm 1$ & $0.68 \pm 0.04$ & \nodata & \nodata \\
RECX-11 & $56 \pm 4$ & $0.8 \pm 0.1$ & $52 \pm 2$ & $0.92 \pm 0.08$ & $14 \pm 1$ & $13 \pm 2$ \\
RECX-15 (2010) & $35 \pm 4$ & $0.9 \pm 0.2$ & $41 \pm 1$ & $0.63 \pm 0.02$ & $6 \pm 3$ & $26 \pm 21$ \\
RECX-15 (2013) & $47 \pm 2$ & $0.48 \pm 0.04$ & $38 \pm 4$ & $0.7 \pm 0.1$ & $14 \pm 4$ & $6 \pm 3$ \\
RY Lupi & $51 \pm 10$ & $1.9 \pm 0.8$ & $48 \pm 5$ & $2.1 \pm 0.4$ & \nodata & \nodata \\
T Cha & $58 \pm 2$ & $1.4 \pm 0.1$ & $54 \pm 10$ & $1.7 \pm 0.6$ & \nodata & \nodata \\
UX Tau A & $31 \pm 5$ & $1.5 \pm 0.5$ & $30 \pm 3$ & $1.7 \pm 0.4$ & $7.7 \pm 0.5$ & $25 \pm 3$ \\
V4046 Sgr & $51 \pm 3$ & $0.63 \pm 0.07$ & $47 \pm 1$ & $0.74 \pm 0.03$ & $13 \pm 3$ & $6 \pm 3$ \\
\hline
\\
\textbf{Median} & \textbf{51 km s$^{-1}$} & \textbf{0.75 AU} & \textbf{48 km s$^{-1}$} & \textbf{0.78 AU} & \textbf{9 km s$^{-1}$} & \textbf{20 AU} 
\enddata
\tablenotetext{a}{Individual rotational states within the UV-CO bands are too close to each other in wavelength space to spectrally resolve with \textit{HST}-COS. Instead of measuring FWHMs directly, we fit multi-Gaussian profiles to states with $J'' > 10$, where the separation between rotational states becomes wider. The FWHMs quoted here correspond to the best-fit parameters from the multi-Gaussian models.}
\tablenotetext{b}{Targets with missing values of $\rm{FWHM}_{\rm{CO}, (14-4)}$ and $\left<R_{CO} \right>$ did not show UV-CO bands with distinct separations between sets of three high $J''$ states ($J'' > 10$; see Figure \ref{line_example}), so the multi-Gaussian fitting procedure described in Section 3 could not be applied to their spectra. Since individual UV-CO emission lines are not spectrally resolved, the radii derived here are not as robust as the locations measured from the UV-H$_2$ features.}
\end{deluxetable*}    

We also find that the UV-CO emission in the 2013 spectrum of AA Tau may originate from larger radii $\left(R_{\rm{UV-CO}} = 23 \pm 6 \text{ AU} \right)$ than the same feature in the 2011 spectrum of AA Tau $\left(R_{\rm{UV-CO}} = 15 \pm 6 \text{ AU} \right)$. These empirical $R_{\rm{UV-CO}}$ values are the medians of all radii where emitting gas is present. The differences in the AA Tau spectra therefore imply that contributions from hot gas in the inner disk appear more prominently in the emission lines from 2011 than in those from 2013, shifting the empirical emitting radius to a smaller value in 2011. This effect was also seen in the UV-H$_2$ emission lines from the two sets of spectra, with the 2011 spectrum showing an empirical UV-H$_2$ emitting radius of $R_{\rm{UV-H_2}} = 0.7 \pm 0.2$ and the 2013 observations showing $R_{\rm{UV-H_2}} = 1.5 \pm 0.4$. 

Changes between the 2011 and 2013 AA Tau UV-H$_2$ spectra were attributed to extra absorbing material that moved into the line-of-sight after 2011 \citep{Schneider2015, Hoadley2015}, perhaps caused by an increase in the inner disk scale height \citep{Covey2021}. In the case of the UV-CO, the occulting material masks the contribution from the innermost emitting gas, leading to narrower line profiles that are dominated by more distant molecules. This shifts the empirical radius $\left(R_{\rm{UV-CO}} \right)$ further out in the disk in 2013, with the dominant contribution coming from closer to the CO ice line. Future \textit{HST} and \textit{JWST} observations of AA Tau's mysterious geometry will provide a more detailed picture of interactions between the occulting material, stellar radiation field, and molecular gas disk.   

\section{2-D Radiative Transfer Models of UV-H$_2$ and UV-CO}

In order to derive more detailed radial flux distributions from the fluorescent emission lines, we fit 2-D radiative transfer models to the UV-H$_2$ and UV-CO spectra. We build on the models developed by \citet{Hoadley2015}, which create UV-H$_2$ emission line profiles by distributing Ly$\alpha$ photons to a population of hot gas in a thin surface layer of the disk. This version efficiently accounts for UV-H$_2$ lines from four different progressions, pumped by photons with four different wavelengths along the red side of the Ly$\alpha$ profile. Since the [1,4] progression has the strongest signal-to-noise across the full sample of disks, we focus on fitting the radiative transfer models to those features. This allows us to extract the most reliable information on the underlying radial distributions of fluorescent gas while maintaining consistency with \citet{Hoadley2015} and minimizing the required computation time to find the best-fit models. 

The $\left(v'-v'' \right) = (14-3)$ and $\left(v'-v'' \right) = (14-4)$ bands of UV-CO each consist of lines pumped by $>$60 different Ly$\alpha$ wavelengths $\left( \lambda = 1214.2-1218.8 \text{ \AA} \right)$. Rather than adding the UV-CO radiative transfer properties to the existing code for UV-H$_2$, the models were adapted into an object-oriented framework that can accommodate multiple molecular species. A comparison of UV-H$_2$ models from the two versions (see Appendix A) demonstrates that both yield similar results that are consistent with those presented in \citet{Hoadley2015}.    

\subsection{Description of the Models}

Each radiative transfer model is built on the same 2-D disk structure described by \citet{Hoadley2015}, which starts with a radial temperature gradient and pressure scale height. The temperature gradient is normalized to $T_{1 AU}$ and decays with power law exponent $q$:
\begin{equation}
T \left( r \right) = T_{1 AU} \left( \frac{r}{1 \, \text{AU}} \right)^{-q}
\end{equation}
The pressure scale height follows as:
\begin{equation}
H_p = \sqrt{ \frac{k T \left( r \right)}{\mu m_H} \times \frac{r^3}{G M_{\ast}}},
\end{equation}
where $k$ and $G$ are the Boltzmann and gravitational constants, $\mu = 2.33$ is the mean molecular weight of the gas, $m_H$ is the mass of a single hydrogen atom, and $M_{\ast}$ is the stellar mass. The 2-D mass density distribution of gas is then calculated as
\begin{equation}
\rho \left(r, z \right) = \frac{\Sigma \left( r \right)}{\sqrt{2 \pi} H_p} \times \exp \left[-0.5 \left( \frac{z}{H_p} \right)^2 \right]
\end{equation}
with radial surface density distribution
\begin{equation}
\Sigma \left( r \right) = \Sigma_c \left( \frac{r}{r_c} \right)^{-\gamma} \exp \left[ -\left( \frac{r}{r_c} \right)^{2-\gamma} \right].
\end{equation}
The characteristic radius $\left( r_c \right)$ is the location in the disk where the density distribution begins to fall off exponentially. A power-law with coefficient $\gamma$, which describes the radial surface density distribution $\Sigma \left(r \right)$ for a disk with kinematic viscosity $\nu \propto r^{-\gamma}$ \citep{LyndenBell1974}, is used to map the distribution interior to $r_c$.  

Number densities of both H$_2$ and CO in ground states with vibrational level $v$ and rotational level $J$ $\left( \left[v, J \right] \right)$ are calculated from the mass density distribution under LTE conditions:
\begin{equation}
n_{[v, J]} \left(r, z \right) = X_{H_2, CO} \times \frac{ \rho \left(r, z \right)}{\mu m_H} \times g_{[v, J]} \times \frac{\exp{ \left[ -E_[v, J] / k T \left(r \right) \right]}}{Z_{[v, J]} \left(T \right)},
\end{equation}  
where $X_{H_2, CO}$ represents the fraction of the total gas density consisting of H$_2$ or CO, respectively, $g_{[v, J]}$ is the statistical weight of state $[v, J]$, and $Z_{[v, J]}$ is the partition function. Optical depths in each ground state are then calculated as
\begin{equation}
\tau_{\lambda} \left(r, z \right) = \sum_{z}^{z-H_p} z \sigma n_{[v, J]} \left(r, z \right),
\end{equation}
where $\sigma$ is the cross section for absorbing a Ly$\alpha$ photon at the pumping wavelength. However, overlap between cross sections at adjacent pumping wavelengths can lead to an overestimate of the total optical depth, by allowing the same photon to be absorbed by molecules in multiple rovibrational levels of the ground electronic state. To correct for this effect, we scale each $\tau$ by the fractional optical depth expected for that transition at the temperature $\left(T \right)$ and column density $\left( N \right)$ corresponding to its grid point \citep{McJunkin2014, Liu1996, Wolven1997}, such that 
\begin{equation}
\tau_{eff} \left(r, z \right) = \tau_{\lambda} \left(r, z \right) \times \frac{\tau_{\lambda} \left(r, z \right)}{\tau_{all} \left( T, N \right)}.
\end{equation}
The total flux for each transition out of the Ly$\alpha$-pumped upper level is then calculated as
\begin{equation}
F_{\lambda} \left(r, z \right) = \eta F_{\ast, Ly\alpha} \left( \frac{R_{\ast}}{r} \right)^2 \left( \frac{r \times \cos i_{disk}}{s \left(r, z \right)} \right)^2 B_{mn} \sum_{\tau_{eff}} \left(1 - \text{e}^{- \tau_{eff}} \right),
\end{equation}
where $\eta$ is a geometric filling factor describing the fraction of gas exposed to Ly$\alpha$ photons \citep{Herczeg2004}, $F_{\ast, \rm{Ly}\alpha}$ is the relevant Ly$\alpha$ pumping flux at grid point $\left(r, z \right)$, $s \left(r, z \right)$ is the sightline from the observer to a gas parcel in the disk, and $B_{mn}$ describes the probability of transitioning from $\left[v', J' \right] \rightarrow \left[v'', J'' \right]$. We do not fit for $\eta$ directly in this analysis; rather, the value is folded into the total Ly$\alpha$ pumping flux reaching the gas disk. Emission line profiles are calculated by collapsing this 2-D grid in Keplerian velocity space and convolving the models with the \emph{HST}-COS LSF. We compare the UV-H$_2$ models to the data in velocity space, but the model UV-CO line profiles are first superimposed in wavelength space to produce a band of emission lines. The observed UV-H$_2$ emission lines are all shifted to radial velocities of zero before fitting the radiative transfer models to the data, based on the central wavelengths from the Gaussian models. For the UV-CO features, we shift the observed bandheads $\left(J'' = 0 \right)$ to radial velocities of zero. The impact of this shift on the modeling results is discussed further in Section 5.4.

\subsection{Reconstructed Ly$\alpha$ Profiles}

Although the Ly$\alpha$ line profile is included in the \emph{HST}-COS bandpass, the core of the feature is always attenuated by circumstellar and interstellar absorption (see e.g., \citealt{McJunkin2014}). The line is further contaminated by geocoronal emission, making the high-velocity wings the only portions that are directly observable \citep{McJunkin2014}. However, the line profile can be reconstructed using the emission from fluorescent UV-H$_2$ lines \citep{Herczeg2004, Schindhelm2012}. Progression fluxes, calculated by integrating over all UV-H$_2$ features from the same upper level $i$ $\left(F_{H_2, i} \right)$, are treated as ``y'' data points. Ly$\alpha$ pumping wavelengths act as corresponding ``x'' values, such that $\left(x_i, y_i \right) = \left( \lambda_{Ly\alpha, i}, F_{H_2, i} \right)$. The data points are then fitted with a model profile, assumed to be an outflow-absorbed Gaussian emission line ($I_{\rm{Ly}\alpha}$; \citealt{Schindhelm2012}):
\begin{equation}
I_{\rm{abs}} \left( \lambda \right) = I_{\rm{Ly}\alpha} \left( \lambda \right) \times \exp^{-\tau_{\rm{out}}} \\  
\end{equation}
with $\tau_{\rm{out}}$ calculated as a Voigt profile centered at outflow velocity $v_{out}$ with H I column density $N_{out}$. 

To convert the progression fluxes to intensities that can be fit with the Gaussian model, each $y_i$ data point is divided by an equivalent width calculated for Ly$\alpha$-absorbing H$_2$ with temperature $T_{H_2}$ and column density $N_{H_2}$
\begin{equation}
W_{\lambda_{\rm{pump}}} = \int_{\lambda_{\rm{pump}}} \left(1 - \exp^{-\tau_{\lambda_{\rm{pump}}} \left[\rm{T_{H_2}}, \rm{N_{H_2}} \right]} \right).
\end{equation} 
We applied this Ly$\alpha$ reconstruction method to obtain profiles for CW Tau, T Cha, and the 2013 AA Tau spectrum, which had not yet been modeled in the literature. The remaining model profiles were taken from previous work that utilized the same procedure \citep{Schindhelm2012, France2014, Arulanantham2018}. Uncertainties in the best-fit reconstruction parameters typically make the total reconstructed Ly$\alpha$ fluxes accurate to within $\sim$20\% \citep{Schindhelm2012}.

\subsection{UV-H$_2$ Model Fitting Procedure}

UV-H$_2$ models were fit to emission lines from the [1,4] progression by allowing six parameters to vary \citep{Hoadley2015}: 
\begin{itemize}
\item height of the emitting layer $\left(2 \, H_p < z / r < 7 \, H_p \right)$
\item temperature at 1 AU $\left(500 \, \rm{K} < T_{1 \, \rm{AU}} < 5000 \, \rm{K} \right)$
\item power law exponent for the temperature distribution $\left(-2.5 < q < 2.5 \right)$
\item power law exponent for the surface density distribution $\left( 0 < \gamma < 2 \right)$
\item radius where the surface density distribution turns over from a power law to an exponential decline $\left(0.1 \, \rm{AU} < r_{\rm{char}} < 20 \, \rm{AU} \right)$
\item total mass of H$_2$ $\left(1 \times 10^{-5} \, M_{\odot} < M_{\rm{H_2}} < 0.1 \, M_{\odot} \right)$ 
\end{itemize} 
Although the model can also generate emission lines from the [1,7], [0,1], and [0,2] progressions, we follow the same approach as \citet{Hoadley2015} and focus on the [1,4] features with the strongest signal-to-noise across the full sample of targets. Since the UV-H$_2$ line shapes do not change significantly between progressions (see e.g., \citealt{Hoadley2015, Arulanantham2018}), we are able to extract sufficient information about the underlying distributions of emitting gas from this single progression.

We obtain a best-fit UV-H$_2$ model for each disk by minimizing the mean square error ($MSE$):  
\begin{equation}
MSE = \frac{1}{N} \sum \limits_{i = 0}^{N} \left(y_i - \tilde{y_i} \right)^2,
\end{equation}  
which describes the goodness-of-fit of the model profiles ($\tilde{y_i}$) to the observed UV-H$_2$ emission lines $\left(y_i \right)$. We choose the $MSE$ over test statistics that include the measurement uncertainties, which are estimated by the \textit{HST}-COS pipeline and are not necessarily derived from a distribution that represents the true errors. This makes it especially difficult to find robust fits for targets with weak but distinct UV-H$_2$ features, since anomalously small uncertainties in the continuum can bias test statistics (e.g. $\chi^2$) toward models that predict non-detections. Uncertainties on the best-fit model parameters, which contribute more to the total error than the measurement errors, were estimated by using a Markov chain Monte Carlo (MCMC) sampler to explore the parameter space defined above \citep{emcee2013}. 

Models that provide the closest match to the observed UV-H$_2$ emission lines are representative of the underlying radial distributions of flux from hot, Ly$\alpha$-pumped H$_2$ in the inner disk (see Table \ref{best_fit_H2}). However, the surface gas masses and scale heights are often degenerate, as are the temperatures at 1 AU and the temperature power-law exponents \citep{Hoadley2015}. Furthermore, several targets in our sample have UV-H$_2$ line profiles with unusually broad wings relative to the narrow widths of the line cores (e.g., DF Tau, RY Lupi; see \citealt{Herczeg2006, Banzatti2015, Arulanantham2018}, Hoadley et al. in prep). This makes it difficult for a single model to simultaneously capture both the broad wing emission from material close to the star and the narrow core emission from more distant gas. Both the degenerate model parameters and occasionally non-standard line shapes make it difficult to directly extrapolate uncertainties in the radial flux distributions from the uncertainties in the best-fit model parameters. 

Instead, we analyze the radial flux distributions through the 100 models with the smallest $MSE$ values for each target, which represent the top 1\% of all models generated during the MCMC resampling. Taken together, this group of models includes combinations of parameters that fit both the broad and narrow emission line components in DF Tau and RY Lupi. For all targets, the sets of 100 best-fit models include the degenerate parameter combinations of $z/r$, $M_{H_2}$, $T_{1 \, \text{AU}}$, and $q$. We produce a final radial flux distribution for each system by calculating the median flux value from the set of 100 models at each radial grid point. This method does not necessarily capture a perfect best-fit model, but it allows us to compare the flux distributions while accounting for degenerate model parameters. All 100 models are also overplotted against the observed emission lines in Appendix A, demonstrating that the set collectively accounts for nearly all of the UV-H$_2$ emission from each target. Although we used a different mass parameterization than \citet{Hoadley2015}, the UV-H$_2$ flux distributions presented here follow the evolutionary trends discovered in that work (see Appendix A). 

Finally, we note that the radiative transfer models always produce double-peaked Keplerian line profiles, while the data generally show single-peaked emission lines. Previous work has attributed this to slow, uncollimated winds that fill in the low velocity emission and can be treated as extensions of the disk surfaces (see e.g., \citealt{Pontoppidan2011, Hoadley2015}). However, the contributions from multiple scale heights likely impose some velocity signatures on the observed emission lines (see e.g., \citealt{Weber2020}) that are not accounted for in our modeling approach.          

\begin{deluxetable*}{ccccccc}
\tablecaption{Best-Fit Parameters for UV-H$_2$ Radiative Transfer Models \label{best_fit_H2}
}
\tablewidth{0 pt}
\tabletypesize{\scriptsize}
\tablehead{ \colhead{Target} & \colhead{$T_{1 \text{ AU}}$} & \colhead{$q$} & \colhead{$r_{char}$} & \colhead{$r_{in}$\tablenotemark{a}} & \colhead{$r_{peak}$\tablenotemark{b}} & \colhead{$r_{out}$\tablenotemark{c}} \\
 & \colhead{[K]} & & \colhead{[AU]} & \colhead{[AU]} & \colhead{[AU]} & \colhead{[AU]} \\
}
\startdata
AA Tau (2011) & $2700^{+900}_{-1200}$ & $0.3^{+0.6}_{-0.26}$ & $14 \pm 6$ & $0.2^{+0.2}_{-0.1}$ & $0.34^{+0.01}_{-0.1}$ & $2^{+5}_{-1}$ \\
AA Tau (2013) & $2400^{+1200}_{-900}$ & $0.45^{+0.55}_{-0.4}$ & $17^{+3}_{-6}$ & $0.5 \pm 0.4$ & $0.7^{+0.3}_{-0.06}$ & $4^{+8}_{-2}$ \\
CS Cha & $3600 \pm 900$ & $0.15^{+0.7}_{-0.04}$ & $19.3^{+0.6}_{-8}$ & $0.2^{+0.2}_{-0.15}$ & $0.4 \pm 0.1$ & $3^{+6}_{-1}$ \\
CW Tau & $3200^{+800}_{-1200}$ & $0.6^{+0.4}_{-0.5}$ & $7.3^{+5}_{-7}$ & $0.1^{+0.1}_{-0.06}$ & $0.26^{+0.03}_{-0.1}$ & $2^{+3}_{-1}$ \\
DF Tau & $1300^{+1200}_{-700}$ & $-0.3^{+0.5}_{-1.7}$ & $17^{+3}_{-5}$ & $0.04^{+0.02}_{-0}$ & $0.04^{+0.03}_{-0}$ & $2^{+5}_{-1}$ \\
DM Tau & $3300 \pm 800$ & $0.7^{+0.5}_{-1}$ & $10 \pm 6$ & $0.19 \pm 0.08$ & $0.26 \pm 0.05$ & $3^{+2}_{-1}$ \\
LkCa 15 & $2400^{+800}_{-1300}$ & $0.14^{+1.1}_{-0.1}$ & $15 \pm 5$ & $0.2^{+0.2}_{-0.15}$ & $0.3^{+0.3}_{-0.07}$ & $2^{+8}_{-1}$ \\
RECX-11 & $2400^{+600}_{-800}$ & $0.3^{+0.5}_{-1.1}$ & $6 \pm 5$ & $0.1^{+0.2}_{-0.05}$ & $0.25^{+0.02}_{-0.09}$ & $5^{+6}_{-3}$ \\
RECX-15 (2010) & $4100^{+800}_{-1100}$ & $0.9^{+0.4}_{-0.7}$ & $15 \pm 5$ & $0.06^{+0.1}_{-0.02}$ & $0.2^{+0.02}_{-0.1}$ & $5^{+6}_{-3}$ \\
RECX-15 (2013) & $2100 \pm 1200$ & $0.2^{+0.7}_{-1}$ & $5^{+5}_{-4.8}$ & $0.06^{+0.1}_{-0.02}$ & $0.15^{+0.05}_{-0.07}$ & $4^{+5}_{-2}$ \\
RY Lupi & $1900^{+1200}_{-800}$ & $0.3^{+0.5}_{-0.7}$ & $11 \pm 6$ & $0.3 \pm 0.2$ & $0.5^{+0.05}_{-0.15}$ & $11^{+6}_{-5}$ \\
T Cha & $4000^{+900}_{-1300}$ & $-0.4^{+1.8}_{-1}$ & $12^{+7}_{-5}$ & $0.7^{+0.2}_{-0.6}$ & $1^{+0}_{-0.05}$ & $3^{+4}_{-1}$ \\
UX Tau A\tablenotemark{d} & $3200^{+700}_{-1400}$ & $-0.7^{+1.8}_{-0.6}$ & $15 \pm 5$ & $0.5^{+0.1}_{-0.2}$ & $0.8^{+0}_{-0.2}$ & $3 \pm 1$ \\
V4046 Sgr & $2200^{+1000}_{-700}$ & $0.5^{+0.8}_{-0.3}$ & $17^{+3}_{-7}$ & $0.3^{+0.1}_{-0.2}$ & $0.4^{+0}_{-0.05}$ & $2^{+3}_{-1}$ \\
\hline
\\
\textbf{Median} & \textbf{2550 K} & \textbf{0.3} & \textbf{14.5 AU} & \textbf{0.195 AU} & \textbf{0.3 AU} & \textbf{3.3 AU} 
\enddata
\tablenotetext{a}{5\% of the total UV-H$_2$ flux is contained within this radius.}
\tablenotetext{b}{Radial location of UV-H$_2$ flux peak.}
\tablenotetext{c}{95\% of the total UV-H$_2$ flux is contained within this radius.}
\tablenotetext{d}{Bi-modal posterior distributions are retrieved for $T$, and $q$. The reported best-fit values are taken from the mode with the highest peak, and the uncertainties are extended to include the degenerate set of parameters.}
\end{deluxetable*}

\subsection{UV-CO Model Fitting Procedure}

The main difference in the treatment of UV-H$_2$ versus UV-CO is the temperature distribution within the fluorescent gas. \citet{Herczeg2004} found that the UV-H$_2$ features from TW Hya were consistent with emission from a gas layer with $T \sim 2000-3000$ K and $\log N \left( \rm{H_2} \right) = 18-20$, so the models presented here allow the temperature of the H$_2$ distribution to reach as high as the 5000 K dissociation limit \citep{Hoadley2015}. Assuming a CO/H$_2$ ratio of $10^{-4}$ \citep{France2014_COH2}, the corresponding column densities of CO would be $\log N \left( \rm{CO} \right) \sim 14-16$ in the 2500 K layer. 

However, cooling via CO ro-vibrational emission becomes particularly efficient in regions where H I abundances reach the threshold for molecule formation to begin $\left(\log N \left( \rm{H} \right) \sim 21 \right)$. The gas temperature is then quickly reduced to $\sim$1000 K in denser surface layers of the disk \citep{Adamkovics2014}. This implies that hotter layers, where the UV-H$_2$ emission originates, do not have significant enough populations of CO to either cool the gas or produce detectable UV-CO fluorescence. 

These column density and temperature conditions are also consistent with 4.7 $\mu$m CO rovibrational emission in surface layers of the inner disk. The rotational temperatures of these features never exceed $T \sim 1800$ K and are typically much cooler than the $T \sim 1500$ K threshold required for Ly$\alpha$ pumping of H$_2$ \citep{Salyk2011CO}. Since the IR-CO features originate much closer to the star than the UV-CO $\left(R_{IR-CO} \sim 0.1-5 \text{ AU; see Figure \ref{Gaussian_Radii_Comp}, \citealt{France2012}} \right)$, where gas heating via stellar irradiation is highest, the temperature of this warm gas provides a rough upper limit for the temperature distribution of UV-CO. The range of temperatures within the UV-CO emitting region is therefore smaller than what is expected for UV-H$_2$, including gas between 20-1800 K to also avoid temperatures where CO is expected to be frozen out of the gas phase (see e.g., \citealt{Bosman2018}). 

Since the UV-CO and UV-H$_2$ emission lines trace radially separated regions of the gas disk, we expect the Ly$\alpha$ radiation field to look different to each molecular species. To account for this, we add two additional variable parameters to the UV-CO models: the column densities and radial velocities of H I between the UV-H$_2$ and UV-CO emitting regions. We restrict the H I column densities to range between the values that best fit the UV-H$_2$ progression fluxes \citep{Schindhelm2012} and those estimated for the total ISM absorption \citep{McJunkin2016}. A new Ly$\alpha$ absorption feature is then superimposed on the intrinsic Gaussian emission line generated by the H$_2$ reconstruction procedure. 

In order for Ly$\alpha$ photons to reach warm CO at radii between $r = 15$ and $r = 30$ AU, the flared disk geometry forces the UV-CO emitting region to higher values of $z/r$ and larger characteristic radii than the inner disk UV-H$_2$. The variable parameters for UV-CO are then:   
\begin{itemize}
\item height of the emitting layer $\left(7 \, H_p < z / r < 11 \, H_p \right)$
\item temperature at 1 AU $\left(20 \, \rm{K} < T_{1 \, \rm{AU}} < 1800 \, \rm{K} \right)$
\item power law exponent for the temperature distribution $\left(-2.5 < q < 2.5 \right)$
\item power law exponent for the surface density distribution $\left( 0 < \gamma < 2 \right)$
\item radius where the surface density distribution turns over from a power law to an exponential decline $\left(0.1 < r_{\rm{char}} < 100 \, \rm{AU} \right)$
\item total mass of CO $\left(10^{-20} \, M_{\odot} < M_{\rm{CO}} < 10^{-5} \, M_{\odot} \right)$ 
\item column density of intervening H I (determined individually for each target)
\item radial velocity of intervening H I ($-300 < v < 300$ km s$^{-1}$)
\end{itemize} 

Best-fit models were again obtained by minimizing the $MSE$ test statistic. Uncertainties on the model parameters were derived by carrying out MCMC resampling, using 1000 walkers to explore the full parameter space described above (\citealt{emcee2013}; see Table \ref{best_fit_CO}). Model fits to the UV-H$_2$ and UV-CO emission lines from each individual target are provided in Appendix B, along with the corresponding radial flux distributions. 

\begin{deluxetable*}{ccccccc}
\tablecaption{Best-Fit Parameters for UV-CO Radiative Transfer Models \label{best_fit_CO}
}
\tablewidth{0 pt}
\tabletypesize{\scriptsize}
\tablehead{ \colhead{Target} & \colhead{$T_{10 \text{ AU}}$} & \colhead{$q$} & \colhead{$r_{char}$} & \colhead{$r_{in}$\tablenotemark{a}} & \colhead{$r_{peak}$\tablenotemark{b}} &\colhead{$r_{out}$\tablenotemark{c}} \\
 & \colhead{[K]} & & \colhead{[AU]} & \colhead{[AU]} & \colhead{[AU]} & \colhead{[AU]} \\
}
\startdata
AA Tau (2011) & $950^{+90}_{-400}$ & $-0.8^{+0.2}_{-0.3}$ & $25^{+20}_{-14}$ & $0.16_{-0.12}^{+0.84}$ & $1.8^{+0.5}_{-0}$ & $11^{+4}_{-3}$ \\
AA Tau (2013) & $880 \pm 110$ & $-0.7^{+0.35}_{-0.2}$ & $47^{+50}_{-22}$ & $0.1^{+0.2}_{-0.06}$ & $1.6^{+0.7}_{-1.0}$ & $10^{+2}_{-4}$ \\
CS Cha & $1300 \pm 300$ & $-0.6^{+0.4}_{-0.3}$ & $50^{+10}_{-30}$ & $0.04^{+0.2}_{-0.01}$ & $0.4 \pm 0.1$ & $9^{+30}_{-4}$ \\
CW Tau & $1000 \pm 100$ & $0.15^{+0.03}_{-0.05}$ & $99^{+1}_{-5}$ & $0.04^{+0.06}_{-0}$ & $0.04^{+0.06}_{-0}$ & $7 \pm 2$  \\
DF Tau & $1100^{+150}_{-800}$ & $-0.038 \pm 0.1$ & $38^{+50}_{-10}$  & $0.04^{+0.02}_{-0}$ & $0.04^{+0.02}_{-0}$ & $1.4 \pm 0.4$  \\
DM Tau & $1700^{+100}_{-600}$ & $-1.0^{+0.5}_{-0.8}$ & $40^{+20}_{-30}$ & $0.07^{+1}_{-0.03}$ & $1^{+0.6}_{-0.9}$ & $10^{+14}_{-2}$  \\
LkCa 15 & $400^{+400}_{-350}$ & $-0.6^{+3}_{-0.4}$ & $69^{+30}_{-20}$ & $0.2^{+6}_{-0.15}$ & $2^{+5}_{-1}$ & $14^{+20}_{-5}$  \\
RECX-11 & $1700 \pm 100$ & $-1.1^{+0.7}_{-0.4}$ & $9^{+10}_{-4}$ & $0.2^{+0.2}_{-0.15}$ & $1.4 \pm 1$ & $7 \pm 1$ \\
RECX-15 (2010) & $560^{+500}_{-200}$ & $-0.46 \pm 0.1$ & $62^{+30}_{-10}$  & $0.05 \pm 0.02$ & $0.65^{+0.3}_{-0.05}$ & $12^{+6}_{-5}$  \\
RECX-15 (2013) & $540^{+500}_{-200}$ & $-0.4^{+0.2}_{-0.1}$ & $67^{+30}_{-10}$  & $0.043^{+0.006}_{0.004}$ & $0.39^{+0.4}_{-0.08}$ & $9^{+3}_{-5}$ \\
RY Lupi & $1600^{+200}_{-500}$ & $-0.7^{+0.65}_{-0.3}$ & $7^{+35}_{-6}$ & $0.04^{+0.2}_{-0}$ & $0.06^{+0.2}_{-0.02}$ & $6^{+6}_{-1}$ \\
T Cha & $1100^{+500}_{-700}$ & $-0.96^{+0.4}_{-1.3}$ & $63 \pm 30$ & $0.04^{+0.3}_{-0}$ & $0.1^{+0.3}_{-0.06}$ & $6^{+8}_{-2}$  \\
UX Tau A & $1600^{+200}_{-500}$ & $-1.2^{+1}_{-0.6}$ & $24 \pm 20$ & $0.8^{+8}_{-0.7}$ & $3^{+6}_{-1}$ & $12^{+50}_{-2}$  \\
V4046 Sgr & $1550 \pm 80$ & $-0.9 \pm 0.1$ & $34^{+5}_{-9}$ & $0.08^{+0.6}_{-0.04}$ & $0.75^{+0.25}_{-0.06}$ & $7^{+2}_{-1}$ \\
\hline
\\
\textbf{Median} & \textbf{1100 K} & \textbf{-0.7} & \textbf{44 AU} & \textbf{0.06 AU} & \textbf{0.7 AU} & \textbf{9 AU} 
\enddata
\tablenotetext{a}{5\% of the total UV-CO flux is contained within this radius.}
\tablenotetext{b}{Radial location of UV-CO flux peak.}
\tablenotetext{c}{95\% of the total UV-CO flux is contained within this radius.}
\end{deluxetable*}

We find that the median temperature of fluorescent CO at 10 AU is $\sim$1100 K, in very good agreement with models of the disk surface that predict similar temperatures in layers where CO column densities begin to increase \citep{Adamkovics2014}. We also find that 95\% of the UV-CO flux is typically contained within $r_{out} \sim 9$ AU, demonstrating that the emitting regions extend roughly three times further than the radial distributions of UV-H$_2$ $\left(r_{out} \sim 3 \, \rm{AU} \right)$. However, $r_{out}$ for the UV-CO is only about half the size of the empirical radius presented in Table \ref{H2_measured_props} $\left(\left<R_{CO} \right> \sim 20 \, \rm{AU} \right)$.

The number densities of intervening H I between the UV-H$_2$ and UV-CO emitting regions were also calculated for each target (see Table \ref{HI_checkpoints}). Values range between $n_{\rm{H I}} \sim 0.1-3.4 \times 10^6$ cm$^{-3}$, falling well within constraints on the total densities expected to be driven out by inner disk winds (see e.g., \citealt{Turner1999}). However, we note that the number densities were calculated assuming the difference in $r_{out}$ from the model distributions of UV-H$_2$ and UV-CO flux. Any systematic uncertainties in that analysis will then propagate through the number density calculations.

\begin{deluxetable*}{ccccccc}
\tablecaption{H I Column Densities at Various Disk Checkpoints \label{HI_checkpoints}
}
\tablewidth{0 pt}
\tabletypesize{\scriptsize}
\tablehead{ \colhead{Target} & \colhead{$N \left( \rm{H \, I} \right)$, UV-H$_2$\tablenotemark{a}} & \colhead{$N \left( \rm{H \, I} \right)$, UV-CO\tablenotemark{b}} & \colhead{$\Delta N\left(\rm{H \, I} \right) \left(\rm{CO}-\rm{H}_2 \right)$} & \colhead{$N \left( \rm{H \, I} \right)$, total\tablenotemark{c}} & \colhead{$R_{\rm{CO}} - R_{\rm{H}_2}$\tablenotemark{e}} & \colhead{$n_{\rm{H I}}$\tablenotemark{f}} \\
 & \colhead{[dex]} & \colhead{[dex]} & \colhead{[dex]} & \colhead{[dex]} & \colhead{[AU]} & \colhead{[10$^6$ cm$^{-3}$]} \\
}
\startdata
AA Tau (2011) & $19.05 \pm 0.15$ & $20.37^{+0.36}_{-0.46}$ & 1.3 & $20.73^{+0.03}_{-0.13}$ & 9 & 1.7 \\
AA Tau (2013) & $18.29 \pm 0.3$ & $19.93^{+0.8}_{-0.4}$ & 1.6 & \nodata & 6 & 0.9 \\
CS Cha & $19.05 \pm 0.15$ & $19.4 \pm 0.4$ & 0.35 & $20.4^{+0.03}_{-0.15}$ & 6 & 0.15 \\
CW Tau & $19.2 \pm 0.8$ & $20.29^{+0.08}_{-0.07}$ & 1.09 & \nodata & 5 & 2.4 \\
DF Tau & $19.5 \pm 0.2$ & $19.57^{+0.34}_{-1}$ & 0.07 & $20.93 \pm 0.03$ & -0.6 & 0.62 \\
DM Tau & $18.59 \pm 0.03$ & $19.4^{+0.5}_{-0.6}$ & 0.8 & $20.88^{+0.05}_{-0.09}$ & 7 & 0.20 \\
LkCa 15 & $18.9 \pm 0.3$ & $19.69^{+0.4}_{-0.7}$ & 0.8 & $20.68^{+0.15}_{-0.09}$ & 12 & 0.23 \\
RECX-11 & $18.89 \pm 0.4$ & $19.18^{+0.38}_{-0.15}$ & 0.3 & $19.7^{+0.03}_{-0.2}$ & 2 & 0.25 \\
RECX-15 (2010) & $18.98 \pm 0.22$ & $19.25^{+0.12}_{-0.25}$ & 0.27 & $19.58^{+0.05}_{-0.18}$ & 7 & 0.08 \\
RECX-15 (2013) & $18.98 \pm 0.22$ & $19.23^{+0.13}_{-0.2}$ & 0.25 & \nodata & 5 & 0.1 \\
RY Lupi & $19.1^{+0.4}_{-0.2}$ & $19.1^{+0.3}_{-0.1}$ & 0 & 20.34\tablenotemark{d} & -5 & 0 \\
T Cha & $18.9 \pm 0.6$ & $20.20^{+0.7}_{-0.5}$ & 1.3 & \nodata & 3 & 3.4 \\
UX Tau A & $18.26 \pm 0.12$ & $19.4^{+0.5}_{-0.7}$ & 1.1 & $20.9^{+0.05}_{-0.17}$ & 9 & 0.17 \\
V4046 Sgr & $18.81 \pm 0.22$ & $19.83^{+0.01}_{-0.05}$ & 1.02 & $19.85^{+0.03}_{-0.28}$ & 5 & 0.82 \\
\hline
\\
\textbf{Median} & \textbf{18.94} & \textbf{19.49} & \textbf{0.8} & \textbf{20.54} & \textbf{6 AU} & $\mathbf{2.4 \times 10^5}$ \textbf{cm$^{-3}$} 
\enddata
\tablenotetext{a}{H I column densities from \citealt{Schindhelm2012, Arulanantham2020}, corresponding to absorption between the accretion shock and UV-H$_2$ emitting region.}
\tablenotetext{b}{H I column densities measured in this work, corresponding to absorption between the UV-H$_2$ and UV-CO emitting regions.}
\tablenotetext{c}{H I column densities from \citealt{McJunkin2014}; These values are dominated by ISM absorption.}
\tablenotetext{d}{H I column density estimated from \citealt{France2017}}
\tablenotetext{e}{Difference in outer radii of UV-CO and UV-H$_2$ model flux distributions, which contain 95\% of the fluorescent emission. DF Tau and RY Lupi both have negative values, implying that the UV-H$_2$ emitting region is further from the star than the UV-CO. However, this is likely caused by underfitting the high-velocity UV-H$_2$ line wings, which would drop emission at radii close to the star from the model flux distributions.}
\tablenotetext{f}{H I number densities were calculated using $R_{\rm{CO}} - R_{\rm{H}_2}$. Any systematic uncertainties in the estimates for those radii will then propagate through the number density calculations.}
\end{deluxetable*}

Although the 2-D radiative transfer approach provides good fits to the UV-H$_2$ emission lines, we find that even the best-fit models are unable to fully capture the UV-CO bands. The largest discrepancies between the data and models are seen between $\lambda \sim 1316.2-1317$ \AA, where the models significantly underpredict the UV-CO flux from AA Tau, CW Tau, RECX-15, and DF Tau. The models also often appear to miss the line centers in states with $J'' > 10$, showing emission lines that reproduce either the long or short wavelength sides of the features but not both (see e.g., AA Tau). The deviations are caused by the reconstructed, outflow-absorbed Ly$\alpha$ profiles, which do not appear to match the radiation fields seen by the UV-CO. 

\begin{figure*}[t!]
	\begin{minipage}{0.49\textwidth}
	\centering
	\includegraphics[width=\linewidth]{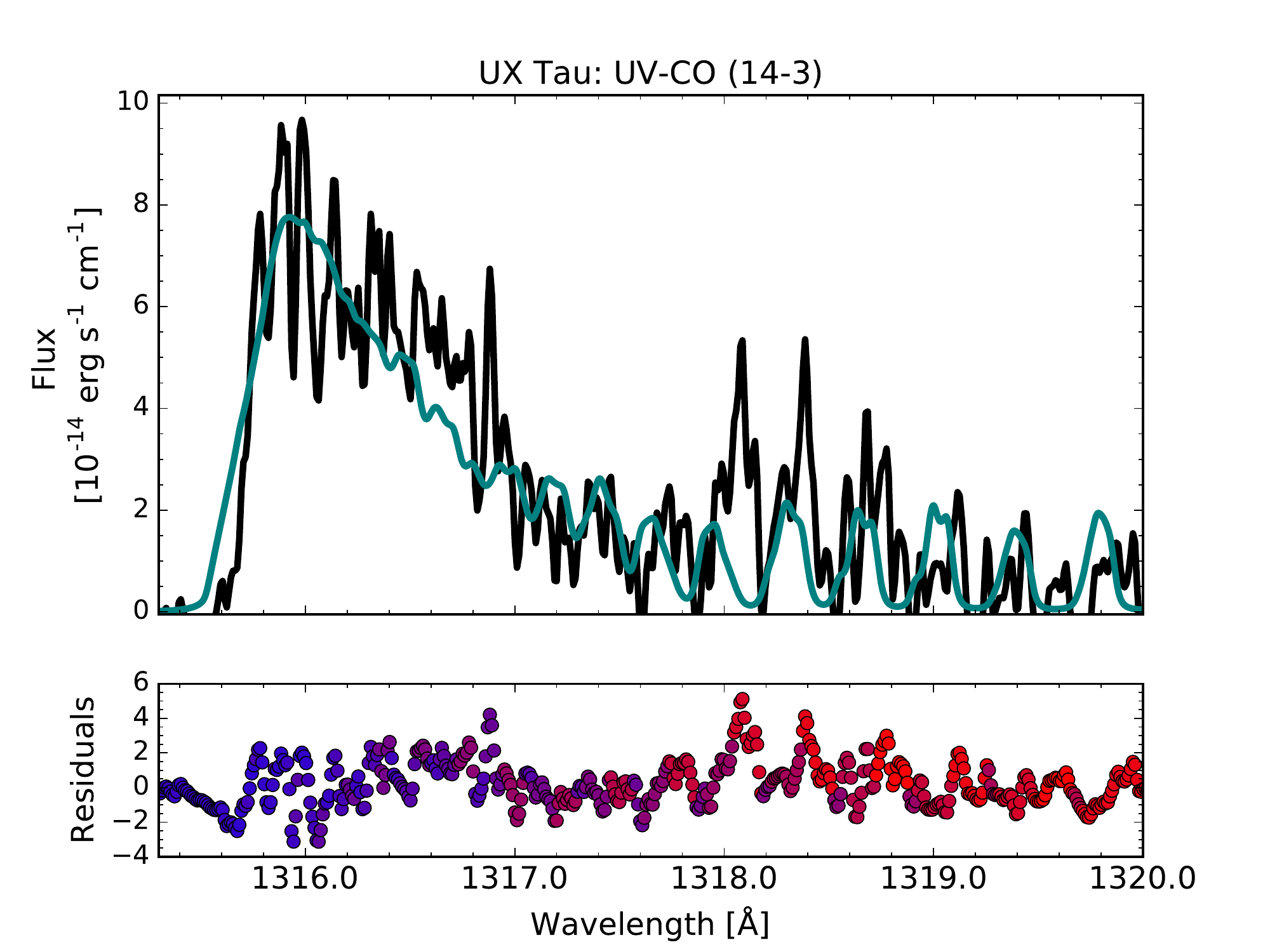}
	\end{minipage}
	\begin{minipage}{0.49\textwidth}
	\centering
	\includegraphics[width=\linewidth]{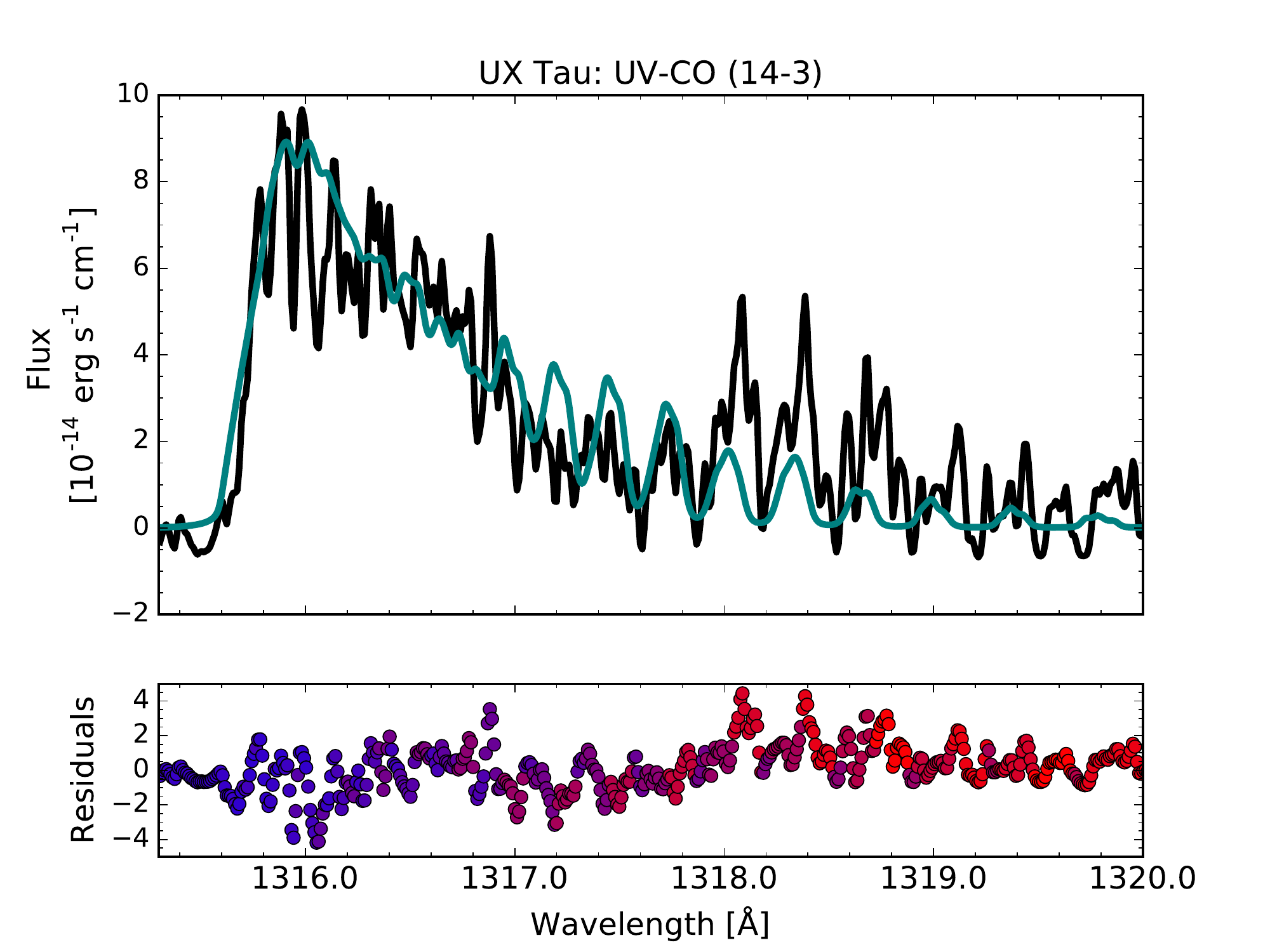}
	\end{minipage}
\caption{Best-fit UV-CO models for UX Tau, with residuals color-coded from blue to red based on the appropriate Ly$\alpha$ pumping wavelength. The model acquired with a Ly$\alpha$ radiation field generated exclusively at the central accretion shock \emph{(left)} underpredicts the data near 1316.5 \AA, which can be corrected by including an extra Gaussian Ly$\alpha$ profile to capture the missing flux near line center \emph{(right)}. However, it is more difficult to reproduce the UV-CO emission from upper levels pumped by redder Ly$\alpha$ photons, implying that some local source of redshifted photons (e.g., a disk wind) may be responsible for exciting gas at larger radii. We note that the spectral resolution of the data is not high enough to determine whether some portion of the UV-CO emission itself originates in a wind, instead of the disk surface.}
\label{LyA_discrepancy_colors}
\end{figure*}

Figure \ref{LyA_discrepancy_colors} shows two different best-fit UV-CO models for UX Tau, with the residuals color-coded from blue to red based on the appropriate Ly$\alpha$ pumping wavelengths. The model generated with an outflow absorbed Ly$\alpha$ profile underpredicts the data from transitions with $J'' \sim 8-12$, which are pumped at Ly$\alpha$ wavelengths between $\lambda \sim 1214.5-1215.7$ \AA. We also find that the models underpredict the data at longer wavelength transitions pumped by redder Ly$\alpha$ photons, causing the modeled emission lines to miss the observed line centers while still managing to reproduce the spacing between transitions. Since these are the high $J''$ features used to derive the empirical radii, the underpredictions are likely responsible for the discrepancies between $\left< R_{CO} \right>$ from the Gaussian fitting procedure (Table \ref{H2_measured_props}) and $r_{peak}$ from the radiative transfer models (Table \ref{best_fit_CO}). Additional sources of Ly$\alpha$ pumping photons, both from line center and red-shifted wavelengths, must be included to more accurately reproduce the UV-CO emission from large radii $\left(r > 15 \, \text{au} \right)$.

This evidence for a missing source of red-shifted Ly$\alpha$ at the disk surface is consistent with previous analyses of UV-H$_2$, since the spectra can show significant emission from progressions excited by pumping photons up to $\sim$1000 km s$^{-1}$ redward of line center (e.g. DF Tau; \citealt{Herczeg2006}). Furthermore, even the best-fit UV-H$_2$ models are not able to fully reproduce both the broad high-velocity wings and narrower line cores in many systems (see e.g., DF Tau, RY Lup, V4046 Sgr). This may also imply that an additional Ly$\alpha$ source is responsible for pumping the more distant gas responsible for the low velocity emission, while the photons generated at the accretion shock reach the gas at smaller radii. Physically, the extra radiation may originate in a disk wind, from which Ly$\alpha$ would appear red-shifted in the frame of the disk. To better capture this missing flux, we added additional low-velocity Gaussian emission components approximating what one might expect to be generated locally in winds or outflows to the model Ly$\alpha$ profiles. Although the excess emission boosted irradiation of the gas by photons near line center (see Figure \ref{LyA_discrepancy_colors} and Appendix B), it still does not fully account for the deviations between the UV-CO models and data. 

A shortcoming of our modeling approach is that we must attempt to simultaneously characterize the strength of Ly$\alpha$ emission at the disk surface, the radial location(s) that receive the strongest irradiation from Ly$\alpha$ sources beyond the accretion shock (e.g., from disk winds), and the physical structure of the fluorescent gas layer. We find that the degeneracies between sets of model parameters, particularly the gas temperature gradients and Ly$\alpha$ pumping fluxes, make it difficult to converge on accurate radial distributions of UV-CO flux as derived for the UV-H$_2$ emission \citep{Hoadley2015}. In the following sections, we instead further analyze the observed Ly$\alpha$ line wings and construct a more detailed picture of interactions between the radiation field and the molecular gas disk. This interpretation lays the ground work for future reconstructions of Ly$\alpha$ profiles from T Tauri stars, which will require more observations of other Ly$\alpha$ sensitive molecules (e.g., HCN, H$_2$O, H$_2$CO; \citealt{Bergin2003, Bethell2011}) in disks with available UV spectra.   

\section{Discussion} 

As described in the previous section, the 2-D radiative transfer models of UV-H$_2$ and UV-CO depend strongly on Ly$\alpha$ irradiation of the gas disk. This is seen most clearly in the best-fit models for the UV-CO bands, which cannot always be reproduced with a reconstructed outflow-absorbed Ly$\alpha$ profile. Since interstellar absorption and geocoronal emission prevent us from directly detecting the Ly$\alpha$ line core, we focus our discussion on what the observed high-velocity wings reveal about the radiation field reaching the molecular gas. We note that the wings were not observed for CW Tau, AA Tau (2013), or RECX-15 (2013), since those spectra were acquired with an orientation of the G130M grating on \textit{HST}-COS that does not include the Ly$\alpha$ transition ($\lambda_{central} = 1222$ \AA, instead of 1291 \AA).  

\subsection{Ly$\alpha$ Irradiation Drives the UV-CO Band Shapes}

\begin{figure*}[t!]
	\centering
	\includegraphics[width=\linewidth]{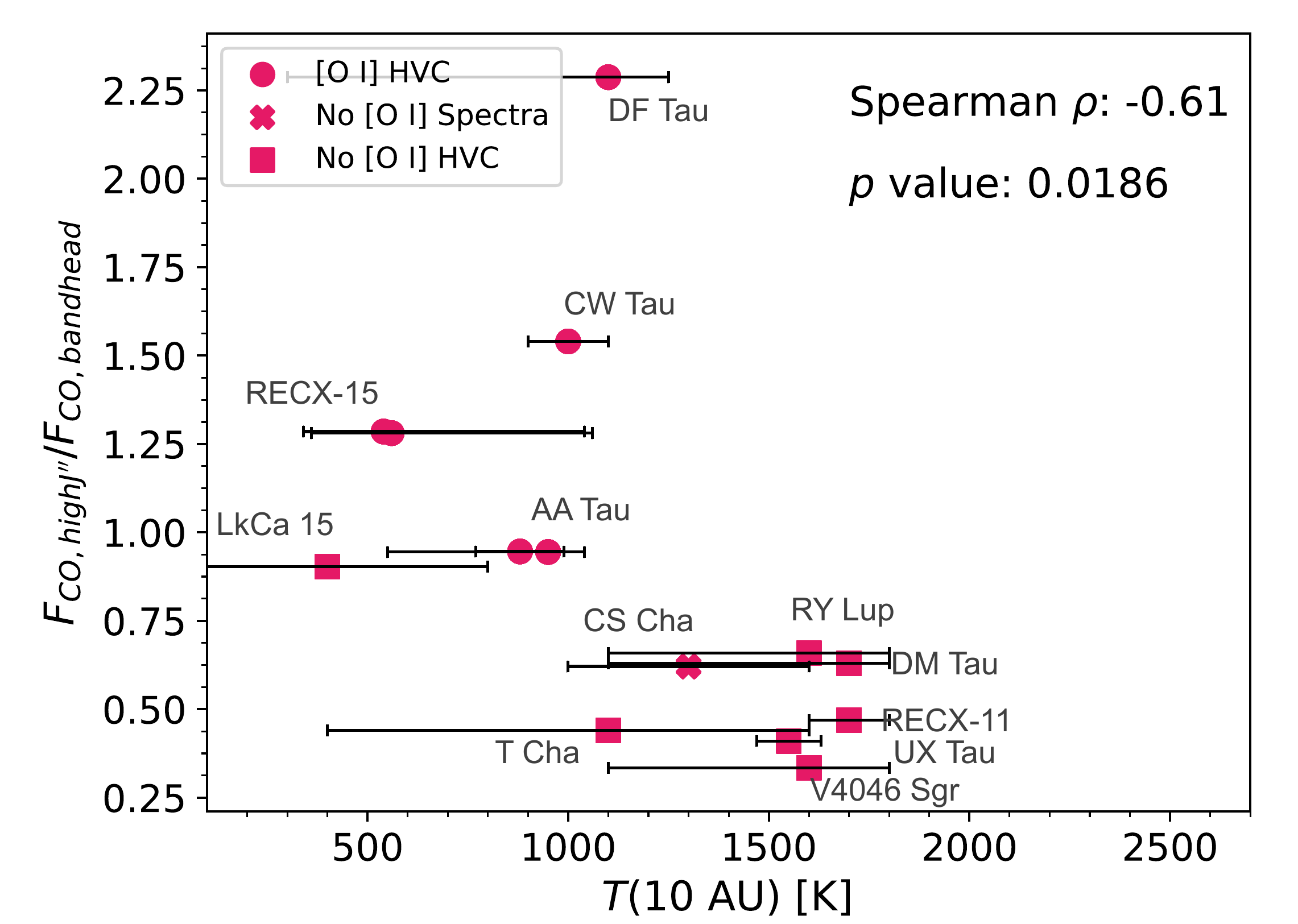}
\caption{UV-CO model temperatures at 10 AU versus observed flux ratios of $J'' > 10$ to $J'' = 0-10$ UV-CO emission. The significant anti-correlation demonstrates that higher gas temperatures do not necessarily lead to increased high $J''$ emission. CS Cha is denoted with a cross (see also Figures \ref{LyA_vs_CO}, \ref{H2_vs_LyA_flux}, and \ref{n1331_vs_LyA}), because it does not have high-resolution [O I] $\lambda$6300 spectra that reveal the presence or absence of a high-velocity component. Uncertainties on the measured flux ratios are smaller than the marker sizes, so vertical error bars are not shown. An interactive version of this figure is provided online, providing disk and stellar properties for each target.}
\label{highJtobandhead_vs_T}
\end{figure*}

Figure \ref{highJtobandhead_vs_T} compares the best-fit model gas temperatures from Table \ref{best_fit_CO} to the ratios of observed UV-CO emission from high $J''$ states relative to the bandhead $\left( F_{J'' > 10} / F_{J'' = 0-10} \right)$. The relationship between the model gas temperatures at 10 AU and the observed UV-CO flux ratios (hereafter referred to as ``band shapes") is an anti-correlation (Spearman rank coefficient $= -0.61$; $p = 0.02$), implying that hotter gas does not necessarily lead to more populated high $J''$ states.

\begin{figure*}[t!]
	\centering
	\includegraphics[width=\linewidth]{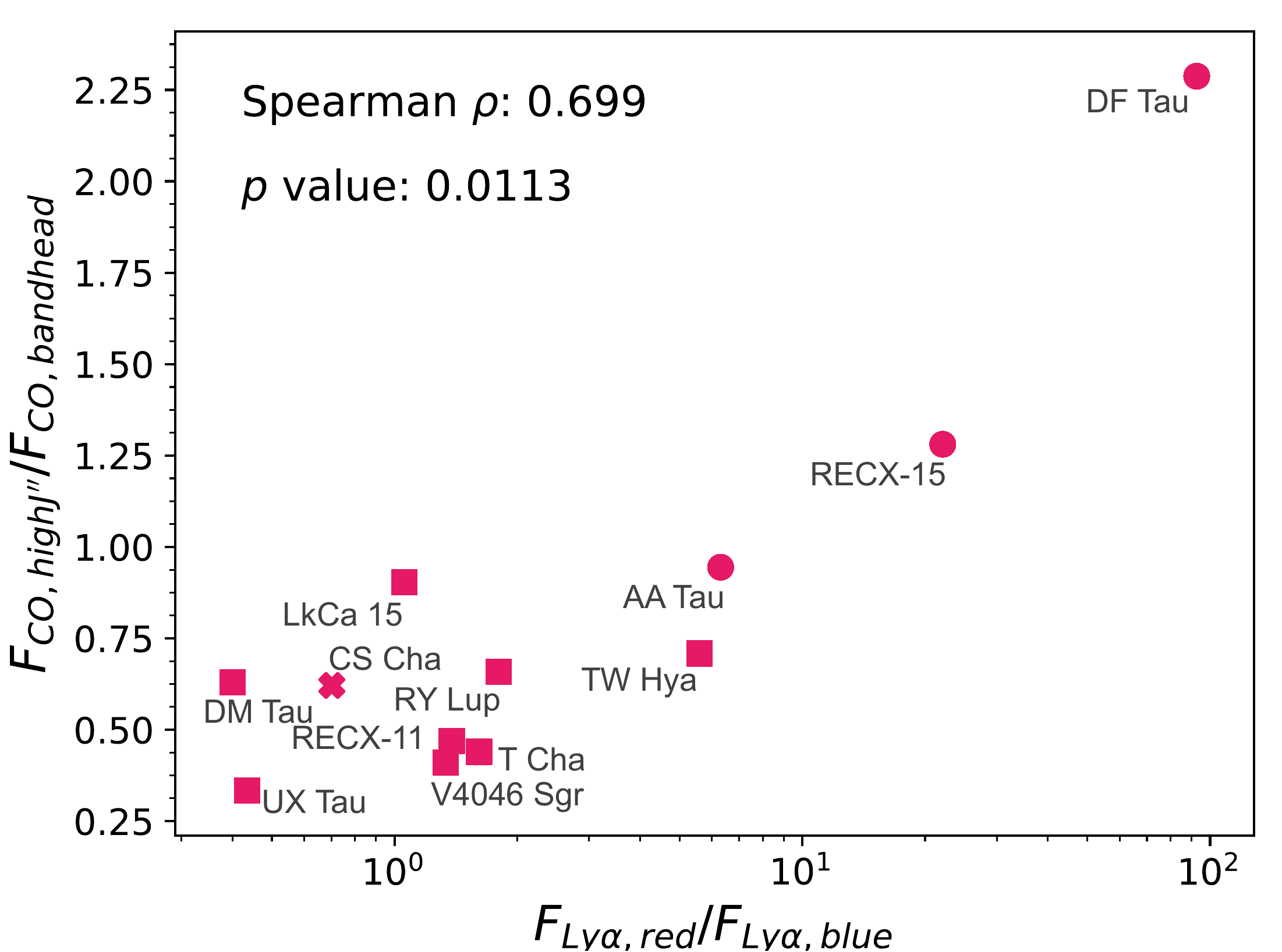}
\caption{Flux ratios of observed red wing (1217-1220 \AA) to blue wing (1210-1214.5 \AA) Ly$\alpha$ emission versus ratios of UV-CO $J'' > 10$ to $J'' = 0-10$ emission. The log-linear relationship implies that the shape of the Ly$\alpha$ profile is the dominant factor in the $J''$ state distribution of UV-CO flux (see also Table 7 of \citealt{Schindhelm2012_CO}). An interactive version of this figure is provided online, providing disk and stellar properties for each target.}
\label{LyA_vs_CO}
\end{figure*}

Since the UV-CO emitting region extends to larger radii than the UV-H$_2$, it is possible (or even likely) that Ly$\alpha$ profiles reconstructed from the UV-H$_2$ emission inside $r < 2$ AU are not representative of the radiation field at $r >15$ AU. To gain a better understanding of the Ly$\alpha$ profiles seen by the UV-CO, we compare the ratios of observed Ly$\alpha$ emission from the red and blue wings of the profiles $\left(F_{1217-1220 \, \rm{\AA}} / F_{1210-1214.5 \, \rm{\AA}} \right)$ to the UV-CO band shapes in Figure \ref{LyA_vs_CO}. Unlike the gas temperatures, the ratios of red wing to blue wing observed Ly$\alpha$ emission show a strong positive correlation with the ratios of UV-CO flux in high $J''$ relative to bandhead states (Spearman rank coefficient $= 0.699$; $p = 0.01$). The UV-CO band shapes are therefore driven by the shape of the illuminating Ly$\alpha$ profile instead of the temperature distribution within the emitting region. 

\begin{figure*}[t!]
	\centering
	\includegraphics[width=\linewidth]{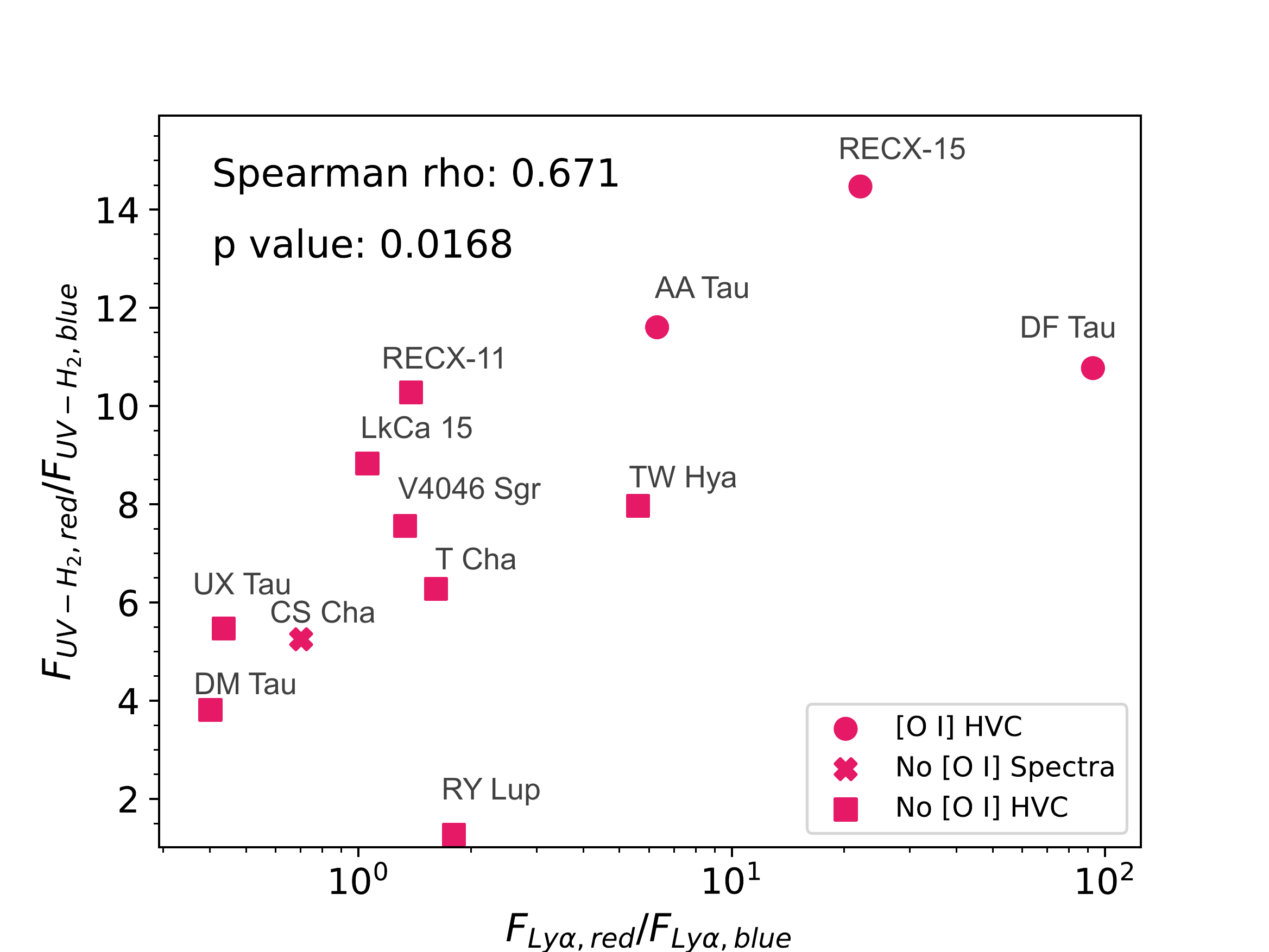}
\caption{Flux ratios of observed red wing to blue wing Ly$\alpha$ emission versus flux ratios of red $\left(\left[1, 4 \right], \left[1, 7 \right] \right)$ to blue $\left[4, 4 \right]$ UV-H$_2$ progressions demonstrate the same trend seen between Ly$\alpha$ and UV-CO in Figure \ref{LyA_vs_CO}. An interactive version of this figure is provided online, providing disk and stellar properties for each target.}
\label{H2_vs_LyA_flux}
\end{figure*}

With this in mind, Figure \ref{H2_vs_LyA_flux} also compares the ratios of UV-H$_2$ fluxes from progressions pumped by red $\left( \left[1, 4 \right], \left[1, 7 \right] \right)$ and blue Ly$\alpha$ photons $\left( \left[4, 4 \right] \right)$ \citep{France2012} to the ratios of red to blue Ly$\alpha$ fluxes. Once again, we find a strong correlation between the flux ratios (Spearman rank coefficient $= 0.89$; $p = 5 \times 10^{-4}$), demonstrating that disks with more blue-wing Ly$\alpha$ emission also show stronger UV-H$_2$ fluxes in the $\left[4,4 \right]$ progression (see also \citealt{McJunkin2016}). Since the shape of the illuminating radiation field clearly has a strong influence on the pumping radiation received at the disk surface, we discuss the mechanisms that dictate the ratios of red to blue Ly$\alpha$ fluxes in the following sections.
 
\begin{figure*}[t!]
	\centering
	\includegraphics[width=\linewidth]{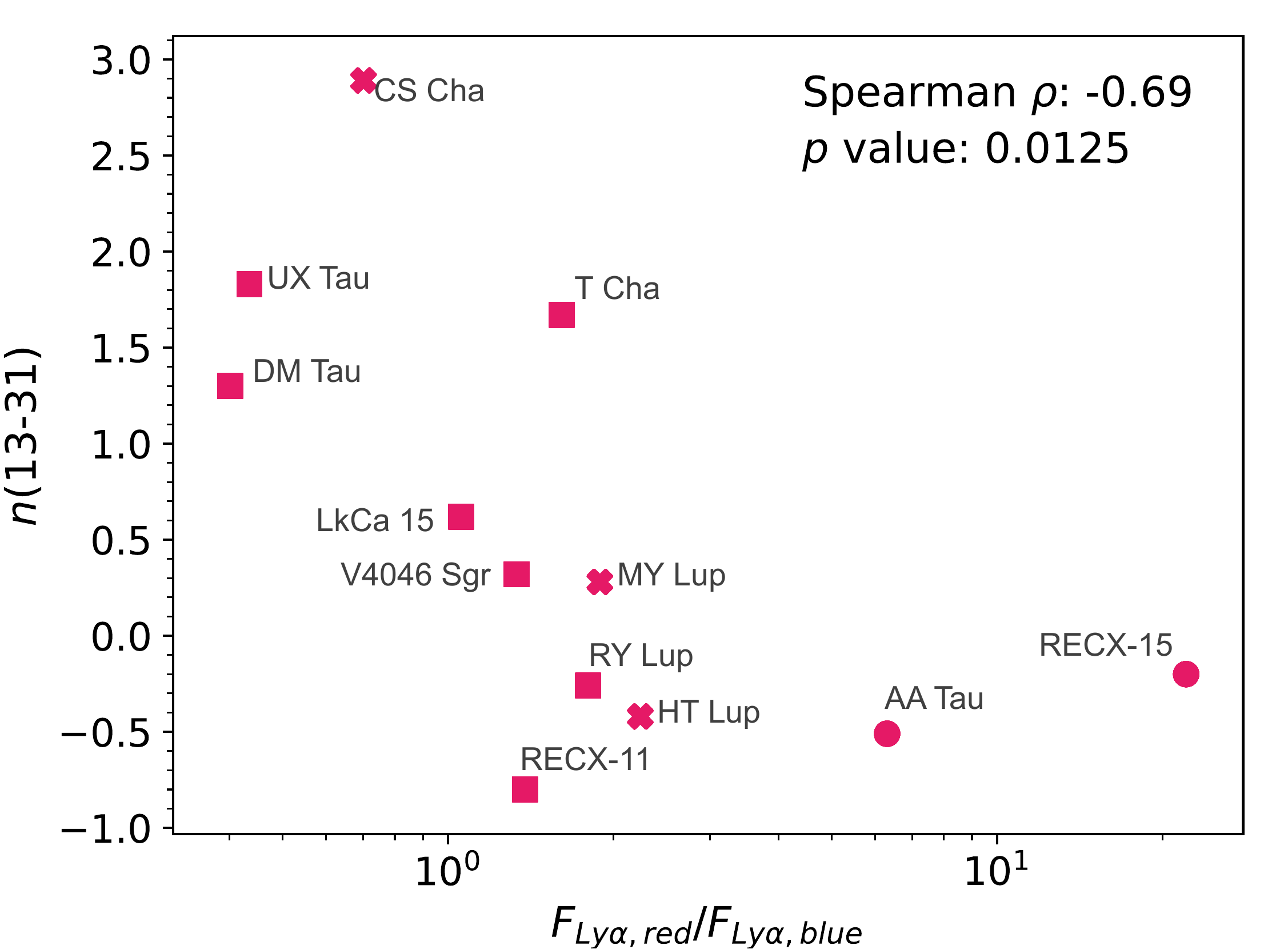}
\caption{Flux ratios of observed red wing to blue wing Ly$\alpha$ emission versus $n \left(13-31 \right)$ for disks with dust cavities, indicating that disks with more blue Ly$\alpha$ flux show less emission from warm grains relative to emission at longer wavelengths from cold dust. We omit DF Tau from the figure, since it is the only system without dust substructure resolved at infrared or sub-mm wavelengths. This means that the slope of its SED is representative of the degree of grain growth and dust settling towards the midplane, and therefore the disk flaring angle \citep{Furlan2009}. Since the remaining targets do have gaps and cavities in their dust distributions, the slopes of the SEDs instead trace the disk temperatures, and therefore radial locations, where grains have been removed from the disk instead of settled to the midplane (see e.g., \citealt{Espaillat2010}). DF Tau should instead be compared to other targets without resolved dust substructure, which will be possible once all new data from the ULLYSES program is available. An interactive version of this figure is provided online, providing disk and stellar properties for each target.}
\label{n1331_vs_LyA}
\end{figure*}

\subsection{Three Categories of UV-CO Band Shapes and Ly$\alpha$ Morphologies}

\begin{figure*}[t!]
	\centering
	\includegraphics[width=0.85\linewidth]{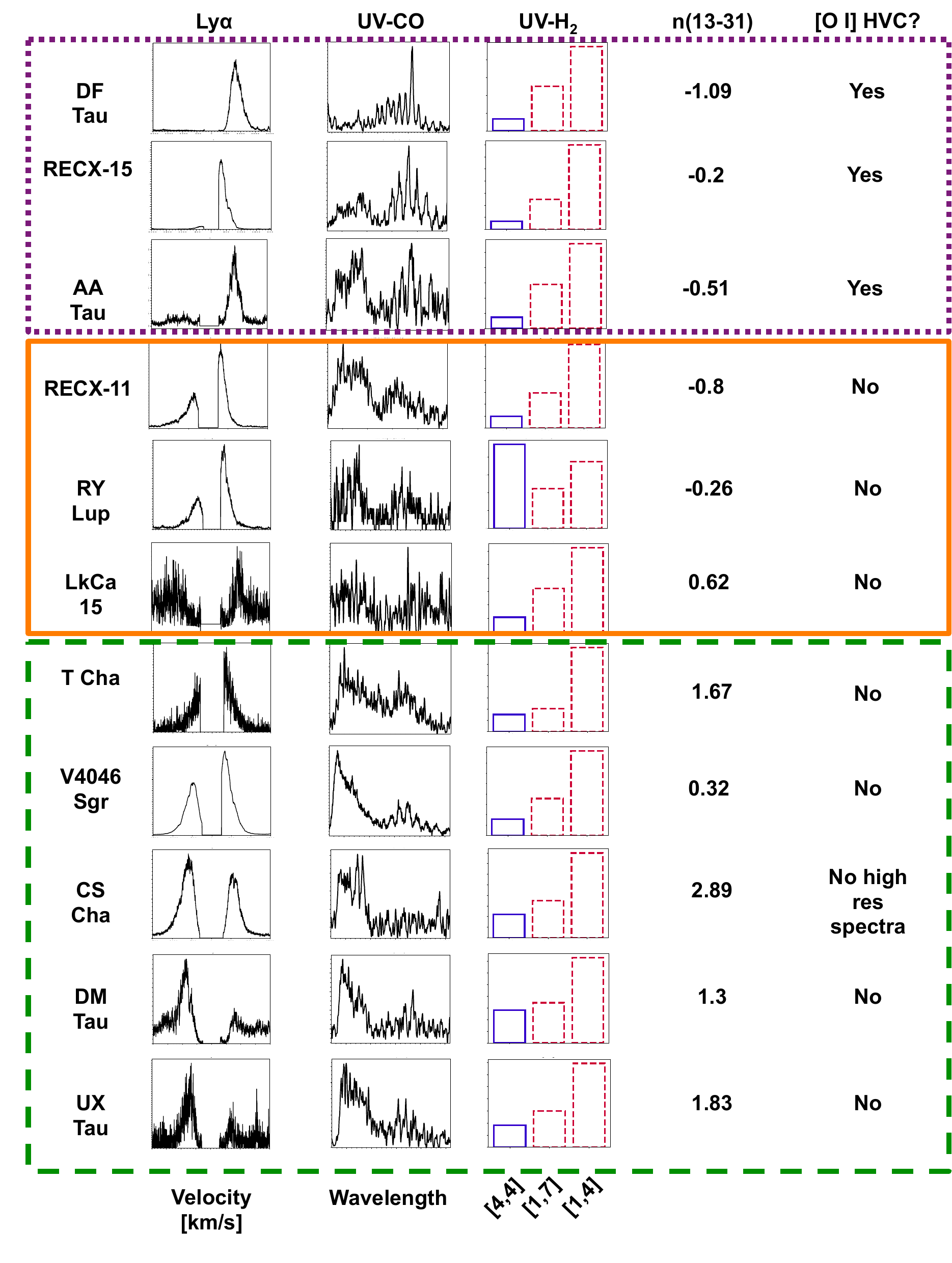}
\caption{Summary of empirical tracers used to categorize the targets in Table \ref{targ_list} (see also Table \ref{disk_categories}, Figure \ref{LyA_cartoon}). The ``UV-H$_2$" column shows bars representing the total UV-H$_2$ flux from each of three progressions, pumped by blue ([4,4]) and red ([1,4], [1,7]) Ly$\alpha$ photons. This demonstrates that the [4,4] progression fluxes become relatively stronger as the blue Ly$\alpha$ emission increases. CW Tau is not included, since its Ly$\alpha$ wings were not observed with \textit{HST}-COS. Systems with similar ratios of observed red to blue Ly$\alpha$ emission have similar UV-CO band shapes, UV-H$_2$ fluxes from progressions pumped by blue ([4,4]) and red ([1,7], [1,4]) Ly$\alpha$ photons, infrared spectral indices, and [O I] $\lambda$6300 emission line components. Outflow-dominated targets are enclosed in the purple, dotted box, intermediate targets in the orange, solid box, and infall-dominated targets in the green, dashed box. The UV-H$_2$ [4,4] flux from RY Lup is unusually high, as driven by the (4-9)P(5) 1526.55 \AA \, feature \citep{Arulanantham2018}.}
\label{empirical_summary}
\end{figure*}

\begin{figure*}
\centering
\includegraphics[width=1.05\linewidth]
{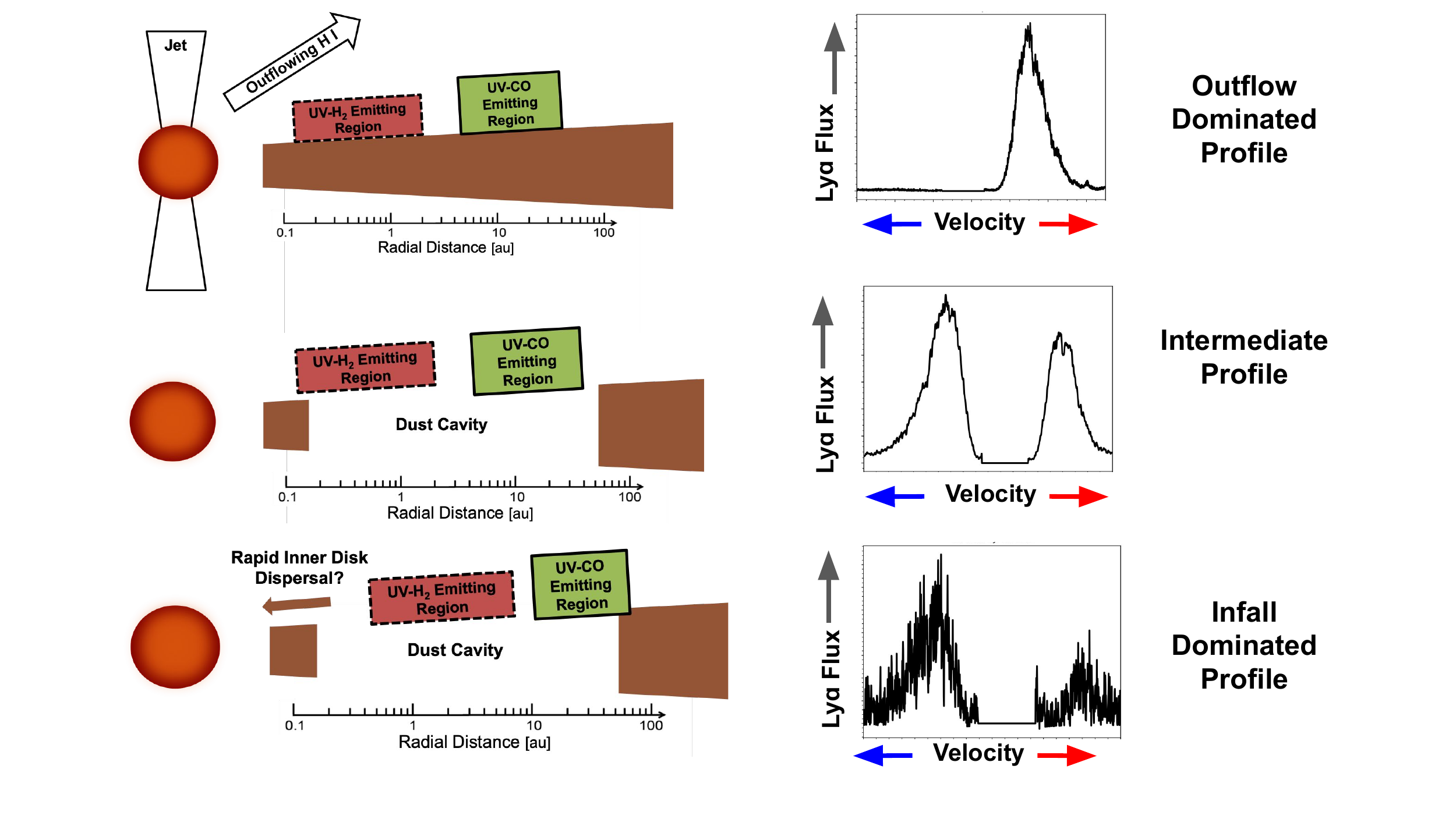}
\caption{Illustration of Ly$\alpha$ evolution over the disk lifetime. High-velocity jets are present in targets with \textit{outflow dominated profiles}, where outflowing H I absorbs the blue side of the emission line and fluorescent emission is seen from as close as $r < 0.1$ AU. When outflowing mass loss rates decline, \textit{intermediate, symmetric} Ly$\alpha$ profiles are observed and the UV-H$_2$ and UV-CO emitting regions shift to larger radii. At this stage, any asymmetries in the observed Ly$\alpha$ wings are present at low velocities that are mostly hidden by geocoronal emission. However, this does not necessarily represent an intermediate phase of disk evolution. In the absence of high-velocity outflowing material, \textit{infall dominated} Ly$\alpha$ profiles are produced as H I absorbs the red sides of the emission lines. UV-H$_2$ and UV-CO emitting regions are represented based on the median empirical radii measured for each group of targets (see Table \ref{disk_categories}).}
\label{LyA_cartoon}
\end{figure*}

Figure \ref{empirical_summary} provides the observed Ly$\alpha$ wings, UV-CO $\left(v'-v'' \right) = \left(14-3 \right)$ bands, UV-H$_2$ progression fluxes, infrared spectral indices, and [O I] $\lambda$6300 line shapes for all targets in our sample. We order the targets based on where they fall in Figures \ref{highJtobandhead_vs_T}, \ref{LyA_vs_CO}, \ref{H2_vs_LyA_flux}, and \ref{n1331_vs_LyA} and find that they can be divided into three categories: disks showing more flux in the UV-CO bandhead than in the high $J''$ states, systems showing roughly similar bandhead and high $J''$ fluxes, and targets with more high $J''$ than bandhead emission. As depicted in Table \ref{disk_categories}, targets that fall within a specific grouping also have similar observed Ly$\alpha$ profiles and infrared spectral indices. For example, disks with more bandhead than high $J''$ UV-CO emission also have stronger blue wing Ly$\alpha$ flux and larger SED slopes between 13 and 31 $\mu$m that indicate more emission from cold, outer disk dust relative to warm, inner disk material (\citealt{Furlan2009}; e.g. UX Tau A). We expand on each of the three UV-CO categories in the following subsections. In particular, we investigate the connection between Ly$\alpha$ emission and the structure of the gas and dust disks, which appears analagous to the evolutionary sequence discovered in optical and infrared emission from forbidden lines in disk winds (see e.g., [O I], [Ne II]; \citealt{Banzatti2019, Pascucci2020}). Figure \ref{LyA_cartoon} illustrates this sequence, demonstrating how the Ly$\alpha$ profile might shift as the disk changes. 

\begin{deluxetable*}{cccccccc}
\tablecaption{Group Properties Based on Empirical Tracers Discussed in the Text \label{disk_categories}
}
\tablewidth{0 pt}
\tabletypesize{\scriptsize}
\tablehead{  & \colhead{$<R_{H_2}>$\tablenotemark{a}} & \colhead{$<R_{CO}>$} & \colhead{UV-CO} & \colhead{Ly$\alpha$} & \colhead{Dust Evolution} & \colhead{Intervening H I\tablenotemark{d}} & \colhead{Targets} \\
 & \colhead{[AU]} & \colhead{[AU]} & \colhead{$\frac{F \left(J'' > 10 \right)}{F \left(J'' < 10 \right)}$} & \colhead{$\frac{F \left(\text{red} \right)}{F \left( \text{blue} \right)}$\tablenotemark{b}} & \colhead{$n \left(13-31 \right)$\tablenotemark{c}} & \colhead{$10^6$ cm$^{-3}$} & \\
}
\startdata
Outflow Dominated & $\sim$0.7 & $\sim$16 & $>\sim 1$ & $>\sim 1$ & $< \sim -0.5$ & $\sim$0.51 & AA Tau, DF Tau, RECX-15 \\
Intermediate & $\sim$0.65 & $\sim$13 & $\sim 1$ & $ \sim 1$ & $\sim -0.5-0.5$ & $\sim$0.23 & RECX-11, LkCa 15, RY Lupi \\
Infall Dominated & $\sim$1.4 & $\sim$25 & $<\sim 1$ & $<\sim 1$ & $>\sim 0.5$ & $\sim$0.20 & CS Cha, DM Tau, UX Tau, \\
& & & & & & & T Cha, V4046 Sgr \\ 
\enddata
\tablenotetext{a}{Given the larger uncertainties in the 2-D radiative transfer models, we report the median empirical radii for both the UV-H$_2$ and UV-CO features from each group. Radii are calculated from the Gaussian emission line widths, under the assumption that all emitting material is in Keplerian rotation.}
\tablenotetext{b}{$F \left(\text{red} \right)/F \left( \text{blue} \right) = F \left( \lambda \sim 1217-1220 \, \rm{\AA} \right) / F \left(\lambda \sim 1210-1214.5 \, \rm{\AA} \right)$}
\tablenotetext{c}{$n \left(13-31 \right) = \frac{\log \left( \lambda_{31} F_{\lambda_{31}} \right) - \log \left(\lambda_{13} F_{\lambda_{13}} \right)}{\log \left( \lambda_{31} \right) - \log \left( \lambda_{13} \right)}$; \citet{Furlan2009}}
\tablenotetext{d}{Number densities of intervening H I were estimated from the column densities and distances for each target presented in Table \ref{HI_checkpoints}. A median value was then calculated for each group. The number of targets in any individual group is small and the scatter is $\sim$40\%, so we urge the reader to interpret these results with caution.}
\end{deluxetable*}  

\subsubsection{Outflow Dominated Profiles}

The disks around AA Tau, RECX-15, and DF Tau show the most prominent UV-CO emission from high $J''$ states, with relatively suppressed emission from the bandhead. All three disks also have observed Ly$\alpha$ line wings with significantly more red than blue flux. For these systems, the shape of the Ly$\alpha$ radiation field reaching both the UV-CO emitting region and the observer appears generally consistent with the model used to reconstruct the Ly$\alpha$ profile from the UV-H$_2$ features (see Section 4.2).      

Ly$\alpha$ photons from young stellar objects are generated at the accretion shock and in heated funnel flows (see e.g., \citealt{Alencar2012}), where high opacity material and Stark broadening produce strong, broad emission lines (FWHM$\sim550-900$ km s$^{-1}$; \citealt{Schindhelm2012}). The photons must then travel through optically thick outflowing H I before reaching the molecular gas disk. In the frames of the gas disk and the observer, the outflowing gas removes flux from the blue side of the Ly$\alpha$ line $\left(\lambda \sim 1214.5-1215.7 \, \text{\AA} \right)$. The resulting Ly$\alpha$ radiation field seen by the molecular gas disk then has stronger red than blue flux.  

In agreement with this picture, all three targets in this category are driving spatially resolved jets traced by high velocity optical forbidden line emission. AA Tau shows a sequence of knots with [S II] $\lambda$6717 emission at $v = -200$ km s$^{-1}$, which are consistent with model predictions of jet velocities \citep{Cox2013}. Prominent jets are also resolved in optical [O I] 6300 \AA \, observations of RECX-15 \citep{Woitke2011} and DF Tau \citep{Uvarova2020}. At high spectral resolution, signatures of high-velocity jets ($|v| \sim$ few hundred km s$^{-1}$) originating near the accretion shock are detected in [O I] 6300 \AA \, spectra from all three targets (see e.g., \citealt{Banzatti2019}). The [O I] features also contain low-velocity ($|v| < 50$ km s$^{-1}$) blue-shifted components, which originate near the base of MHD winds \citep{Banzatti2019}. This outflowing material must also include H I, which further absorbs the Ly$\alpha$ photons generated near the accretion shock. 

\subsubsection{Intermediate Profiles}

Figure \ref{empirical_summary} shows that a handful of targets in our sample have ratios of red wing to blue wing Ly$\alpha$ flux around $\sim$1: LkCa 15, RECX-11, RY Lupi, T Cha, and V4046 Sgr. However, the UV-CO band shapes from T Cha and V4046 Sgr appear more similar to the disks with stronger blue-wing Ly$\alpha$ emission (CS Cha, DM Tau, and UX Tau A), so we have placed them in that category instead. The remaining three targets span a relatively broad range of bandhead to high $J''$ flux ratios. 

RY Lupi and LkCa 15, which host large, dust depleted inner cavities ($r \sim 69$ and 76 AU, respectively; see e.g. \citealt{Francis2020}), have stronger UV-CO emission from the bandhead, while RECX-11 has roughly similar levels of bandhead and high $J''$ flux. The SED slopes between 13 and 31 $\mu$m for these targets are scattered, with RY Lupi and LkCa 15 having $n \left(13-31 \right) = -0.26$ and $n \left(13-31 \right) = 0.62$, respectively \citep{Furlan2009}. Disks in this category seem to represent a middle phase of evolution, where the warm dust producing the 13 $\mu$m emission has not necessarily been cleared from the system and is still able to shield UV-fluorescent gas. 

None of the targets in this group show high-velocity [O I] 6300 \AA \, components from jets \citep{Banzatti2019}, indicating that emission from jets and inner disk outflows is not present. However, accretion is still ongoing in all three disks, with rates ranging from $0.017-1 \times 10^{-8}$ $M_{\odot}$ yr$^{-1}$ \citep{Ingleby2011, Ingleby2013, France2017}. This is supported by the H$\alpha$ emission line from RECX-11, which has a wide blue wing ($v \sim 300$ km s$^{-1}$) and inverse P Cygni (red) absorption characteristic of free-falling gas accreting onto the star \citep{Ingleby2011}.

\subsubsection{Infall Dominated Profiles}

Five disks in our sample have stronger integrated UV-CO fluxes in the bandhead $\left(J'' = 0-10 \right)$ than in the high $J''$ states $\left(J'' > 10 \right)$: CS Cha, DM Tau, T Cha, UX Tau, and V4046 Sgr. All five systems have steeper disk SEDs than targets in the other two groups, as measured from the slopes between 13 and 31 $\mu$m $\left( n \left(13-31 \right); \, \text{\citealt{Furlan2009}} \right)$. This indicates that their dust disks produce more emission from cold, outer disk grains than warm, inner disk material, as confirmed by dust-depleted gaps or cavities resolved at sub-mm wavelengths (see e.g., \citealt{Kudo2018, Hendler2018, RuizRodriguez2019}). 

Within this group, 3/5 targets\footnote{T Cha and V4046 Sgr both display fairly symmetric blue and red wing Ly$\alpha$ profiles. However, the shapes of the UV-CO bands observed from these two systems appear more similar to the disks with stronger blue wing Ly$\alpha$ emission, so we place them in this category instead.} have observed Ly$\alpha$ profiles with stronger blue wing $\left(\lambda < 1214.5 \, \rm{\AA} \right)$ than red wing $\left(\lambda > 1217 \, \rm{\AA} \right)$ emission. The strong blue Ly$\alpha$ wings contrast with the model used to reconstruct the Ly$\alpha$ radiation field, which uses outflowing H I to truncate the blue sides of the line profiles \citep{Herczeg2004, Schindhelm2012}. A model with red-shifted ISM absorption can reproduce the observed line wings by suppressing the red Ly$\alpha$ emission (see e.g., \citealt{McJunkin2014}). However, strong correlations between the Ly$\alpha$ flux ratios, UV-CO band shapes, UV-H$_2$ flux ratios, and $n \left(13-31 \right)$ (see Figure \ref{empirical_summary}) indicate that the red wing suppression occurs before the Ly$\alpha$ photons enter the ISM. 

Similar red-absorbed shapes have long been identified in the hydrogen recombination lines \citep{Edwards1994, Reipurth1996}. The morphologies are attributed to magnetospherically driven mass infall, where high velocity material moving away from both the observer and the gas disk absorbs the red sides of emission lines generated near the accretion shock \citep{Calvet1992, Hartmann1994}. It is possible then that the Ly$\alpha$ profiles with more blue than red emission are also dominated by mass infall. This picture is consistent with the mass accretion rates measured from C IV emission (see e.g., \citealt{Ardila2013}), which are not necessarily slower in targets with more depleted dust distributions.  

Trends in forbidden emission from [O I] and [Ne II] provide some insight into the Ly$\alpha$ profiles for this group of targets. \citet{Pascucci2020} find that CS Cha, DM Tau, T Cha, UX Tau, and V4046 Sgr all have larger ratios of low velocity [Ne II] 12.8 $\mu$m to [O I] 6300 \AA \, luminosities than targets in the category with blue-suppressed Ly$\alpha$. That work demonstrates that low-velocity [Ne II] emission strengthens as the $n \left(13-31 \right)$ spectral index increases and the low-velocity [O I] emitting region migrates to larger radii (see e.g., \citealt{Simon2016, Banzatti2019, Pascucci2020}). Furthermore, none of the targets in this group show high velocity [O I] emission, implying that the jets and other high velocity outflows have turned off. Without these feedback mechanisms, the remaining material in the inner disk may be able to rapidly drain onto the star, as expected from models of inside-out disk dispersal (see e.g., \citealt{Ercolano2011}).  

\begin{figure*}
\centering
\includegraphics[width=\linewidth]
{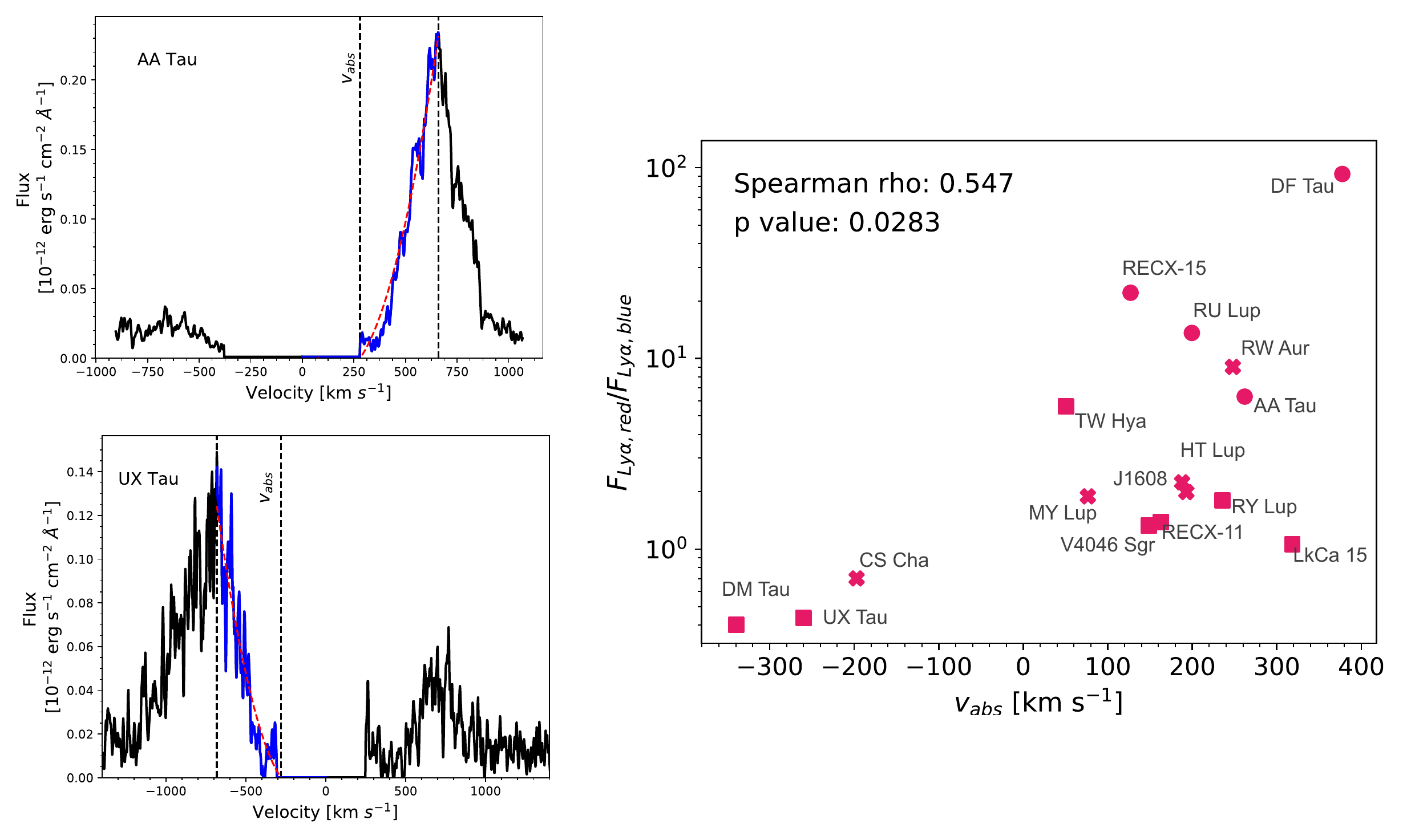}
\caption{\textit{Upper left:} Extrapolated H I absorption velocity for a target with an outflow dominated Ly$\alpha$ profile. \textit{Lower left:} Same as upper left, for a target with an infall dominated Ly$\alpha$ profile. These velocities trace the opacity of intervening H I between the accretion shock and observer, rather than the motion of the material. \textit{Right:} Ratio of observed red-wing to blue-wing Ly$\alpha$ fluxes versus extrapolated H I absorption velocities. The three targets with negative absorption velocities do not have high velocity [O I] $\lambda$6300 \AA \, components, implying that emission from inner disk jets and outflows is not present (see e.g., \citealt{Banzatti2019}). Targets with intermediate Ly$\alpha$ flux ratios $\left(F_{Ly\alpha, red}/F_{Ly\alpha, blue} \sim 1 \right)$ also do not show jet emission.}
\label{HI_abs_vel}
\end{figure*}

In order to connect the shapes of the observed Ly$\alpha$ profiles to the properties of absorbing H I, we estimate the absorption velocities where the fluxes goes to zero (see Figure \ref{HI_abs_vel}). In addition to the targets from our original sample, we also include spectra from systems with observed Ly$\alpha$ wings but noisy UV-CO emission lines. These extra targets are 2MASS J16083070-3828268, HT Lup, and MY Lup \citep{Arulanantham2020}, RU Lup (see e.g., \citealt{Herczeg2005}), RW Aur (see e.g., \citealt{Woitas2002}), and TW Hya (see e.g., \citealt{Herczeg2002}).

The absorption velocities were derived by fitting a 2nd degree polynomial to the strongest observed Ly$\alpha$ wing for each target, between the velocities where the wing emission peaks and the geocoronal emission disrupts the observed spectrum. The best-fit polynomial was then extrapolated to find the velocity where the model flux reaches zero. The three disks with more blue than red Ly$\alpha$ emission (DM Tau, UX Tau, CS Cha) have absorption velocities of $-339$, $-260$, and $-197$ km s$^{-1}$, respectively. The remaining targets, which all have positive absorption velocities and more red than blue Ly$\alpha$ emission, are somewhat scattered, with no clear trend between the ratios of observed Ly$\alpha$ flux ratios and absorption velocities. While mass inflow must be happening in these targets, as evidenced by the high accretion rates, its signature red-shifted absorption may be masked by the ongoing jet and outflow emission. This connection between the absorbing H I and the observed Ly$\alpha$ wings may become more clear when more targets with dominant blue wing emission are observed.            

\section{Summary and Conclusions}

We have presented a joint analysis of Ly$\alpha$ line profiles and fluorescent emission from H$_2$ and CO molecules in a sample of 12 disks around CTTSs, using spectra from \emph{HST}-COS. The features originate in surface layers of the gas disk, where optical depths are low enough for substantial amounts of Ly$\alpha$ flux to reach the molecules. We analyze the spectral features both empirically and with 2-D radiative transfer models, finding that 
\begin{enumerate}
\item The UV-H$_2$ and UV-CO features originate in radially separated regions of the disk, with UV-H$_2$ observed closer to the star $\left(r \sim 0.8 \, \rm{AU} \right)$ than UV-CO $\left(r \sim 20 \, \rm{AU} \right)$ in all targets with partially resolved high $J''$ $\left(J'' > 10 \right)$ UV-CO states. However, these radii were derived assuming the line widths are set entirely by Keplerian broadening in the disk. Emission from molecules in outflows or winds is not accounted for (see e.g., \citealt{Weber2020}), although we note that slow, uncollimated winds can maintain Keplerian rotation and behave as an extension of emission from the disks (see e.g., \citealt{Pontoppidan2011}). 
\item The ratios of red to blue Ly$\alpha$ line wing emission fall into three categories, which correlate significantly with:
\begin{enumerate}
\item The ratios of high $J''$ to bandhead $\left(J'' > 10 / J'' \leq 10 \right)$ UV-CO flux;
\item The ratios of UV-H$_2$ flux from progressions pumped by red to blue Ly$\alpha$ wavelengths ([4,4]/([1,7]+[1,4]);
\item The infrared spectral indices, $n \left(13-31 \right)$, for disks with dust cavities; 
\end{enumerate}
\item However, 2-D radiative transfer models fit to the UV-CO emission lines underpredict the observed flux from progressions pumped by central and red Ly$\alpha$ photons. This causes the radiative transfer models to map the fluorescent gas to smaller radii than the empirical values.
\item It is difficult to map the flux distributions of fluorescent gas beyond the empirically derived radii, since the 2-D radiative transfer modeling results indicate that we are not accounting for some hidden source(s) of Ly$\alpha$ irradiation. 
\end{enumerate}
Furthermore, we find that only the three targets with the largest ratios of red to blue Ly$\alpha$ line wing emission (AA Tau, DF Tau, RECX-15) show high velocity [O I] $\lambda$6300 emission associated with jets. Taken together, these results imply that jets and outflows turn off at some phase of disk evolution, allowing residual inner disk material to rapidly drain onto the stars. Any existing trends will become more clear over the next few years, as data from the \textit{HST} Ultraviolet Legacy Library of Young Stars as Essential Standards (ULLYSES) Director's Discretionary program becomes available.  

\section{Acknowledgements}

We are grateful to Zachary Taylor and Klaus Pontoppidan for enjoyable and insightful discussions in various stages of this work. We also thank the anonymous referee for their very detailed and thoughtful comments that greatly improved the clarity of this manuscript. NA was supported in part by NASA Earth and Space Science Fellowship grant 80NSSC17K0531, and KH is supported by the David \& Ellen Lee Prize Postdoctoral Fellowship in Experimental Physics at Caltech. This paper also made use of data and financial support from HST program GO-15070, and AB acknowledges support from HST grant HST-GO-15128.001-A to the University of Colorado at Boulder. This work has made use of data from the European Space Agency (ESA) mission {\it Gaia} (\url{https://www.cosmos.esa.int/gaia}), processed by the {\it Gaia} Data Processing and Analysis Consortium (DPAC, \url{https://www.cosmos.esa.int/web/gaia/dpac/consortium}). Funding for the DPAC has been provided by national institutions, in particular the institutions participating in the {\it Gaia} Multilateral Agreement. This work utilized the RMACC Summit supercomputer, which is supported by the National Science Foundation (awards ACI-1532235 and ACI-1532236), the University of Colorado Boulder, and Colorado State University. The Summit supercomputer is a joint effort of the University of Colorado Boulder and Colorado State University. This research made use of Astropy,\footnote{http://www.astropy.org} a community-developed core Python package for Astronomy \citep{astropy2013, astropy2018}.  

\begin{appendices}

\section{\\ Verifying UV-H$_2$ Modeling Results with Hoadley et al. (2015)} \label{App:Appendix A}
\setcounter{table}{0}
\renewcommand{\thetable}{A\arabic{table}}

UV-H$_2$ emission lines from almost all targets in the sample presented here were first analyzed by \citealt{Hoadley2015}, who developed the 2-D radiative transfer models. As described in Section 4, we have adapted the original models into an object-oriented framework that accommodates the UV-CO features as well. In this appendix, we compare our results to the previously published work in \citet{Hoadley2015} and demonstrate that the best-fit model temperatures are consistent within $\pm 1 \sigma$ error bounds. We also describe the methods of statistical interpretation used here in greater detail.   

In this work, uncertainties on the best-fit model parameters were derived using MCMC re-sampling \citep{emcee2013}, which extracts the posterior probability for the model parameters, $p \left( \Theta | y_i \right)$:
\begin{equation}
p \left( \Theta | y_i \right) \propto p \left( \Theta \right) p \left(y_i | \Theta \right),
\end{equation} 
where $\Theta = [z/r, \gamma, T_{1 \, \text{AU}}, q, r_{\text{char}}, M_{H_2}]$. We have assumed uniform prior distributions for the model parameters $\left( p \left( \Theta \right) \right)$, with boundaries defined in \citealt{Hoadley2015} (UV-H$_2$) and Section 4.4 of this work (UV-CO). We also adopt a Gaussian likelihood function:
\begin{equation}
p \left(y_i | \Theta \right) = \exp \left[-\sum \limits_{i = 0}^{N} \left( \frac{y_i - \tilde{y_i}}{ \sigma_i} \right)^2 \right],
\end{equation} 
where $y_i$ and $\tilde{y_i}$ represent the observed and model data points and $\sigma_i$ represents the measurement uncertainties in the data. However, we ignore $\sigma_i$ in the algorithm since it doesn't necessarily follow a Gaussian distribution. We find that the chain converges when 1000 walkers are allowed to probe the parameter space over 150 steps.  

\begin{figure*}
\centering
\includegraphics[width=1.0\linewidth]
{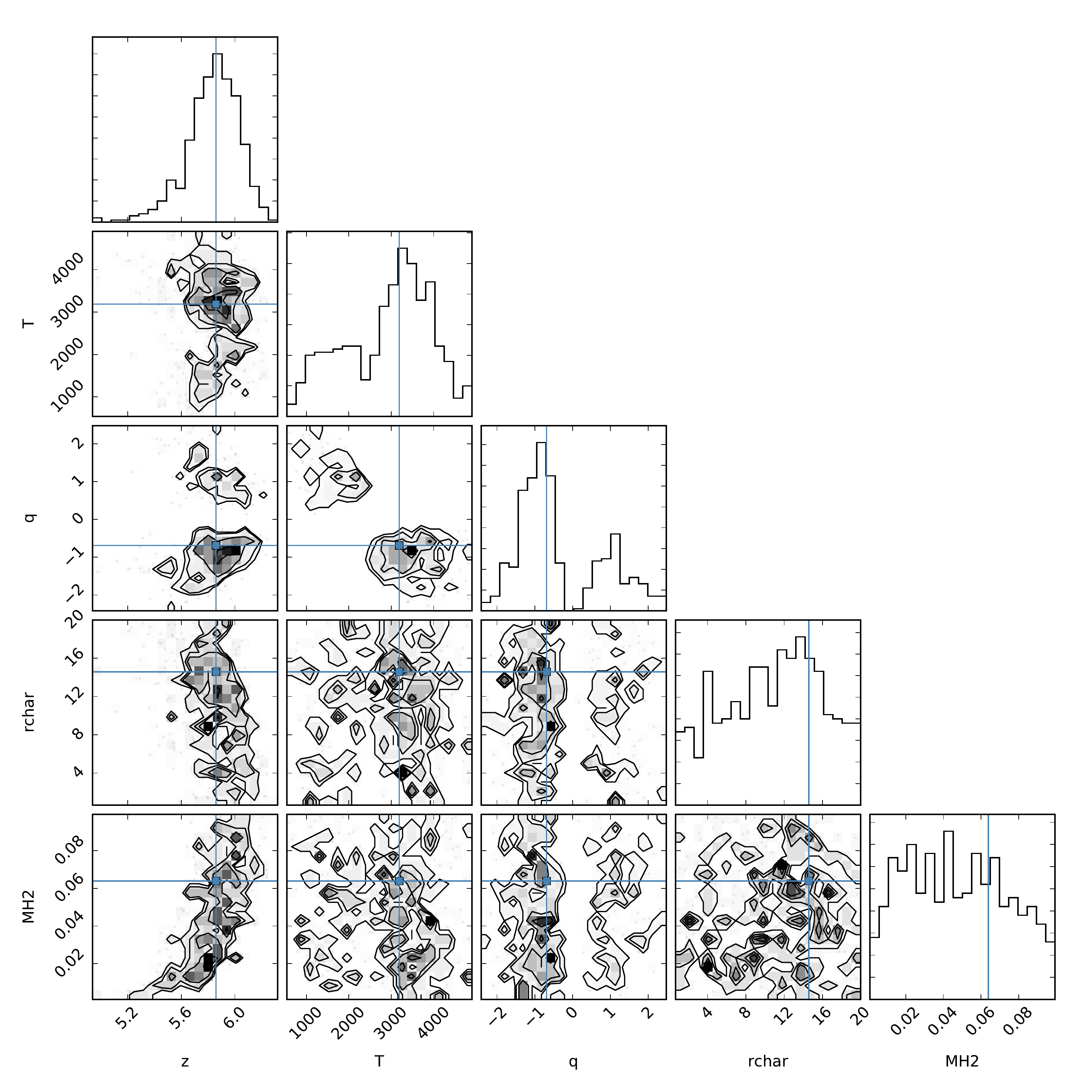}
\caption{Corner plot created from the MCMC chain for UX Tau. Most targets in our sample show double-peaked posterior distributions for $T_{1 \, \text{AU}}$ and $q$ and degeneracies between $z/r$ and $M_{H_2}$ \citep{Hoadley2015}.}
\label{UXTau_corner_plot}
\end{figure*}

The MCMC method allows for finer sampling of the posterior distributions than the discrete grid search method used in \citet{Hoadley2015}. For example, the temperature distributions in \citet{Hoadley2015} are spaced by 500 K, while the MCMC chain can probe temperature differences down to $<$1 K. The finer resolution shows distinct double-peaked posterior distributions on the $T$ and $q$ parameters for most targets, which were not possible to capture with the coarser grid search (see Figure \ref{UXTau_corner_plot}). 

\begin{figure*}
\centering
\includegraphics[width=1.0\linewidth]
{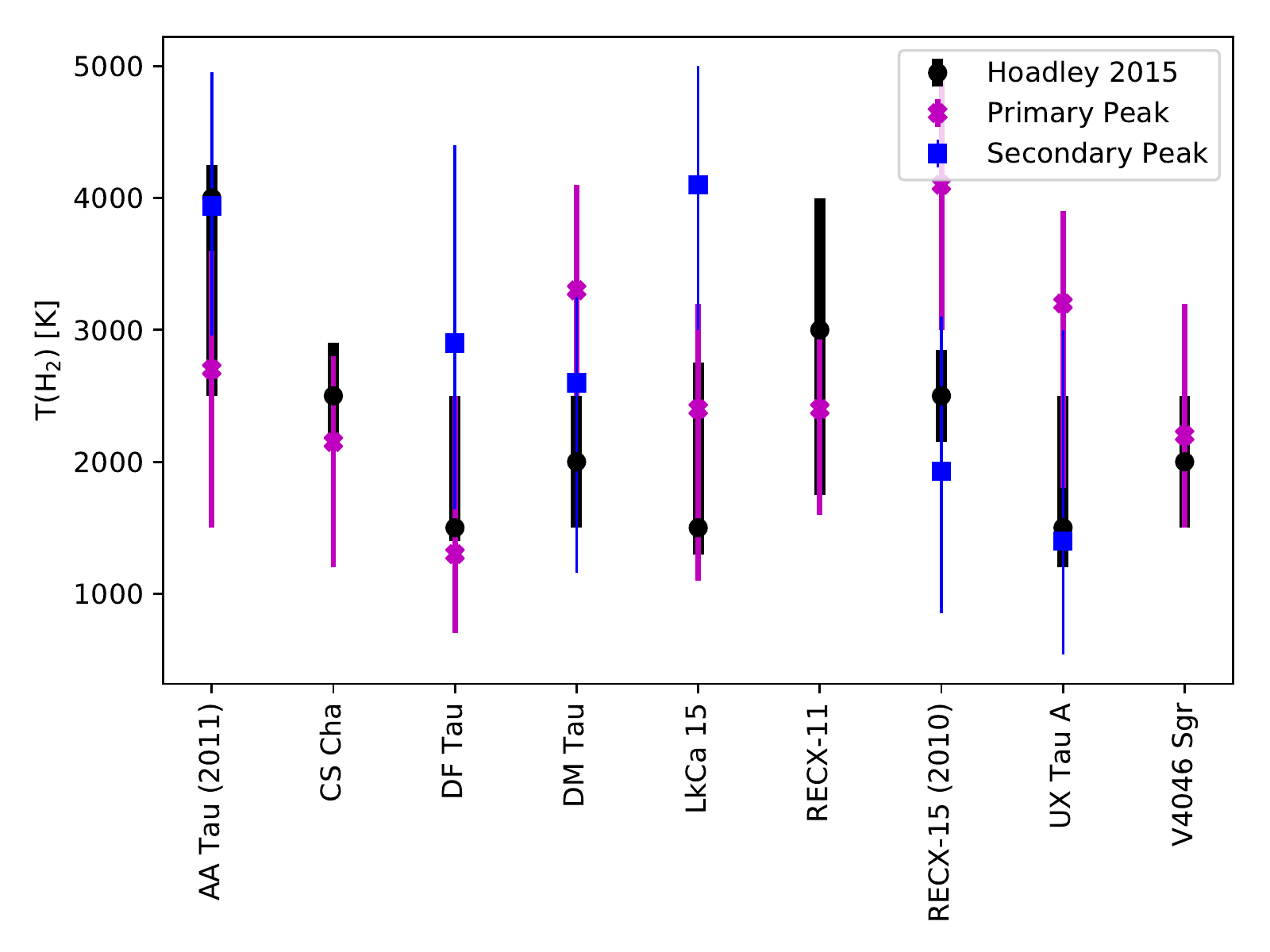}
\caption{Comparison of best-fit temperatures at 1 AU from \citet{Hoadley2015} and this work. The primary peak represents the temperature favored by the MCMC chain, while the secondary peak represents a high-likelihood temperature that was favored less than the primary peak.}
\label{Hoadley_NA_Tcomp}
\end{figure*}

Figure \ref{Hoadley_NA_Tcomp} compares the best-fit UV-H$_2$ temperatures from \citet{Hoadley2015} to two temperatures derived in this work. The first temperature, which we call the primary value, corresponds to the strongest peak in the posterior distribution. The secondary value is extracted from the weaker peak. In Figure \ref{UXTau_corner_plot}, which shows the marginalized posterior distributions for UX Tau, the primary value is $\sim$3200 K, and the secondary value is $\sim$1500 K. CS Cha, RECX-11, and V4046 Sgr do not show significant secondary peaks, so we only plot the primary temperature values in Figure \ref{Hoadley_NA_Tcomp}. 

We find that the temperatures from \citet{Hoadley2015} are consistent with either the primary or secondary temperature value for all targets. The values for AA Tau (2011), DM Tau, RECX-15 (2010), and UX Tau show very close agreement with the secondary peak temperature values, indicating that the primary peak may not have been fully sampled with the grid search method. However, the secondary peak parameters represent equivalent model fits to the data that the MCMC re-sampling simply did not favor as strongly as the primary peak parameters.   

\begin{figure*}
\centering
\includegraphics[width=1.0\linewidth]
{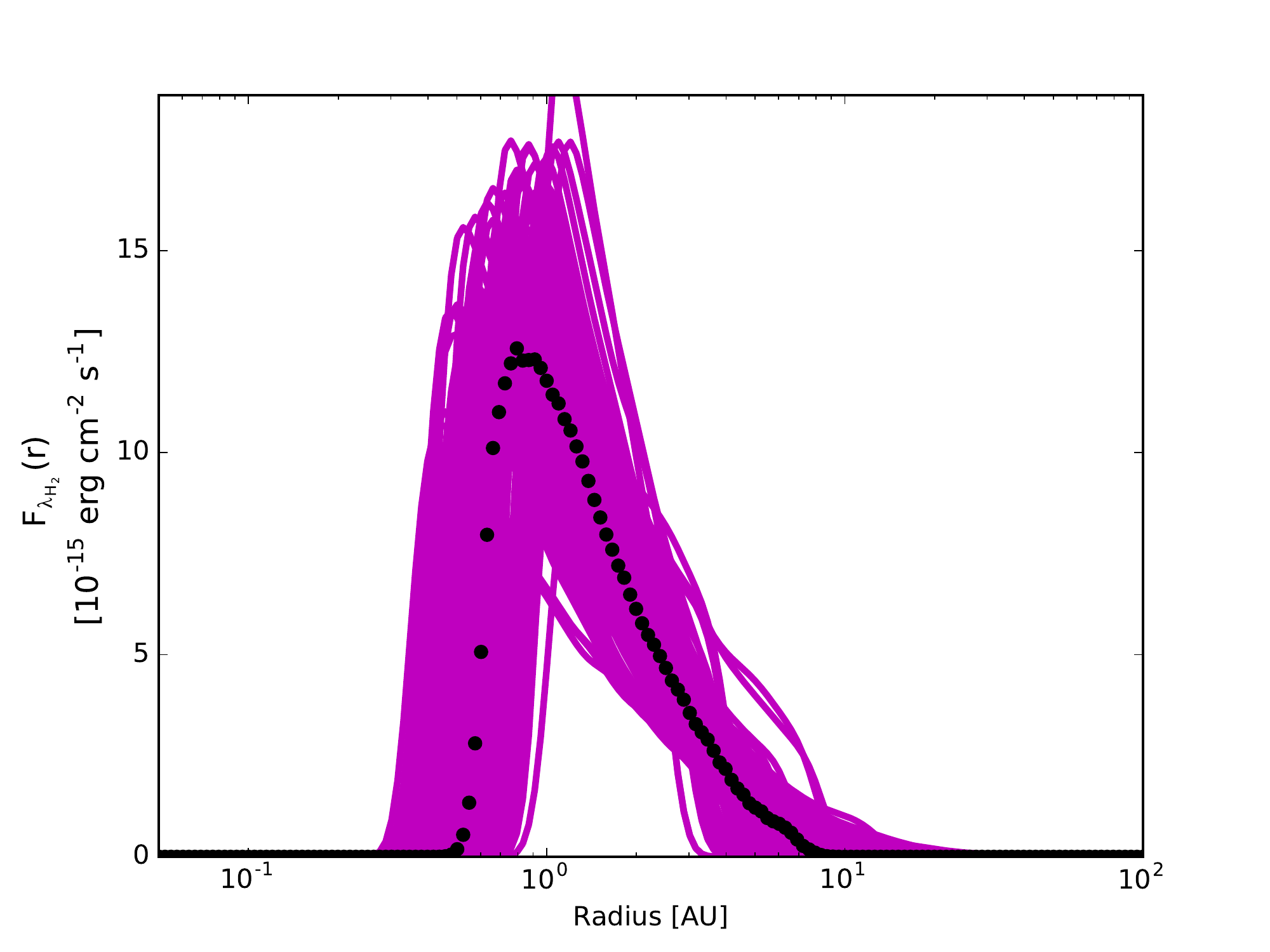}
\caption{Radial distributions of UV-H$_2$ flux from the 100 best-fit models to the UX Tau spectrum (pink). The black dots represent the median flux values at each radial grid point.}
\label{top100_fluxdists}
\end{figure*}

To propagate the uncertainties in the best-fit model parameters to error bounds on the radial flux distributions, including those with double-peaked posterior distributions, we generate models from the 100 sets of parameters with the smallest values of the goodness-of-fit $\left(MSE \right)$ test statistic (see Equation 12). This allows us to capture flux distributions corresponding to both the primary and secondary temperature values, without including intermediate temperatures with significantly lower probabilities. The resulting distributions of UV-H$_2$ flux are shown in Figure \ref{top100_fluxdists}. We obtain final UV-H$_2$ flux distributions by calculating the median flux value at each radial grid point, and error bounds are estimated as the minimum and maximum fluxes. The results represent the uncertainties in the fluorescent flux distributions as a result of the uncertainties in the best-fit model parameters. 

\begin{figure*}
\centering
\includegraphics[width=1.0\linewidth]
{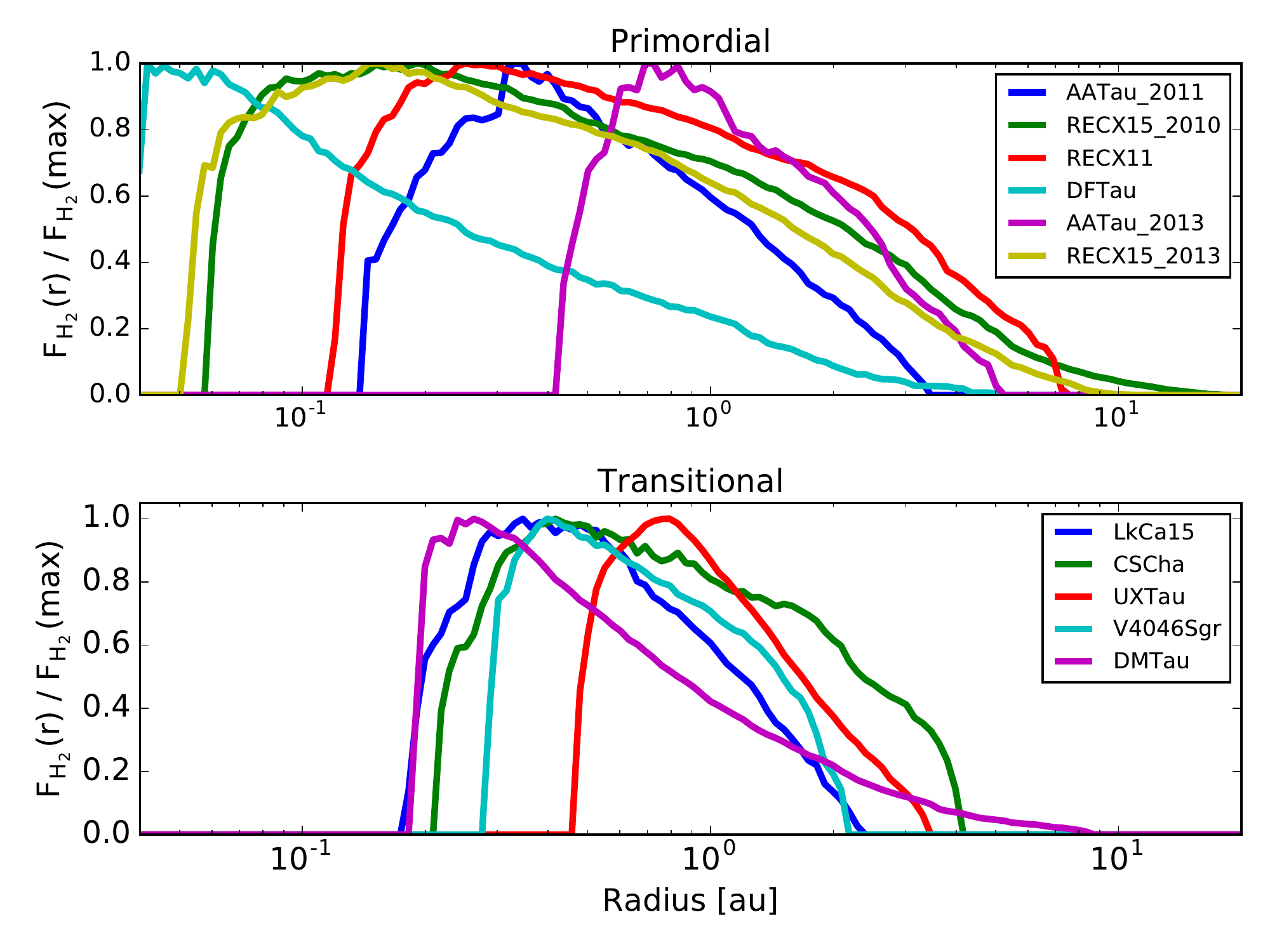}
\caption{Comparison of radial flux distributions from primordial versus transitional disks. We find that the adapted radiative transfer models reproduce the trends discovered by \citet{Hoadley2015}, where primordial systems have more UV-H$_2$ emission at radii close to the star than transitional disks.}
\label{Hoadley_NA_fluxdistcomp}
\end{figure*}

Figure \ref{Hoadley_NA_fluxdistcomp} compares the median UV-H$_2$ flux distributions for the primordial and transitional disks, as classified and presented in the same way as Figure 9 of \citealt{Hoadley2015}. We find that the flux distributions inside $\sim$0.1 AU follow the trends discovered in that work, with transitional disks showing no emitting material this close to the star and primordial disks showing strong peaks at these radii. However, we note a discrepancy with the outer radii of the distributions presented here. This is caused by different methods of calculating the partition functions and effective optical depths in the modeling code, which accounts for reducing the column density of fluorescent gas at radii $>$10 AU. 

Instead of seeing UV-H$_2$ distributions extending to more distant radii in transitional disks than in primordial systems \citep{Hoadley2015}, we observe truncated flux distributions in the transitional disks. This may be attributed to the clearing of warm dust in the inner disk $\left(r < 50 \, \rm{AU} \right)$, which exposes molecules in surface layers of the disks to intense photodissociating radiation fields that also dissipate the gas in warm winds. By contrast, the extended distributions of fluorescent material in the primordial disks may be indicative of higher total gas masses in systems with larger remaining dust reservoirs. However, as noted by \citet{Hoadley2015}, the 2-D radiative transfer models of H$_2$ in a thin disk surface layer are not particularly sensitive to the mass contained in the underlying bulk gas reservoir.        

\section{\\ 2-D Radiative Transfer Modeling Results for UV-H$_2$ and UV-CO} \label{App:Appendix B}
\setcounter{table}{0}
\renewcommand{\thetable}{A\arabic{table}}

\begin{figure*}[t!]
	\begin{minipage}{0.5\textwidth}
	\centering
	\includegraphics[width=\linewidth]{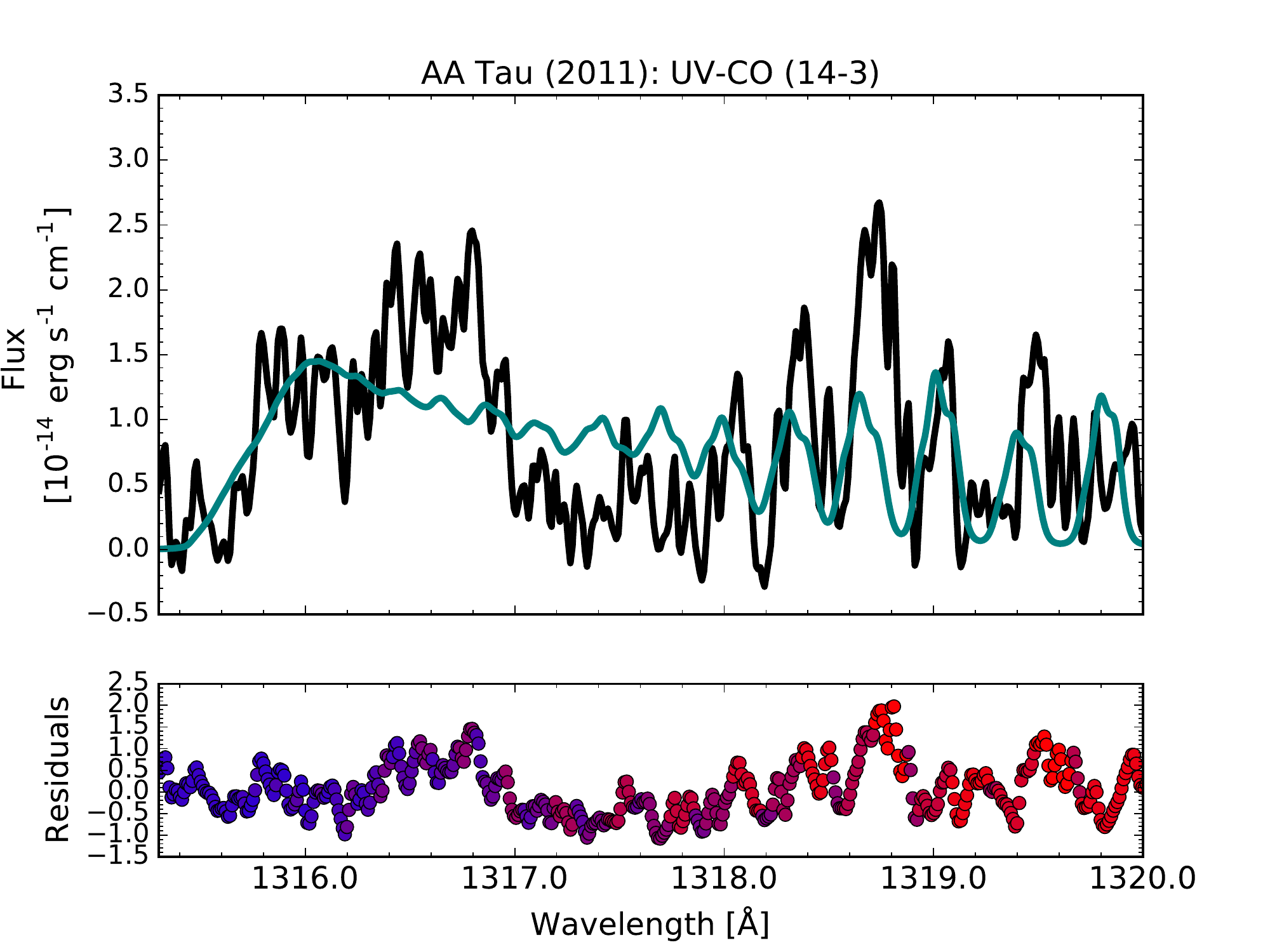}
	\end{minipage}
	\begin{minipage}{0.5\textwidth}
	\centering
	\includegraphics[width=\linewidth]{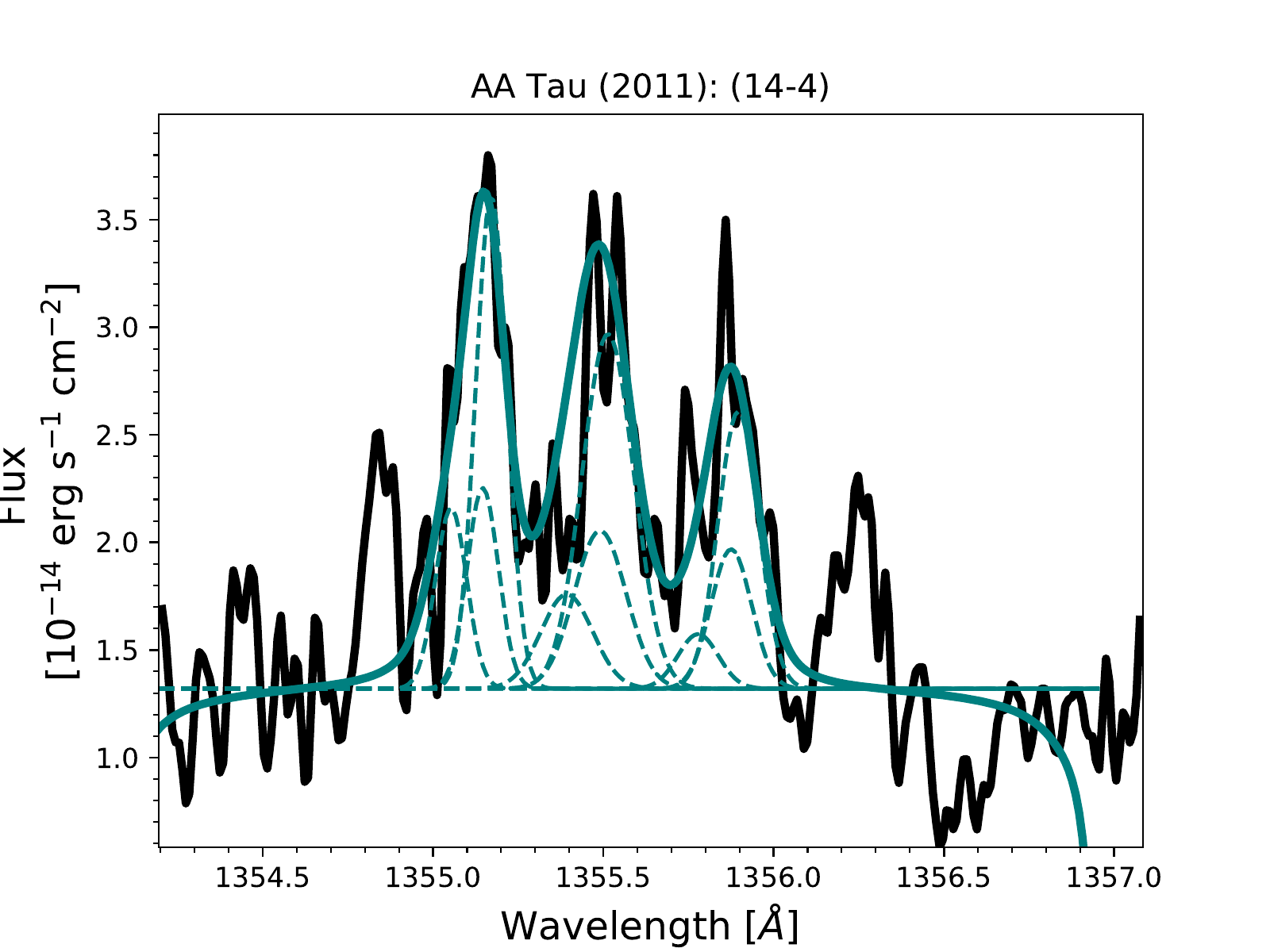}
	\end{minipage}
	\begin{minipage}{0.5\textwidth}
	\centering
	\includegraphics[width=\linewidth]{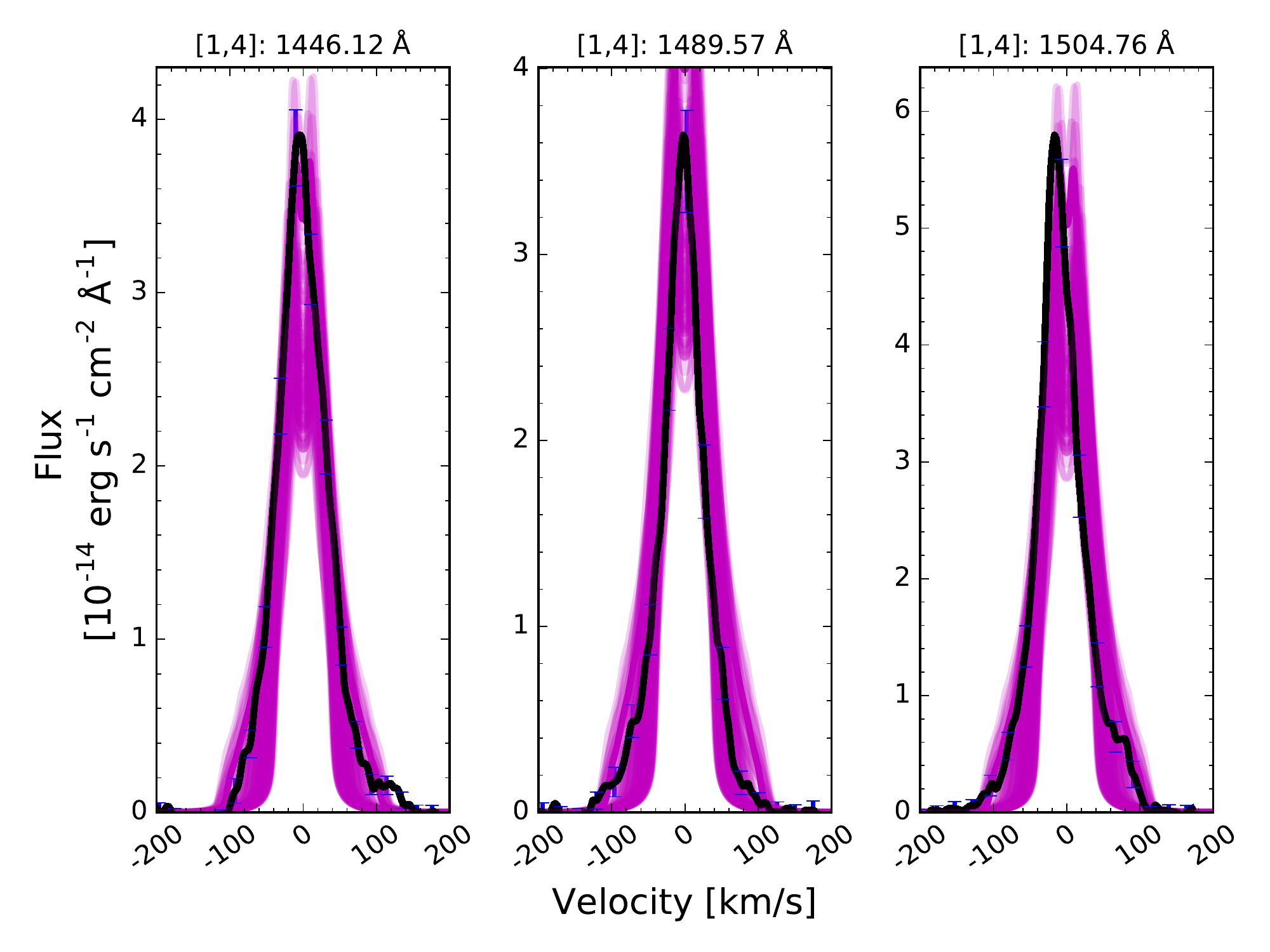} 
	\end{minipage}
	\begin{minipage}{0.5\textwidth}
	\centering
	\includegraphics[width=\linewidth]{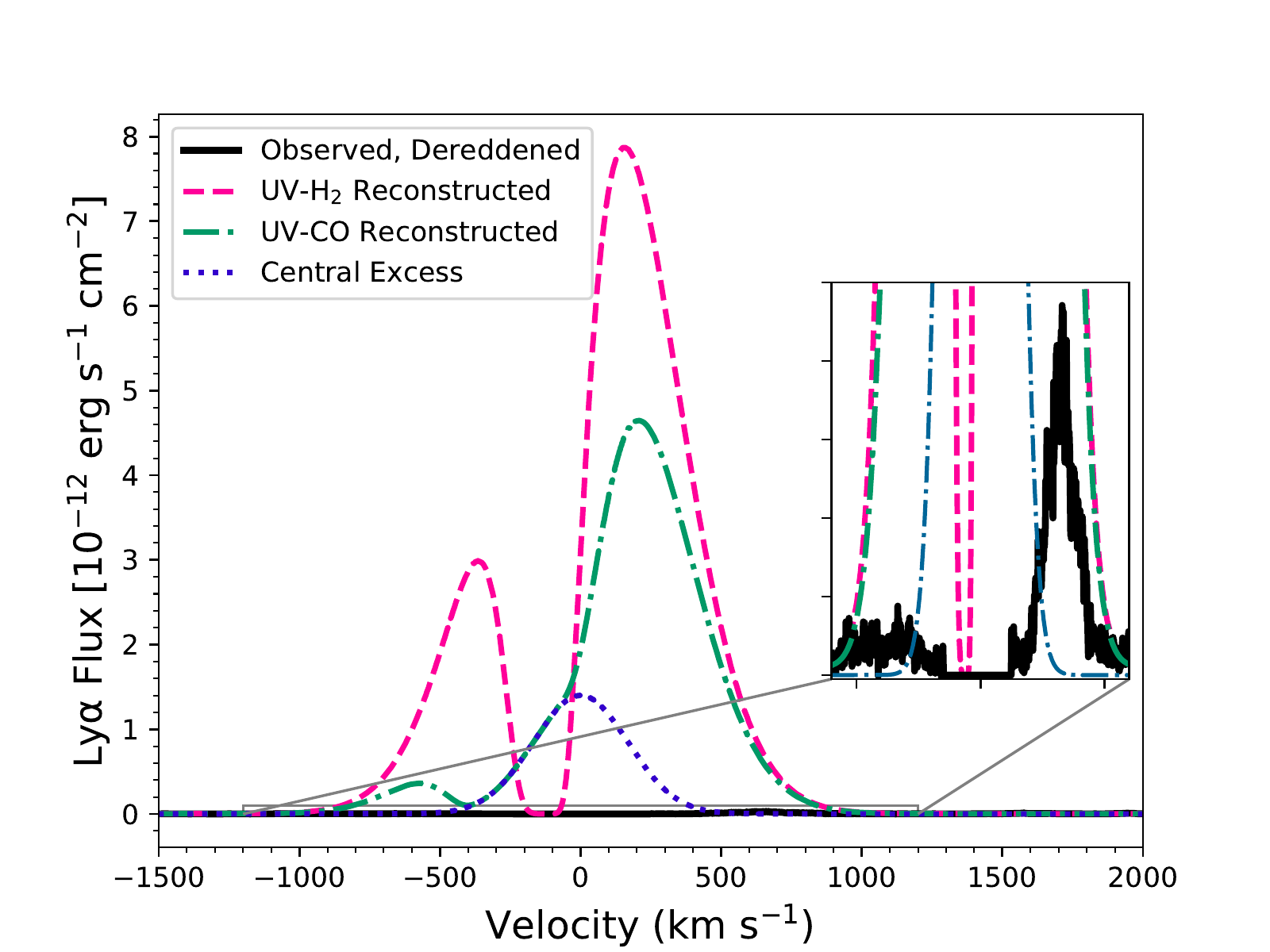} 
	\end{minipage}
	\begin{minipage}{0.5\textwidth}
	\centering
	\includegraphics[width=\linewidth]{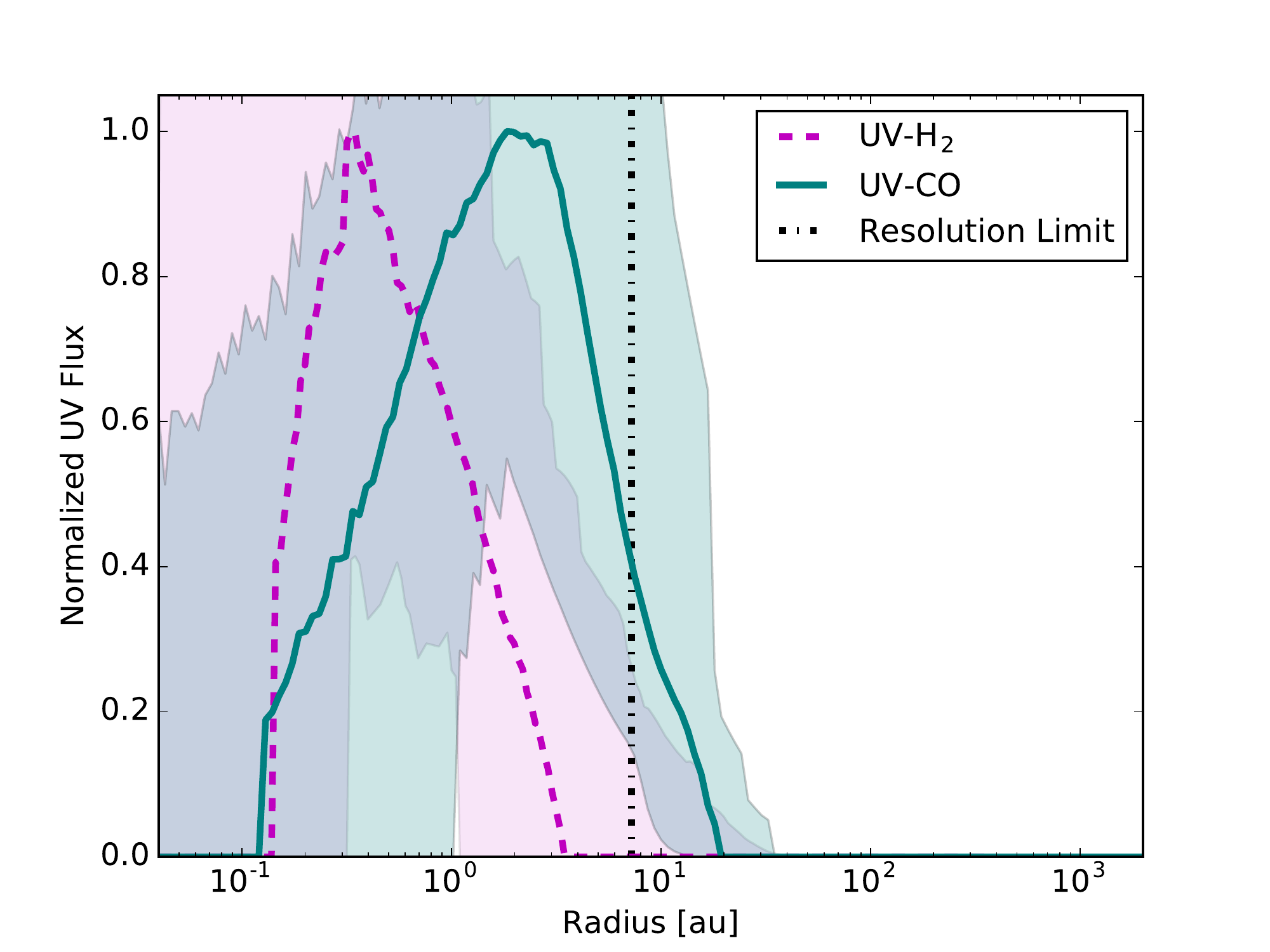} 
	\end{minipage}
\caption{Best-fit UV-CO 2-D radiative transfer (top left) and Gaussian models (top right), 100 UV-H$_2$ radiative transfer models with the smallest \textit{MSE}s (middle left), comparison of UV-H$_2$-/UV-CO-based and observed Ly$\alpha$ profiles (middle right) and radial distributions of flux for both species (bottom left) from the disk around AA Tau in 2011. Residuals on the best-fit UV-CO model are color-coded from blue to red, based on the Ly$\alpha$ pumping wavelength required to excite the upper level of the transition. The largest deviations between the models and data correspond to states with upper levels pumped by central and redder Ly$\alpha$ photons, causing the models to underpredict the observed flux. This effect is not caused by radial velocity effects, but rather missing Ly$\alpha$ irradiation, as supported by the model reproducing the observed spacing between features while still missing the amplitudes of the emission lines. The full figure set, showing UV-CO and UV-H$_2$ radiative transfer model results, empirical UV-CO models (when available), and reconstructed and observed Ly$\alpha$ profiles for all targets, is included in the online journal.}
\label{AATau_2011_H2CO}
\end{figure*}

\begin{figure*}[t!]
	\begin{minipage}{0.45\textwidth}
	\centering
	\includegraphics[width=\linewidth]{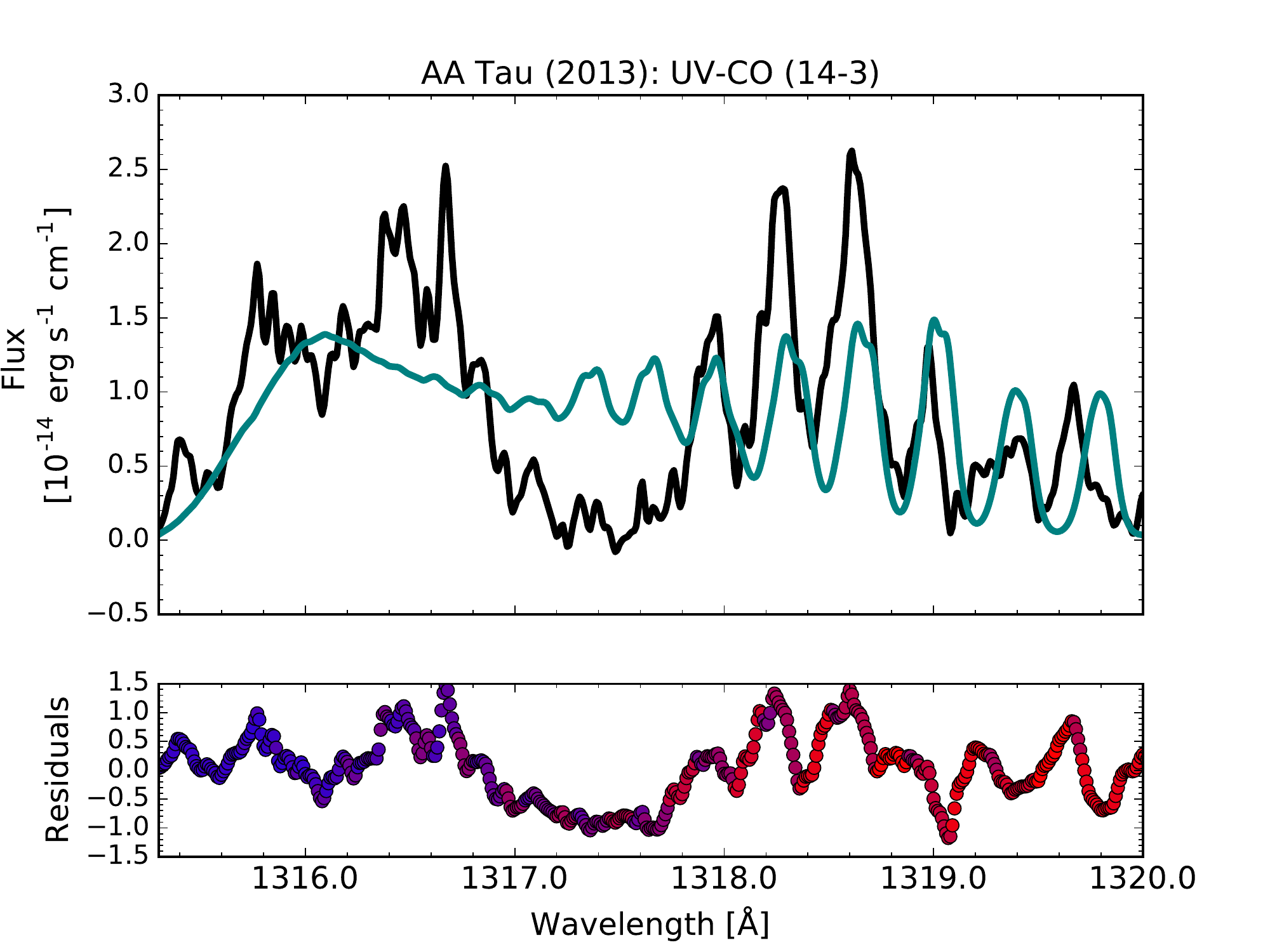}
	\end{minipage}
	\begin{minipage}{0.45\textwidth}
	\centering
	\includegraphics[width=\linewidth]{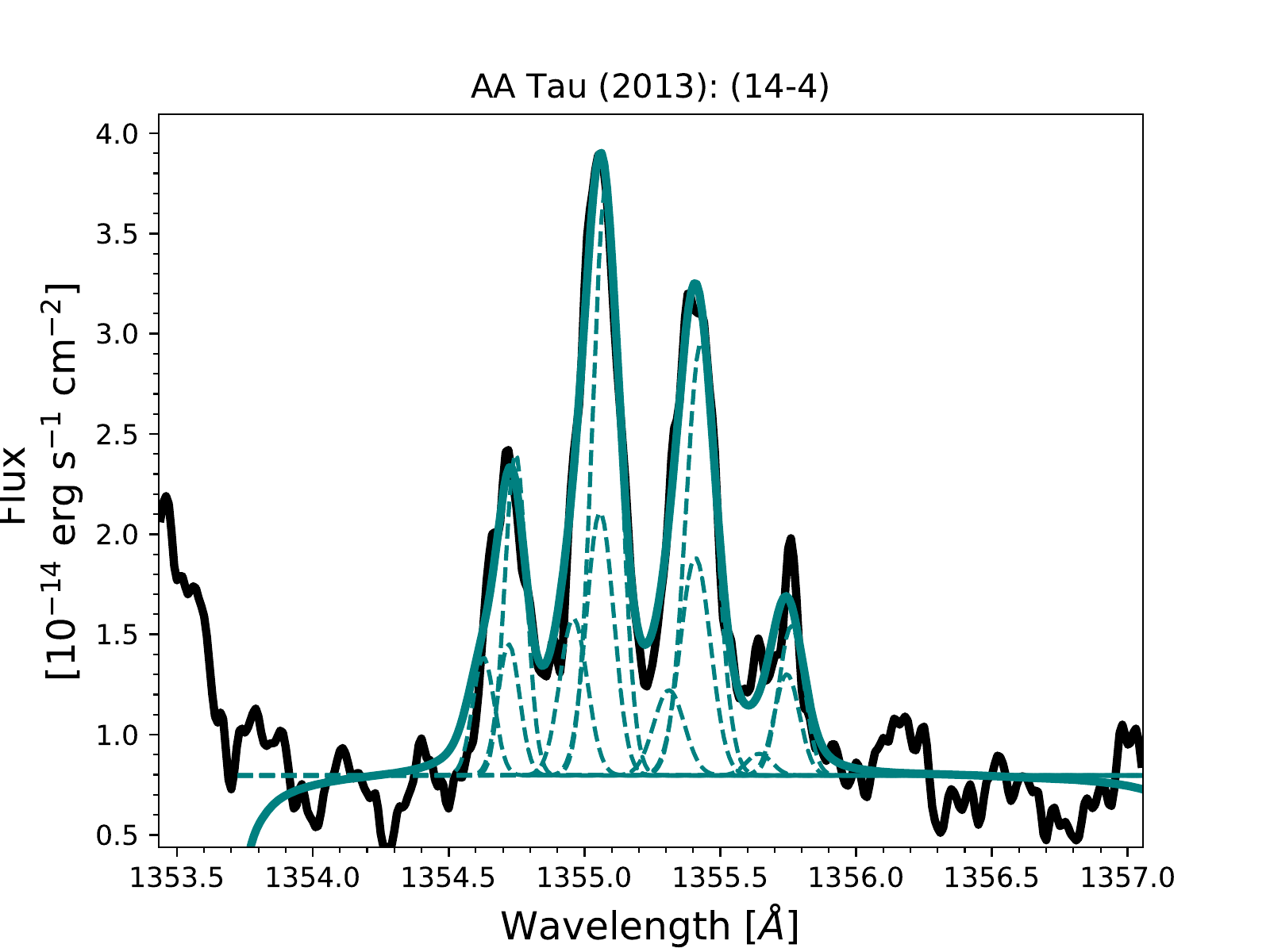}
	\end{minipage}
	\begin{minipage}{0.45\textwidth}
	\centering
	\includegraphics[width=\linewidth]{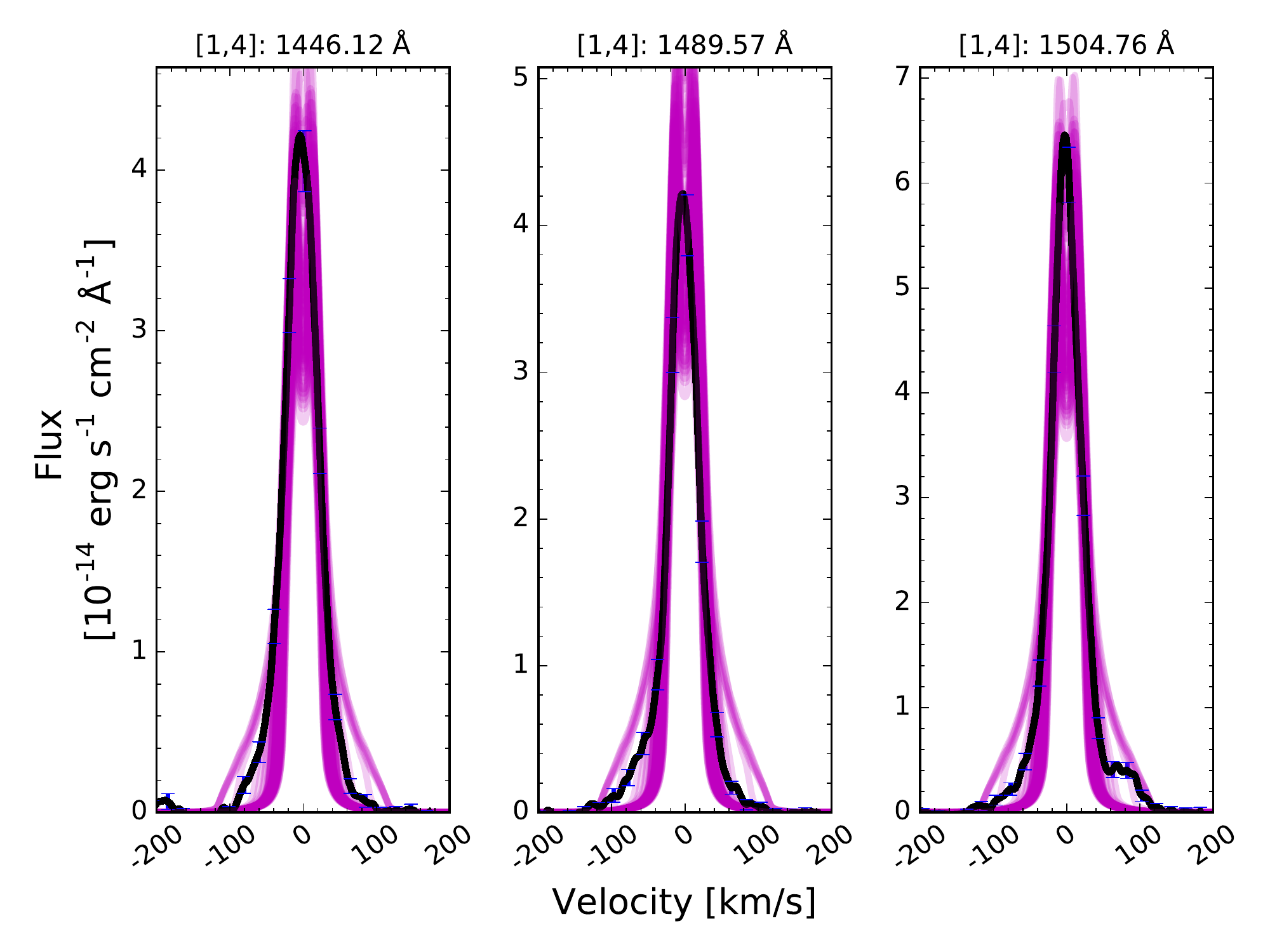} 
	\end{minipage}
	\begin{minipage}{0.45\textwidth}
	\centering
	\includegraphics[width=\linewidth]{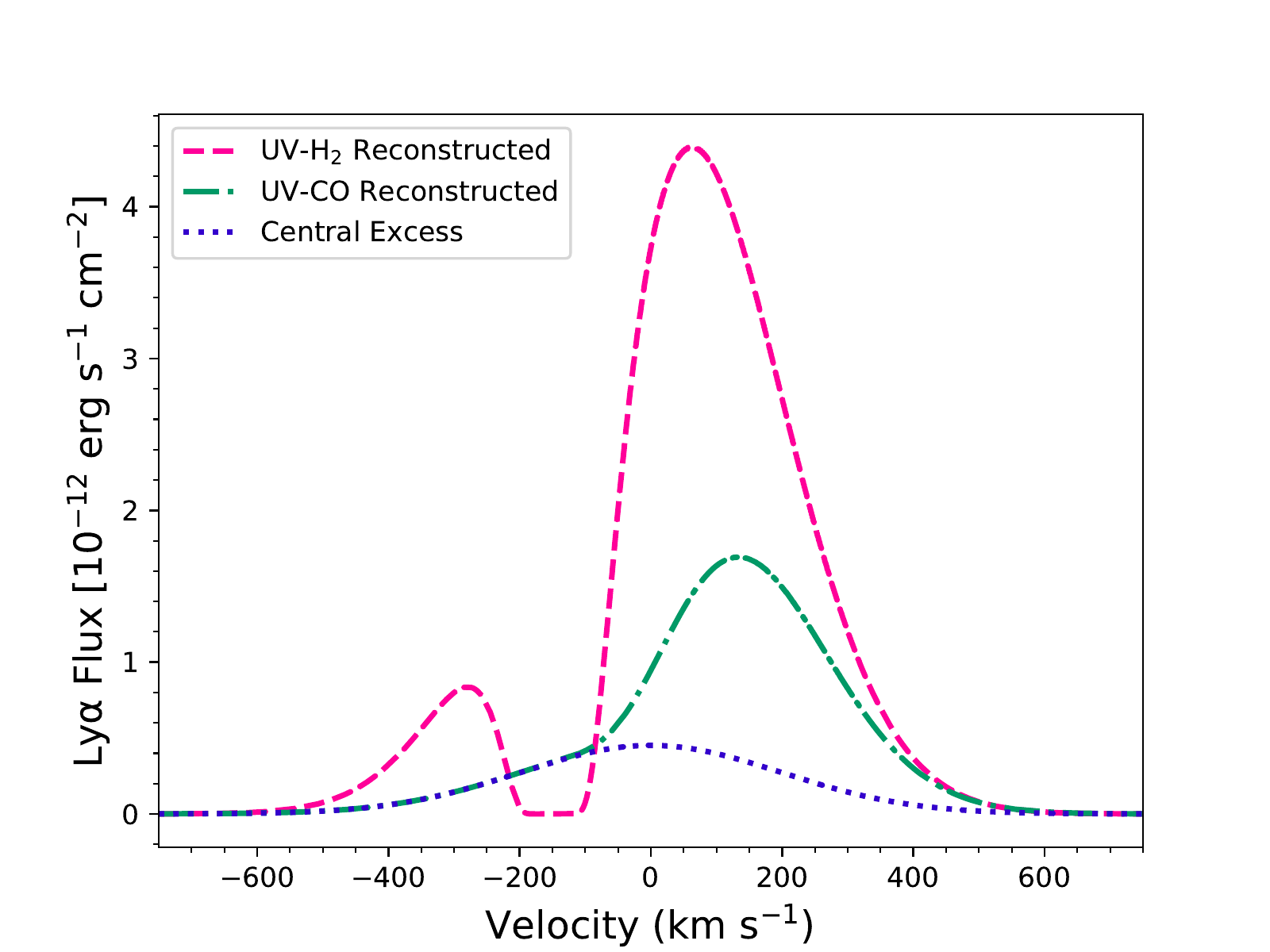} 
	\end{minipage}
	\begin{minipage}{0.45\textwidth}
	\centering
	\includegraphics[width=\linewidth]{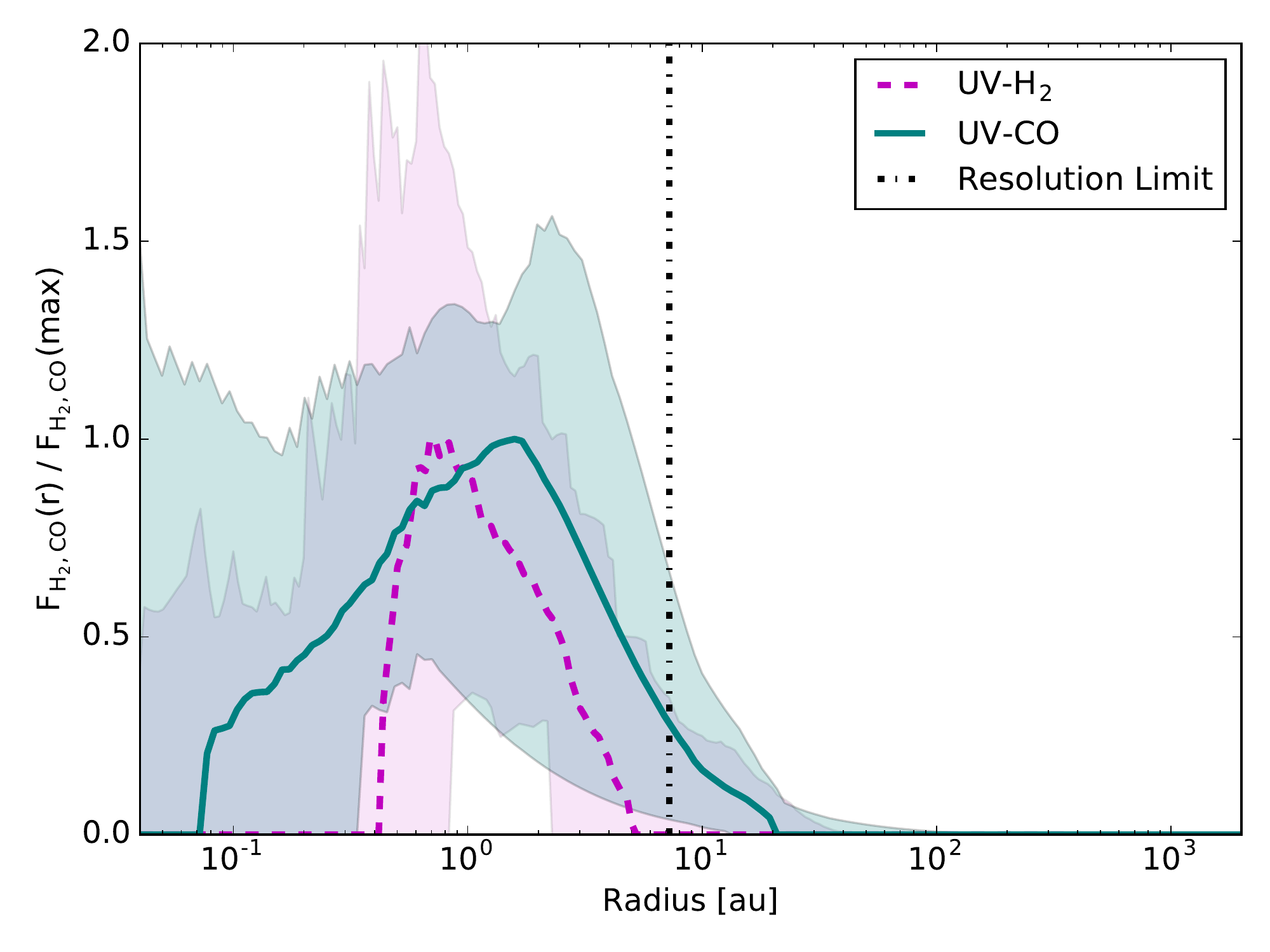} 
	\end{minipage}
	\begin{minipage}{0.45\textwidth}
	\centering
	\includegraphics[width=\linewidth]{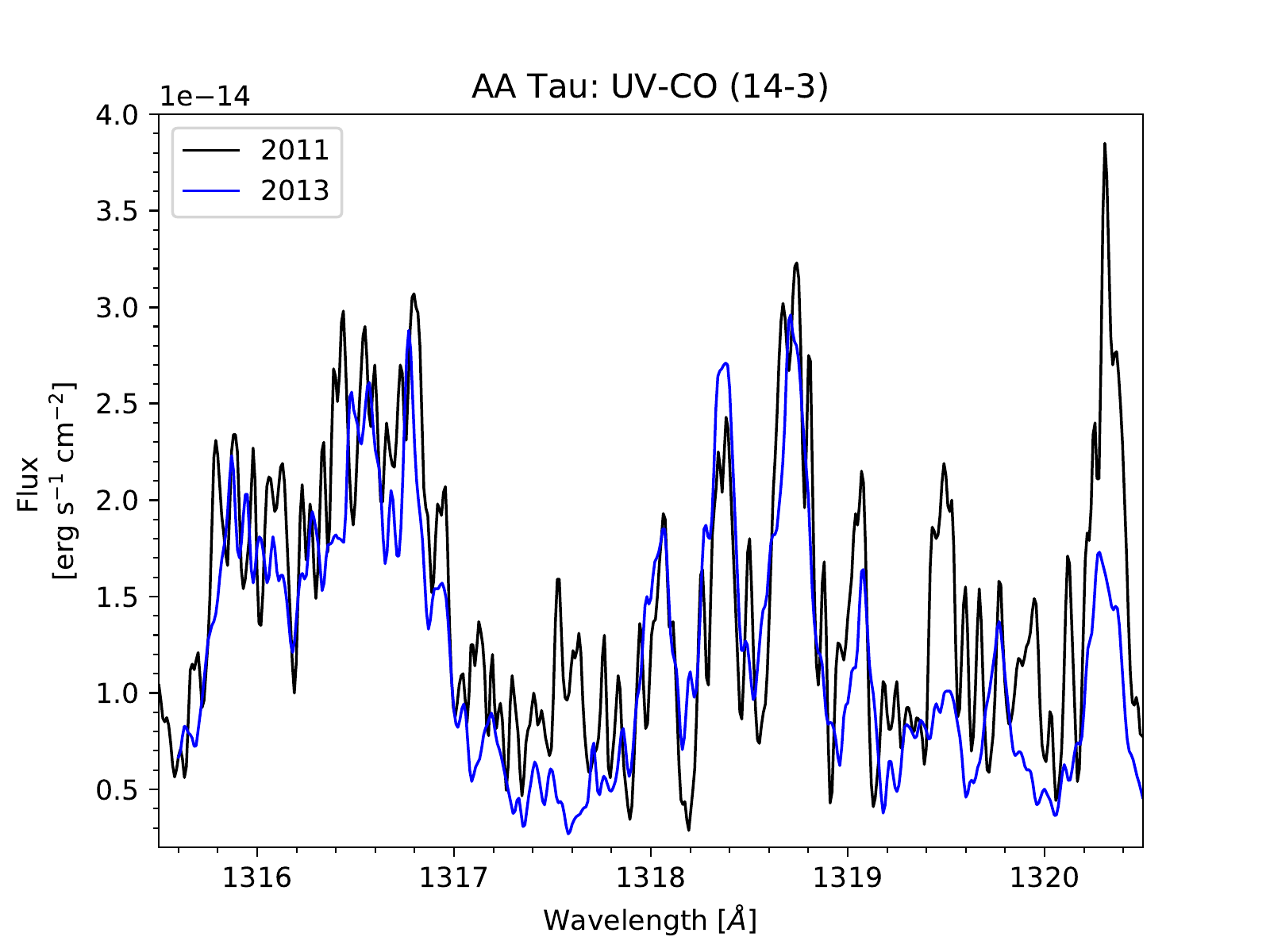} 
	\end{minipage}
\caption{Best-fit UV-CO 2-D radiative transfer (top left) and Gaussian models (top right), 100 UV-H$_2$ radiative transfer models with the smallest \textit{MSE}s (middle left), comparison of UV-H$_2$-/UV-CO-based Ly$\alpha$ profiles (middle right) and radial distributions of flux for both species (bottom left) from the disk around AA Tau in 2013. Residuals on the best-fit UV-CO model are color-coded from blue to red, based on the Ly$\alpha$ pumping wavelength required to excite the upper level of the transition. The two epochs of UV-CO spectra from AA Tau (bottom right) are offset by $\sim$24 km s$^{-1}$. Since we do not see significant wavelength shifts between the individual transitions when the data are overplotted (bottom right), this shift is likely caused by the uncertainty in the \textit{HST}-COS FUV wavelength solution ($\sim$15 km s$^{-1}$). However, the UV-CO and UV-H$_2$ fluxes do change between epochs, and the differences in best-fit models to the two spectra are caused by the uncertainties in the best-fit Ly$\alpha$ profiles required to reproduce the features.}
\label{AATau_2013_H2CO}
\end{figure*}

\begin{figure*}[t!]
	\begin{minipage}{0.5\textwidth}
	\centering
	\includegraphics[width=\linewidth]{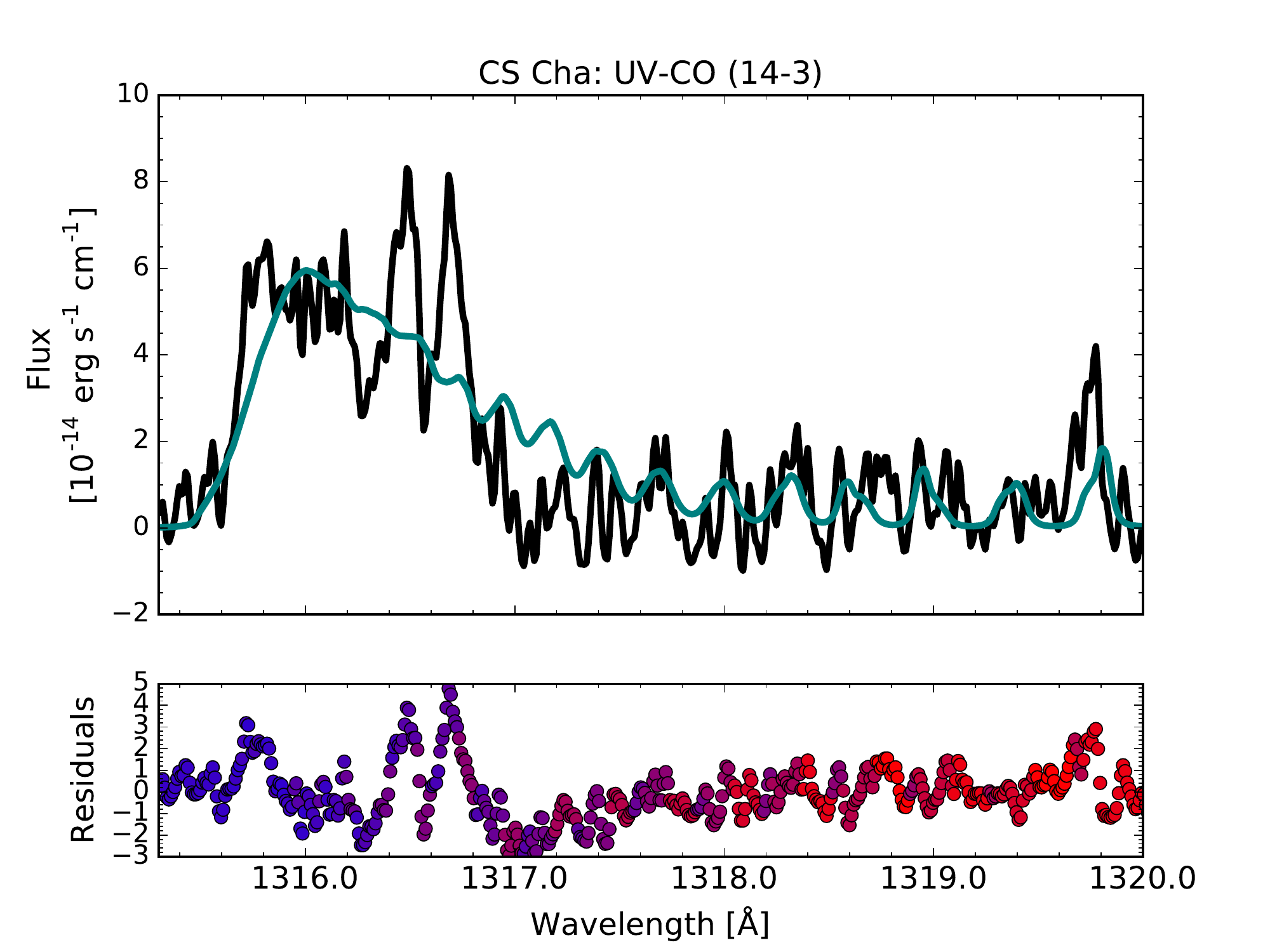}
	\end{minipage}
	\begin{minipage}{0.5\textwidth}
	\centering
	\includegraphics[width=\linewidth]{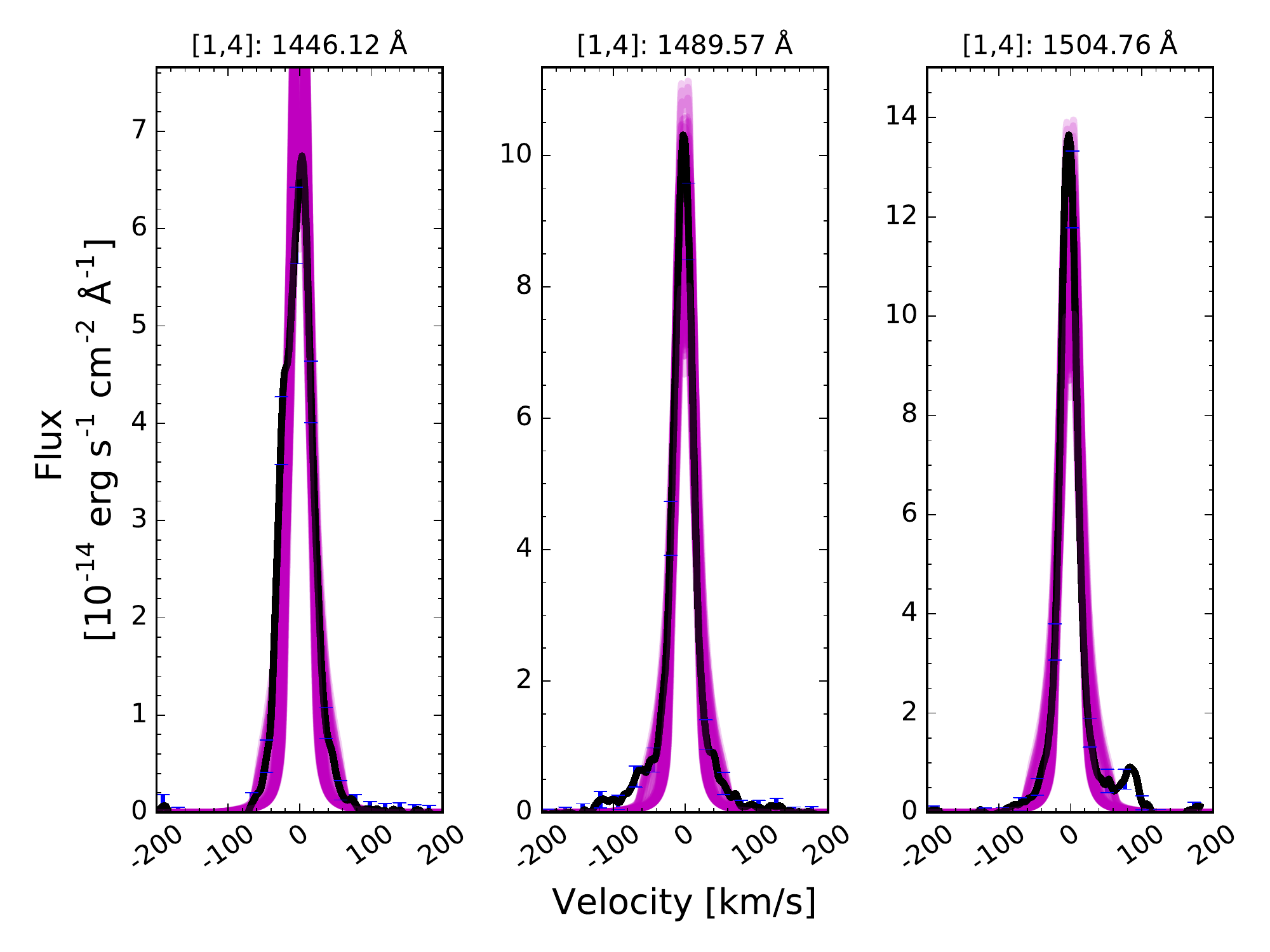} 
	\end{minipage}
	\begin{minipage}{0.5\textwidth}
	\centering
	\includegraphics[width=\linewidth]{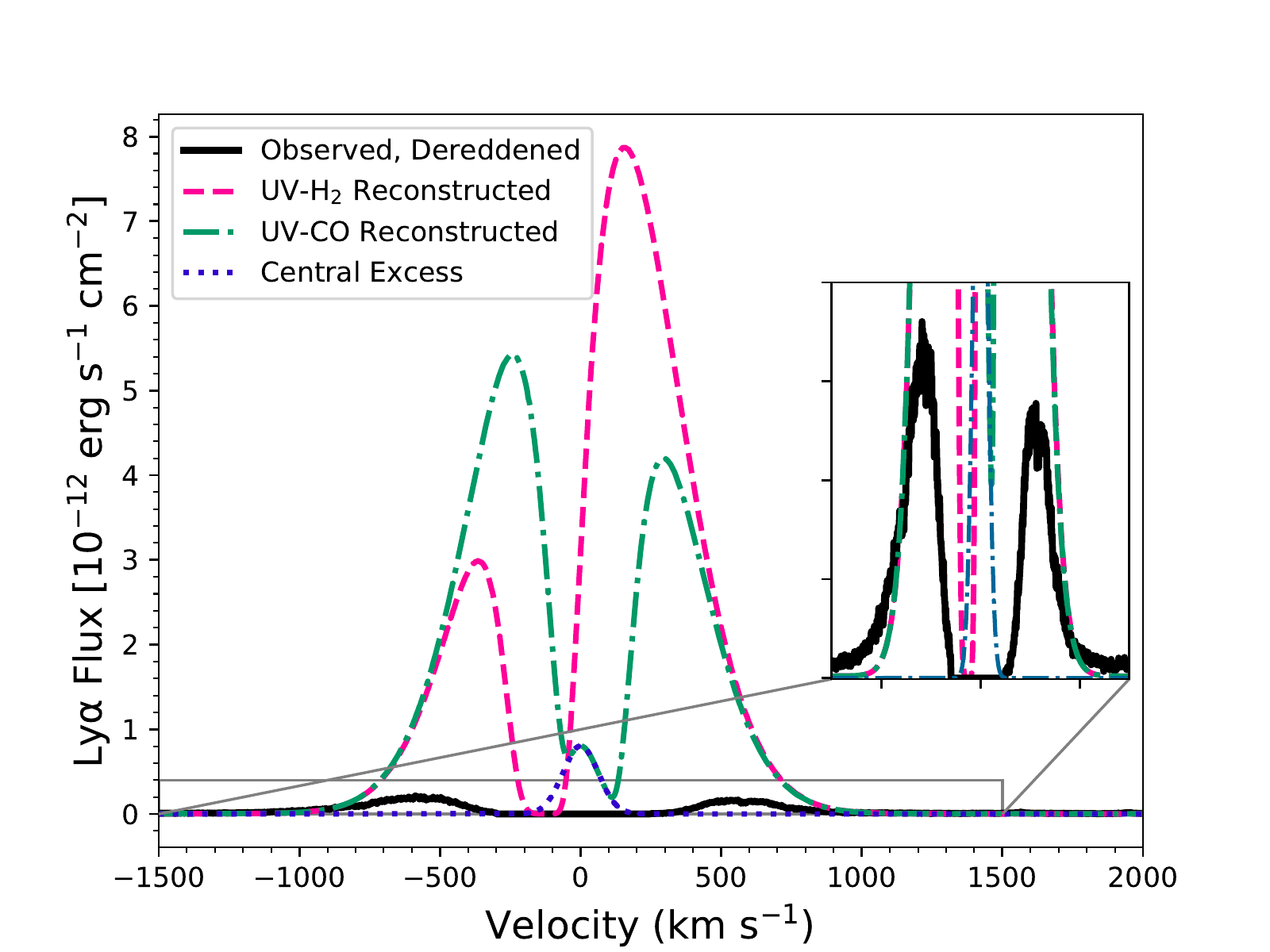} 
	\end{minipage}
	\begin{minipage}{0.5\textwidth}
	\centering
	\includegraphics[width=\linewidth]{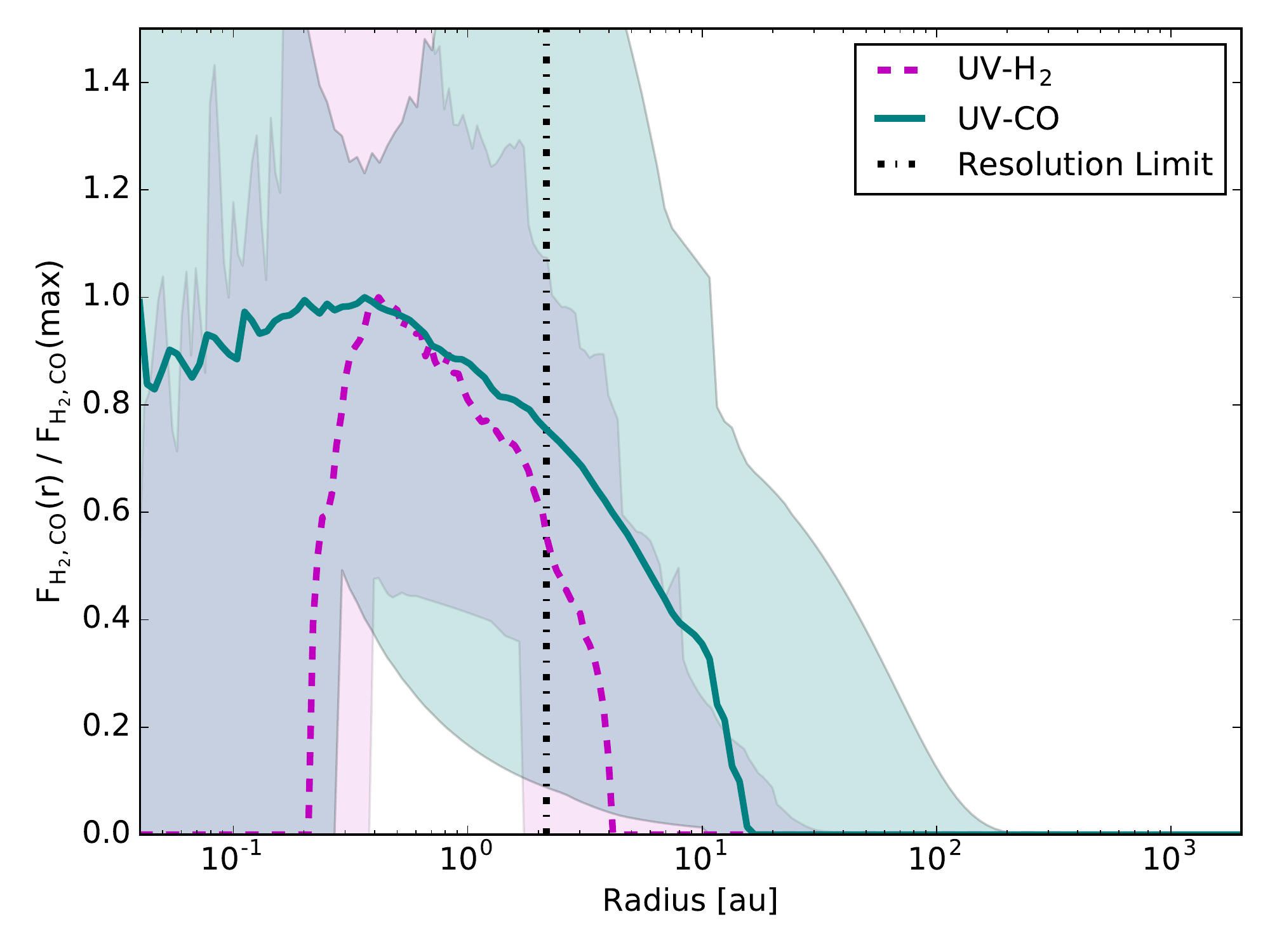} 
	\end{minipage}
\caption{Best-fit UV-CO model (top left), 100 UV-H$_2$ radiative transfer models with the smallest \textit{MSE}s (top right), comparison of UV-H$_2$-/UV-CO-based and observed Ly$\alpha$ profiles (bottom left) and radial distributions of flux for both species (bottom right) from the disk around CS Cha. Residuals on the best-fit UV-CO model are color-coded from blue to red, based on the Ly$\alpha$ pumping wavelength required to excite the upper level of the transition.}
\label{CSCha_H2CO}
\end{figure*}

\begin{figure*}[t!]
	\begin{minipage}{0.5\textwidth}
	\centering
	\includegraphics[width=\linewidth]{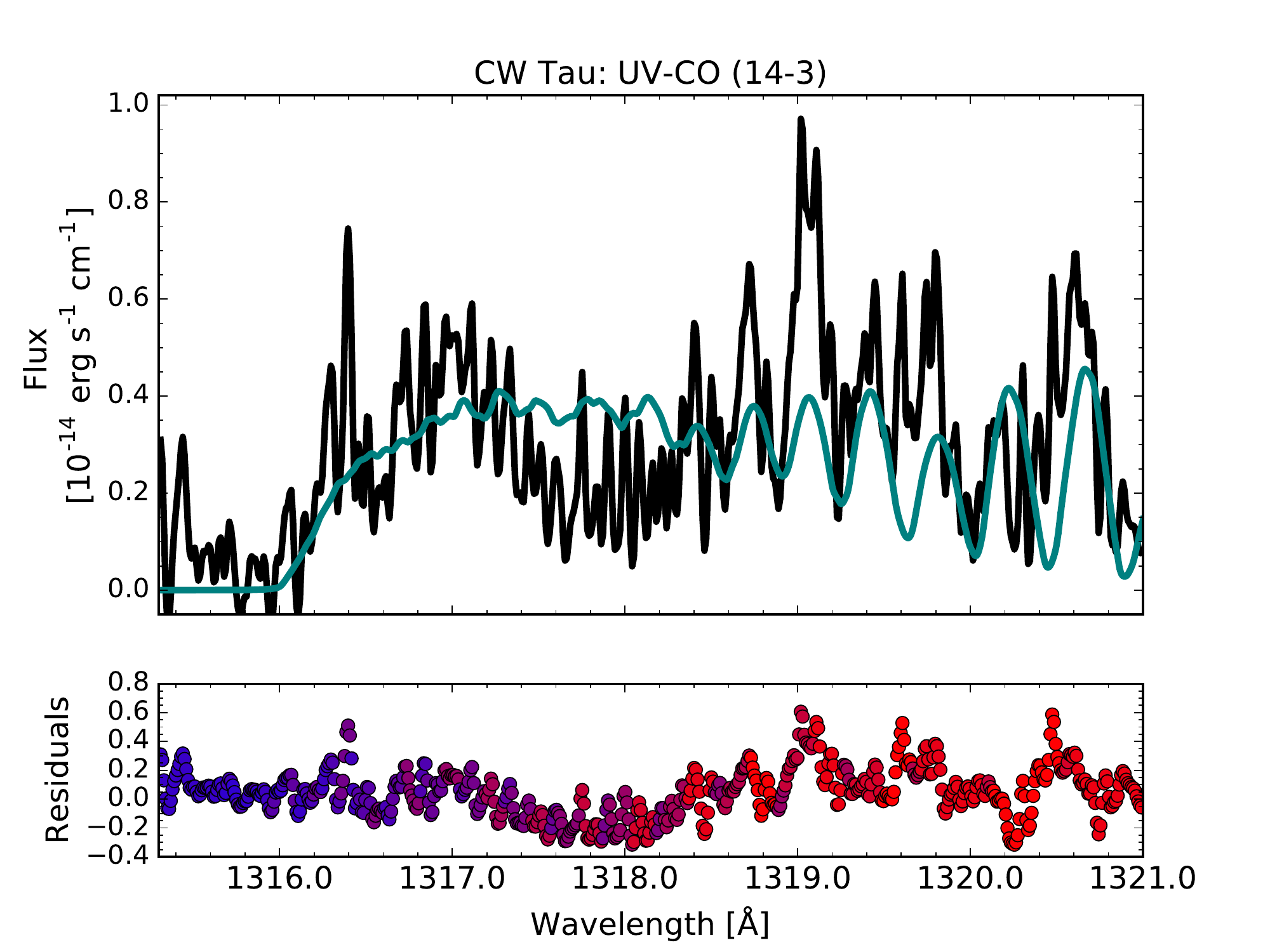}
	\end{minipage}
	\begin{minipage}{0.5\textwidth}
	\centering
	\includegraphics[width=\linewidth]{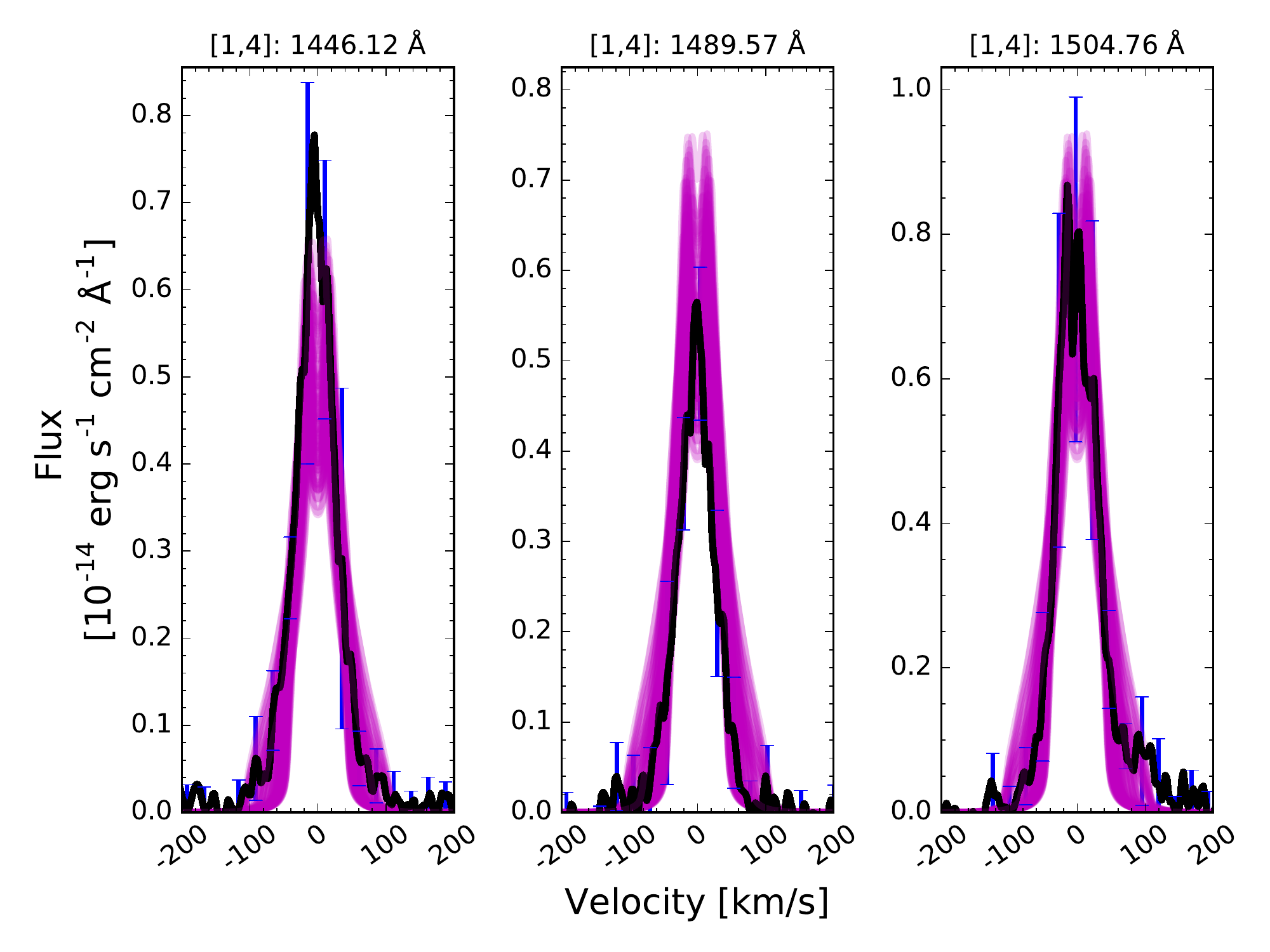} 
	\end{minipage}
	\begin{minipage}{0.5\textwidth}
	\centering
	\includegraphics[width=\linewidth]{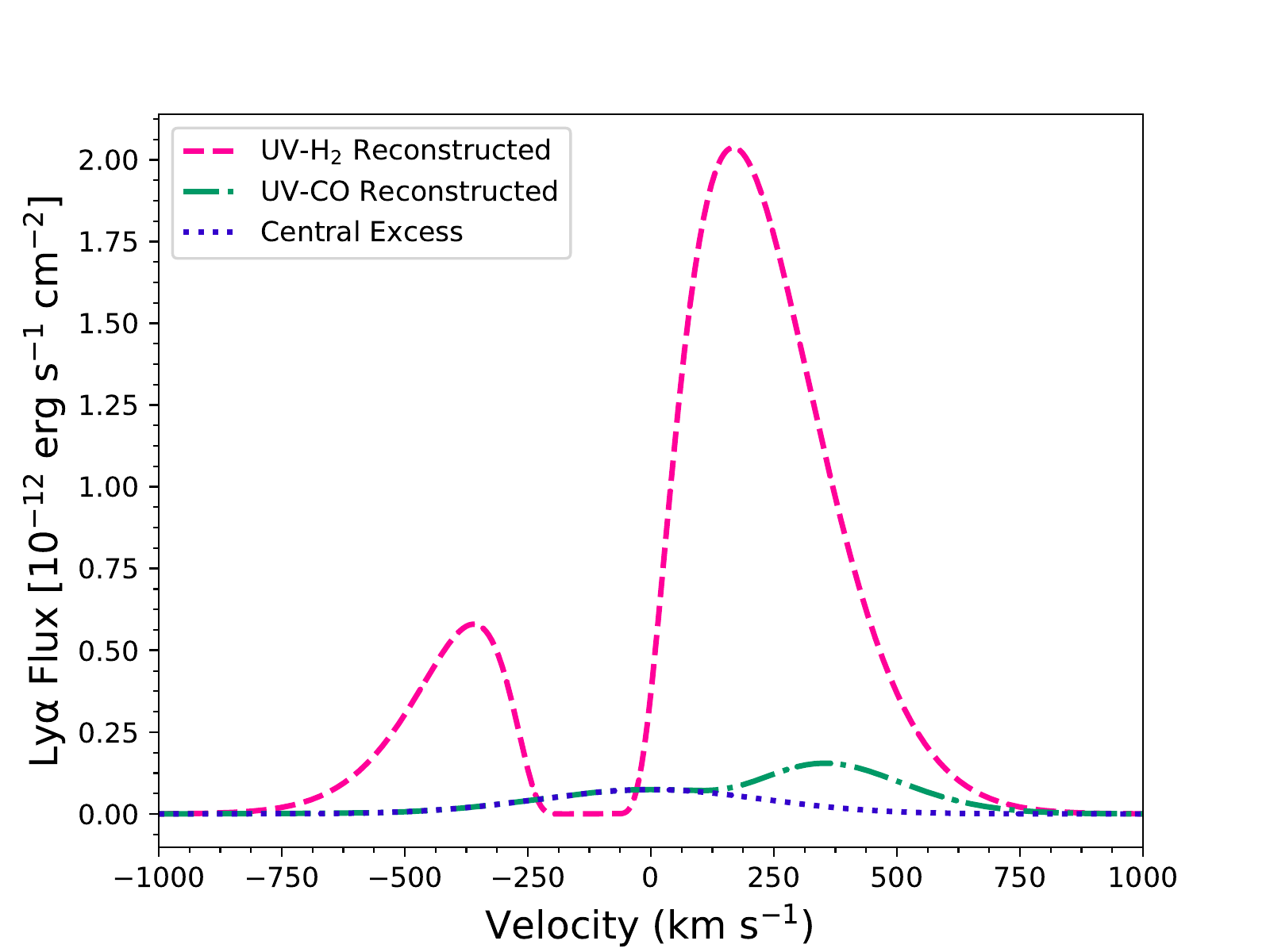} 
	\end{minipage}
	\begin{minipage}{0.5\textwidth}
	\centering
	\includegraphics[width=\linewidth]{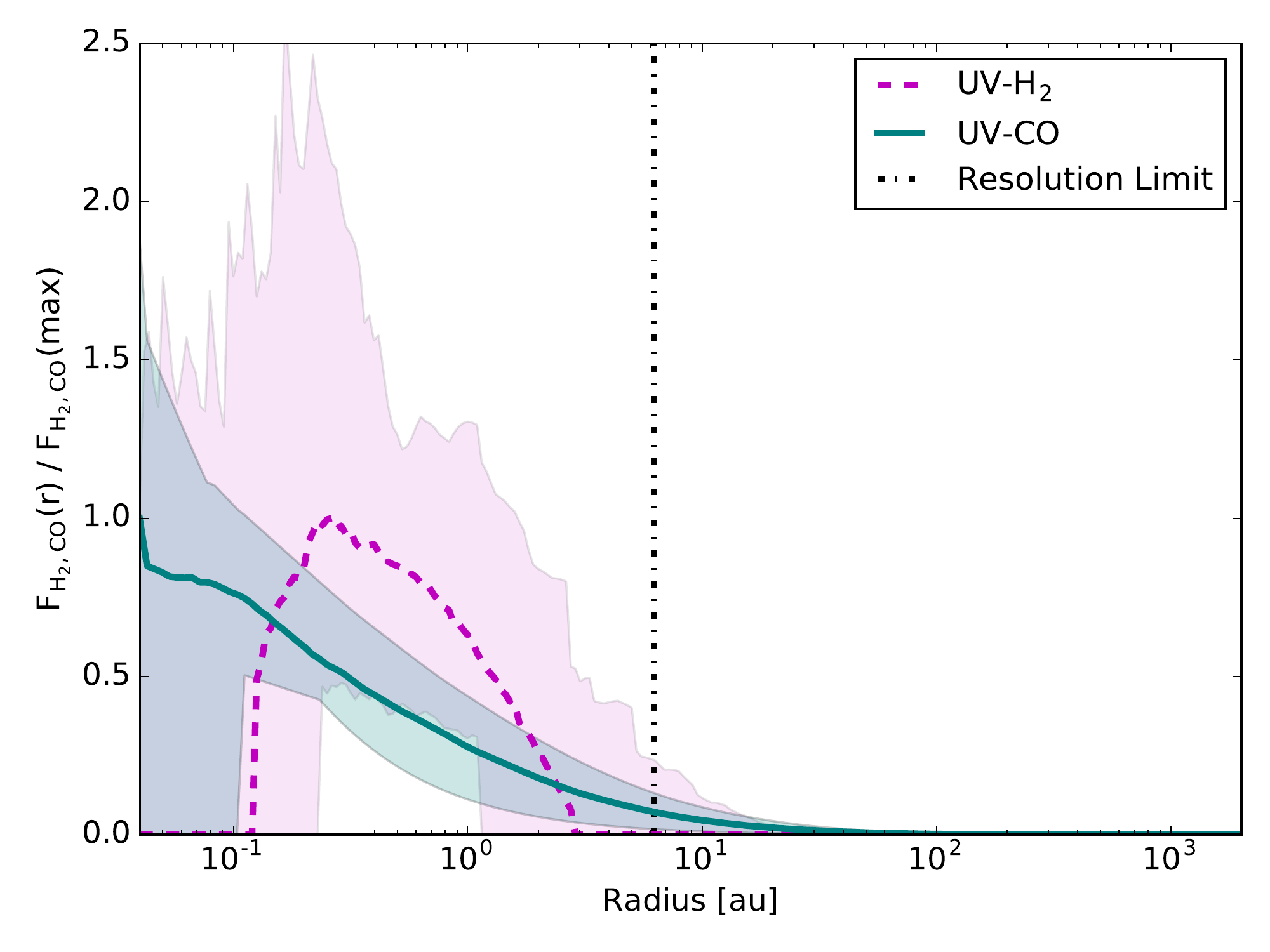} 
	\end{minipage}
\caption{Best-fit UV-CO model (top left), 100 UV-H$_2$ radiative transfer models with the smallest \textit{MSE}s (top right), comparison of UV-H$_2$-/UV-CO-based Ly$\alpha$ profiles (bottom left) and radial distributions of flux for both species (bottom right) from the disk around CW Tau. Residuals on the best-fit UV-CO model are color-coded from blue to red, based on the Ly$\alpha$ pumping wavelength required to excite the upper level of the transition.}
\label{CWTau_H2CO}
\end{figure*}

\begin{figure*}[t!]
	\begin{minipage}{0.5\textwidth}
	\centering
	\includegraphics[width=\linewidth]{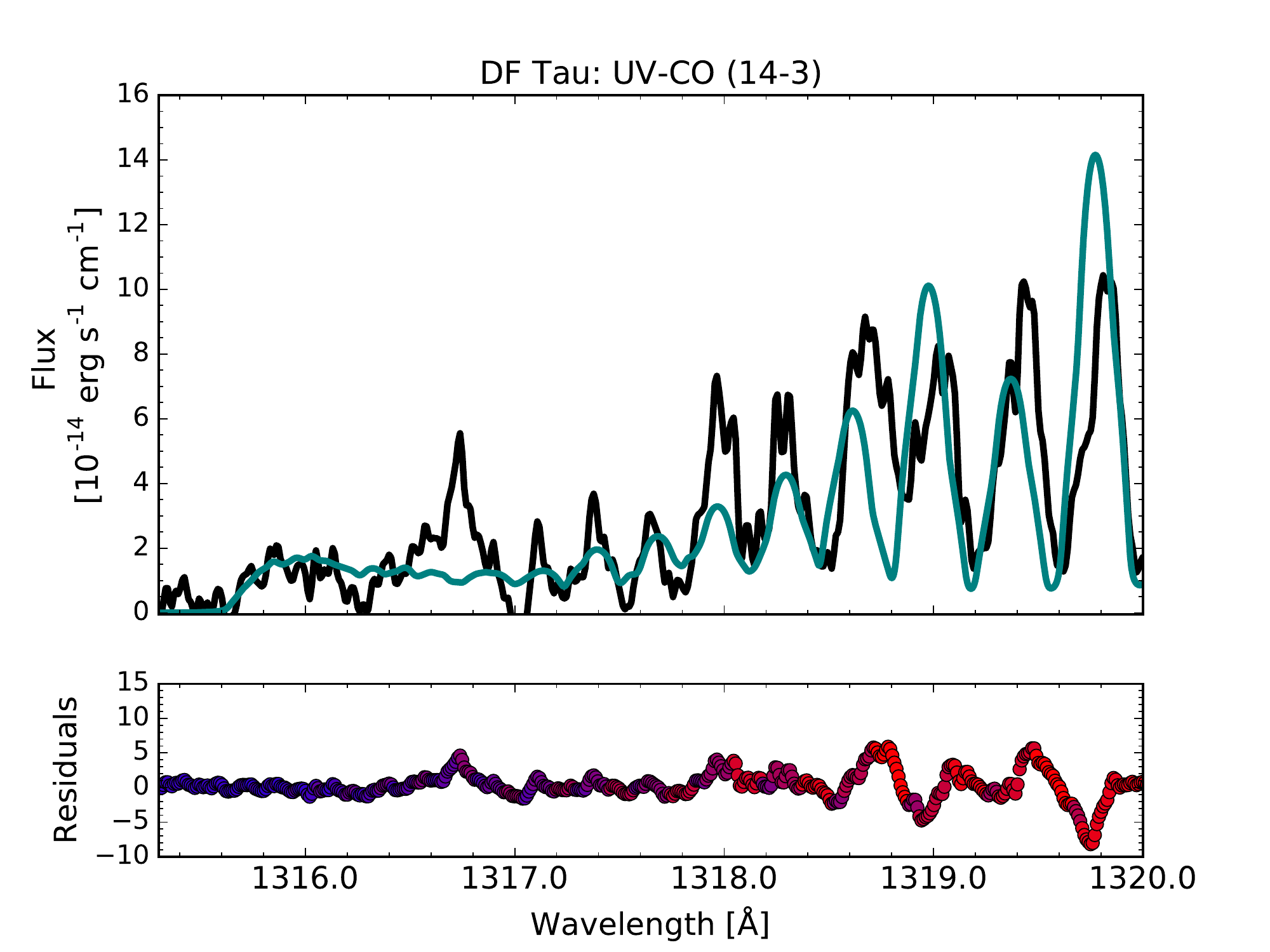}
	\end{minipage}
	\begin{minipage}{0.5\textwidth}
	\centering
	\includegraphics[width=\linewidth]{f15_21.pdf}
	\end{minipage}
	\begin{minipage}{0.5\textwidth}
	\centering
	\includegraphics[width=\linewidth]{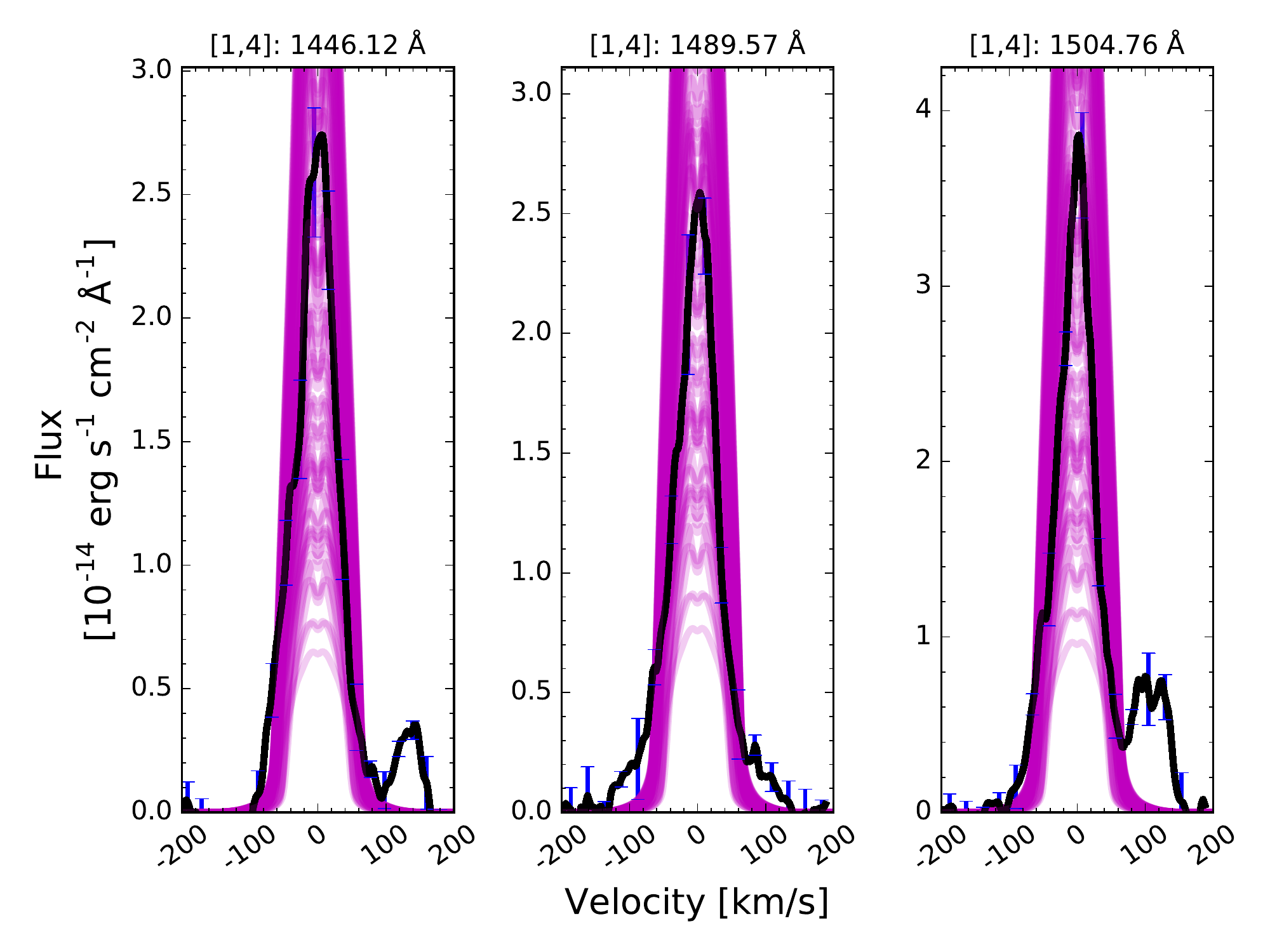} 
	\end{minipage}
	\begin{minipage}{0.5\textwidth}
	\centering
	\includegraphics[width=\linewidth]{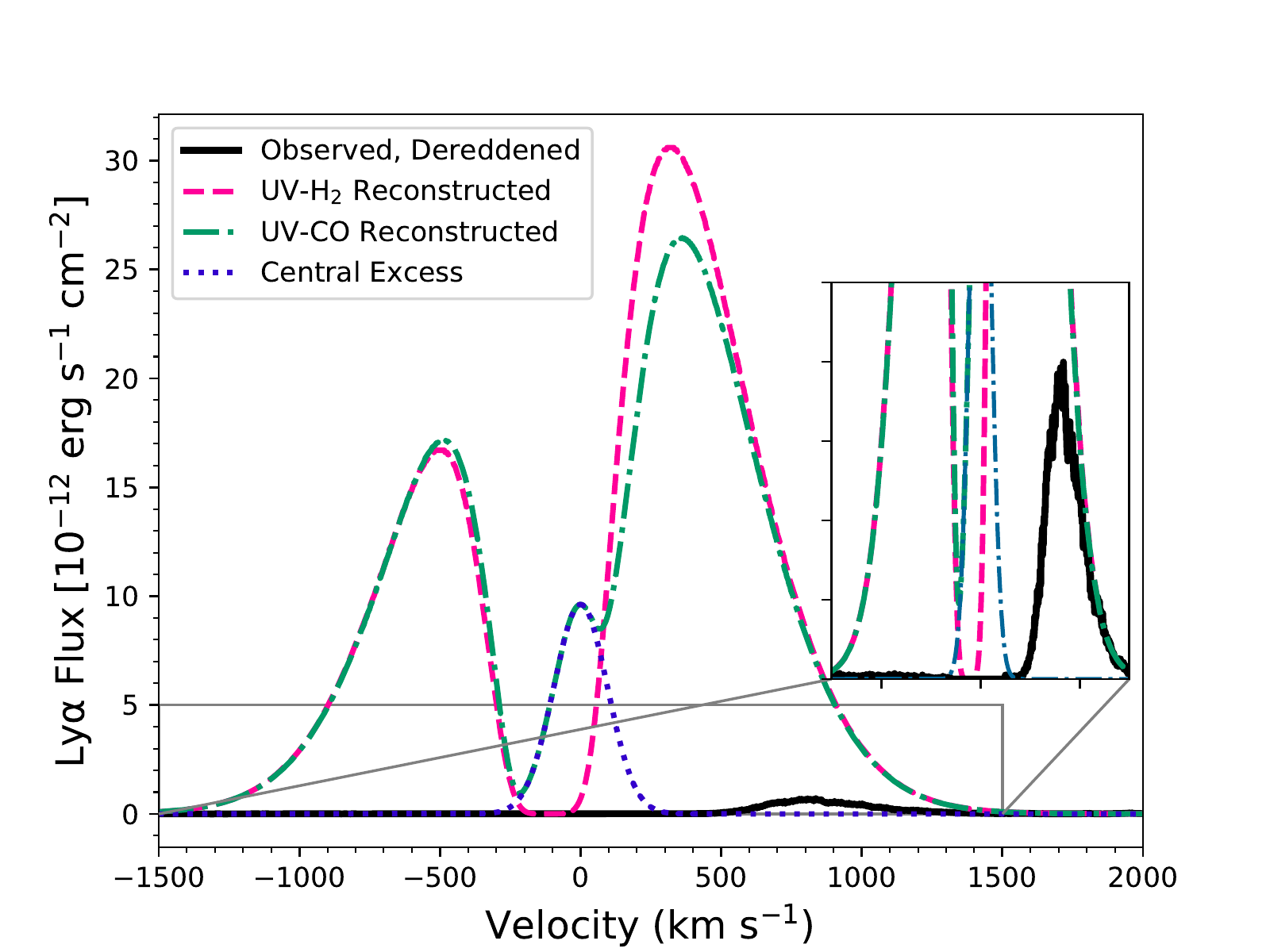} 
	\end{minipage}
	\begin{minipage}{0.5\textwidth}
	\centering
	\includegraphics[width=\linewidth]{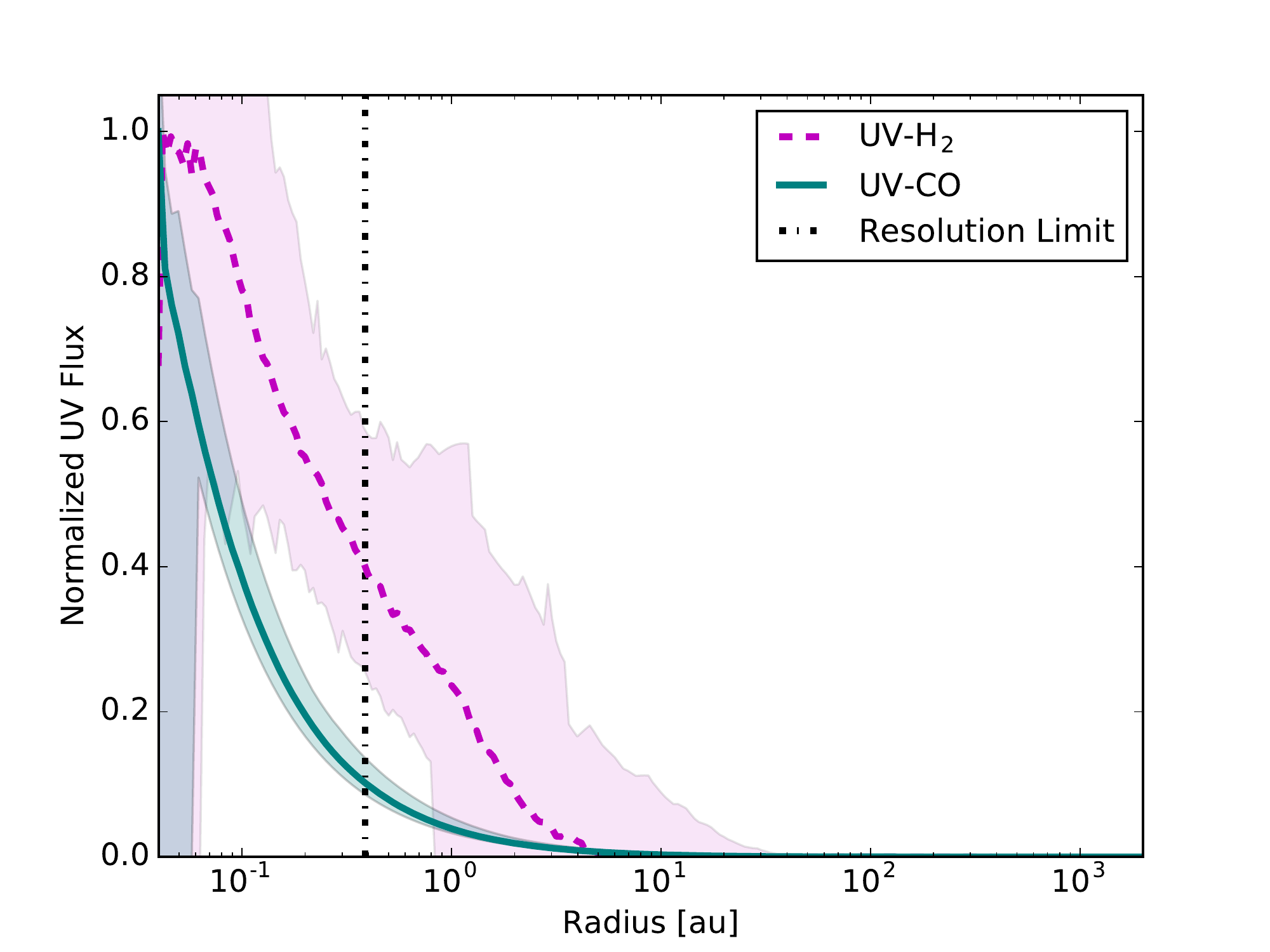} 
	\end{minipage}
\caption{Best-fit UV-CO 2-D radiative transfer (top left) and Gaussian models (top right), 100 UV-H$_2$ radiative transfer models with the smallest \textit{MSE}s (middle left), comparison of UV-H$_2$-/UV-CO-based and observed Ly$\alpha$ profiles (middle right) and radial distributions of flux for both species (bottom left) from the disk around DF Tau. Deviations between the UV-H$_2$ models and data point to a two-component line profile, with pieces originating in radially separated populations of hot gas (see e.g., \citealt{Banzatti2015}, Hoadley et al., in prep). Residuals on the best-fit UV-CO model are color-coded from blue to red, based on the Ly$\alpha$ pumping wavelength required to excite the upper level of the transition.}
\label{DFTau_H2CO}
\end{figure*}

\begin{figure*}[t!]
	\begin{minipage}{0.5\textwidth}
	\centering
	\includegraphics[width=\linewidth]{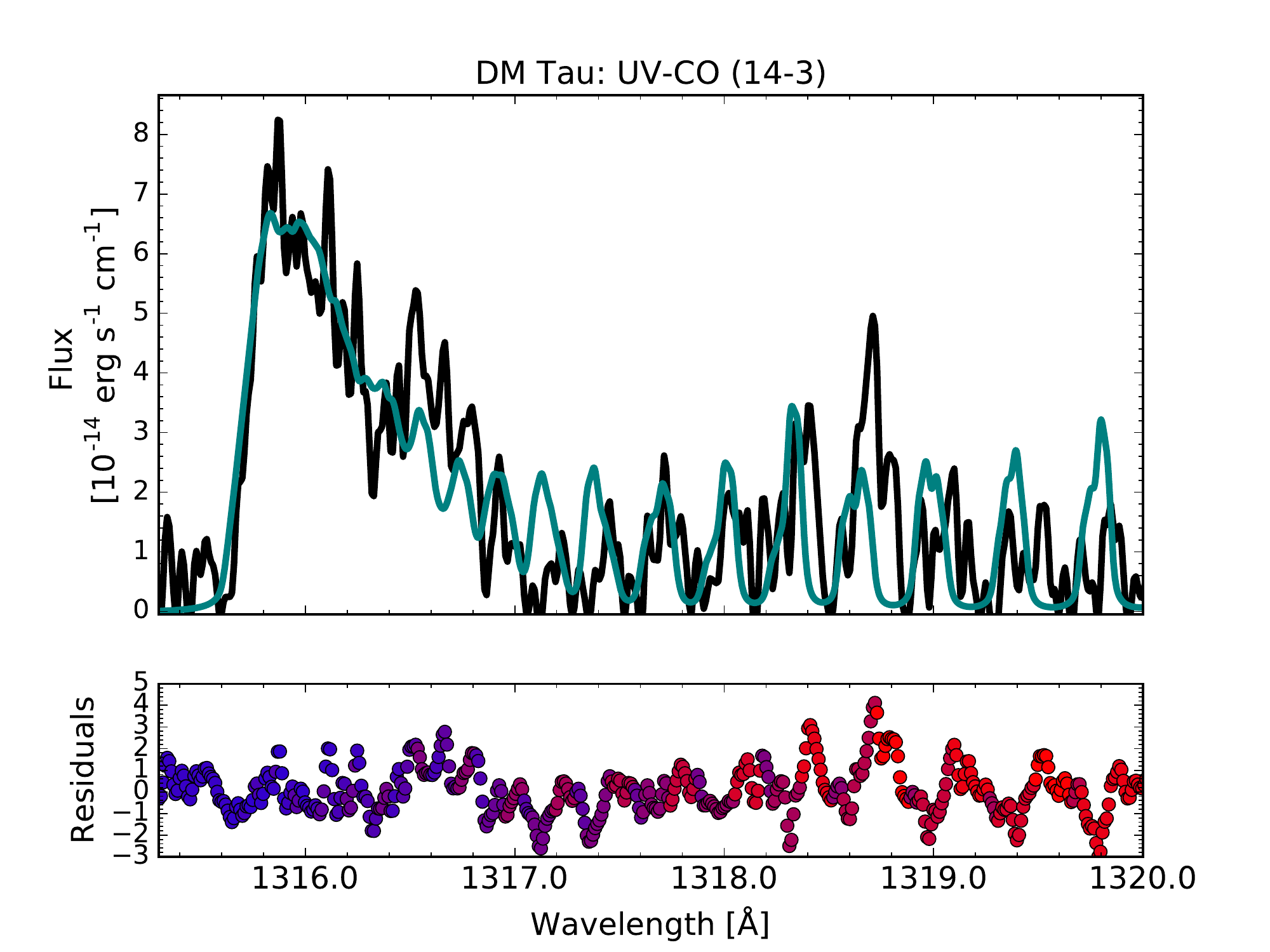}
	\end{minipage}
	\begin{minipage}{0.5\textwidth}
	\centering
	\includegraphics[width=\linewidth]{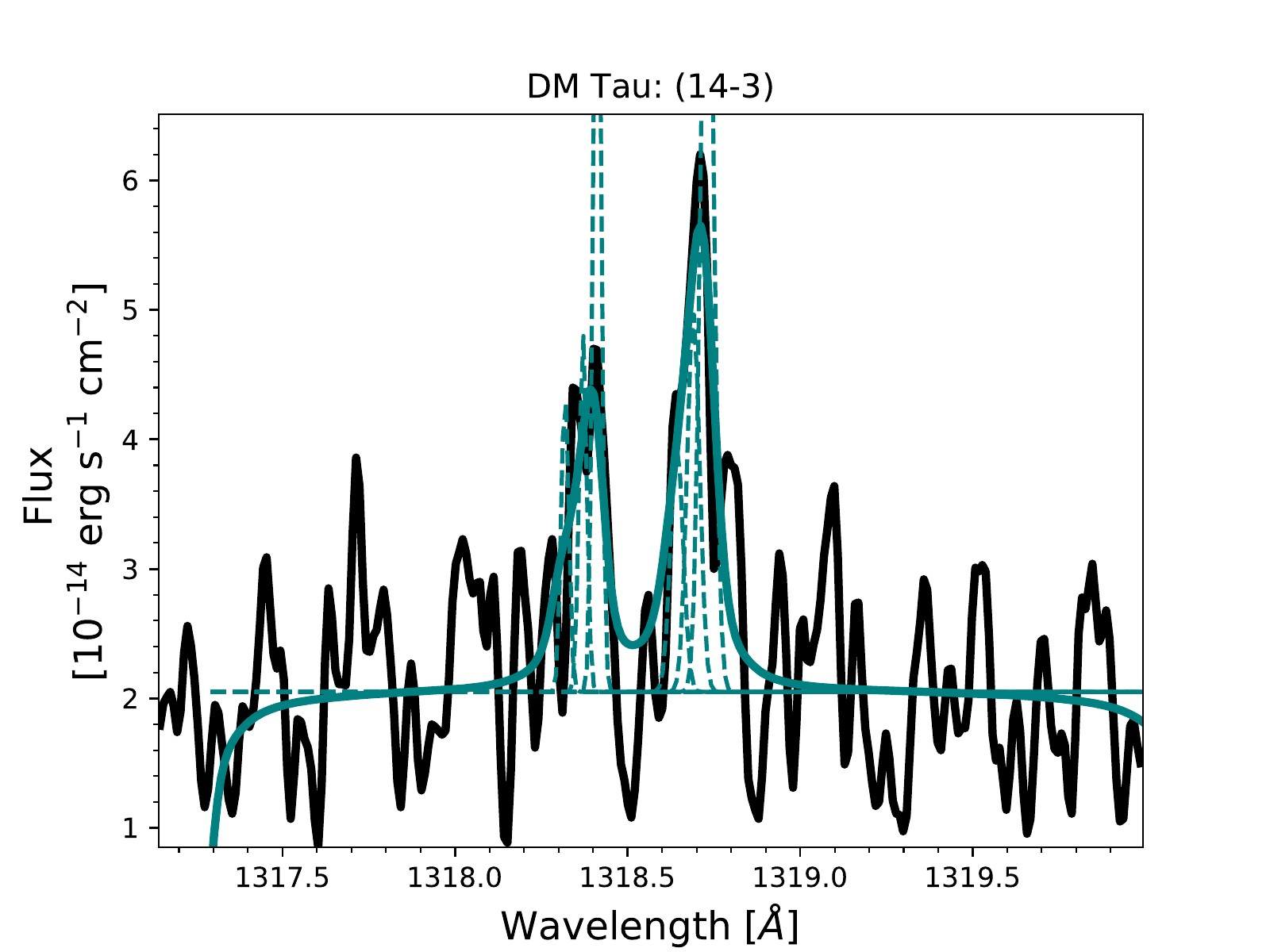}
	\end{minipage}
	\begin{minipage}{0.5\textwidth}
	\centering
	\includegraphics[width=\linewidth]{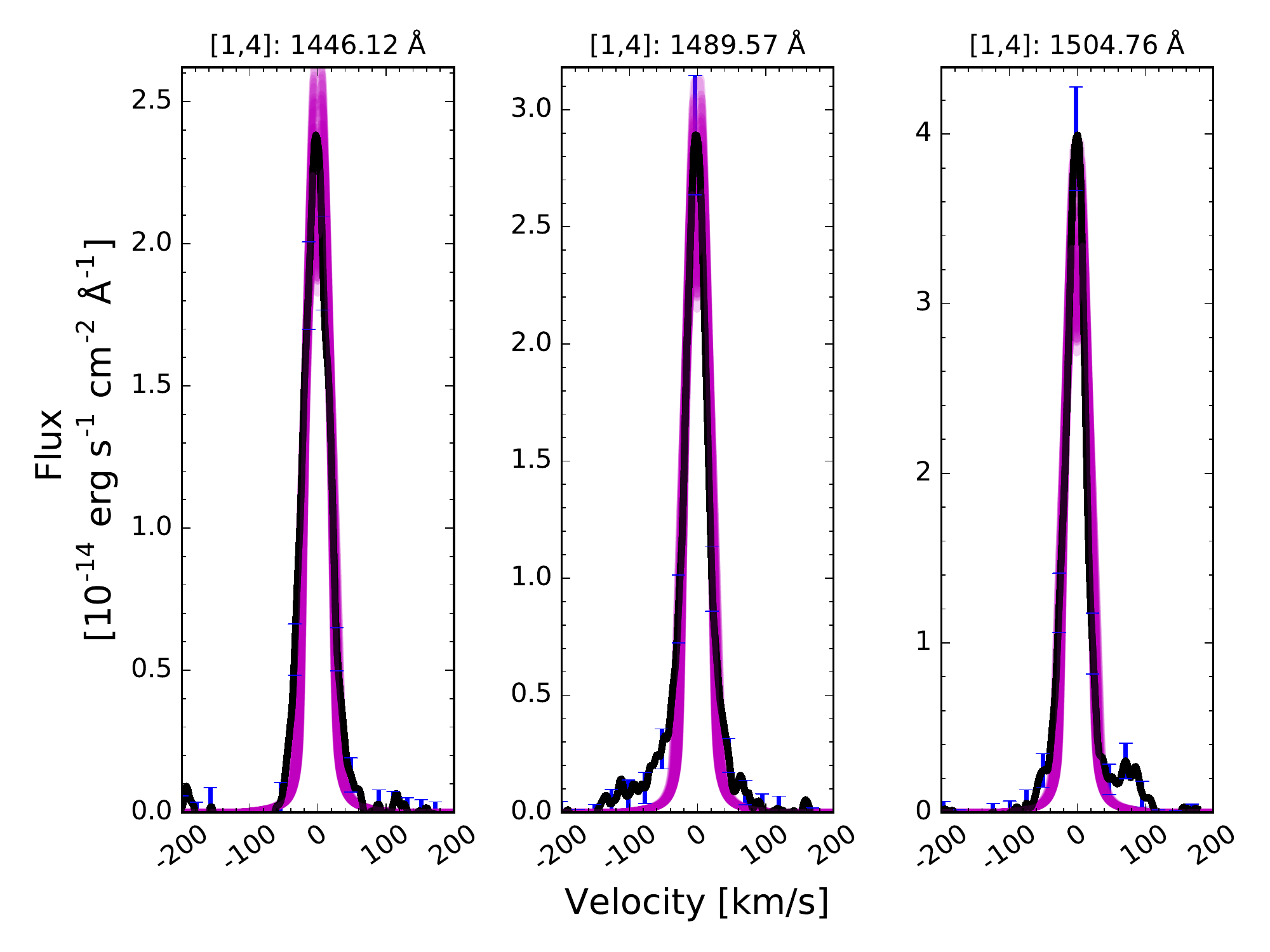} 
	\end{minipage}
	\begin{minipage}{0.5\textwidth}
	\centering
	\includegraphics[width=\linewidth]{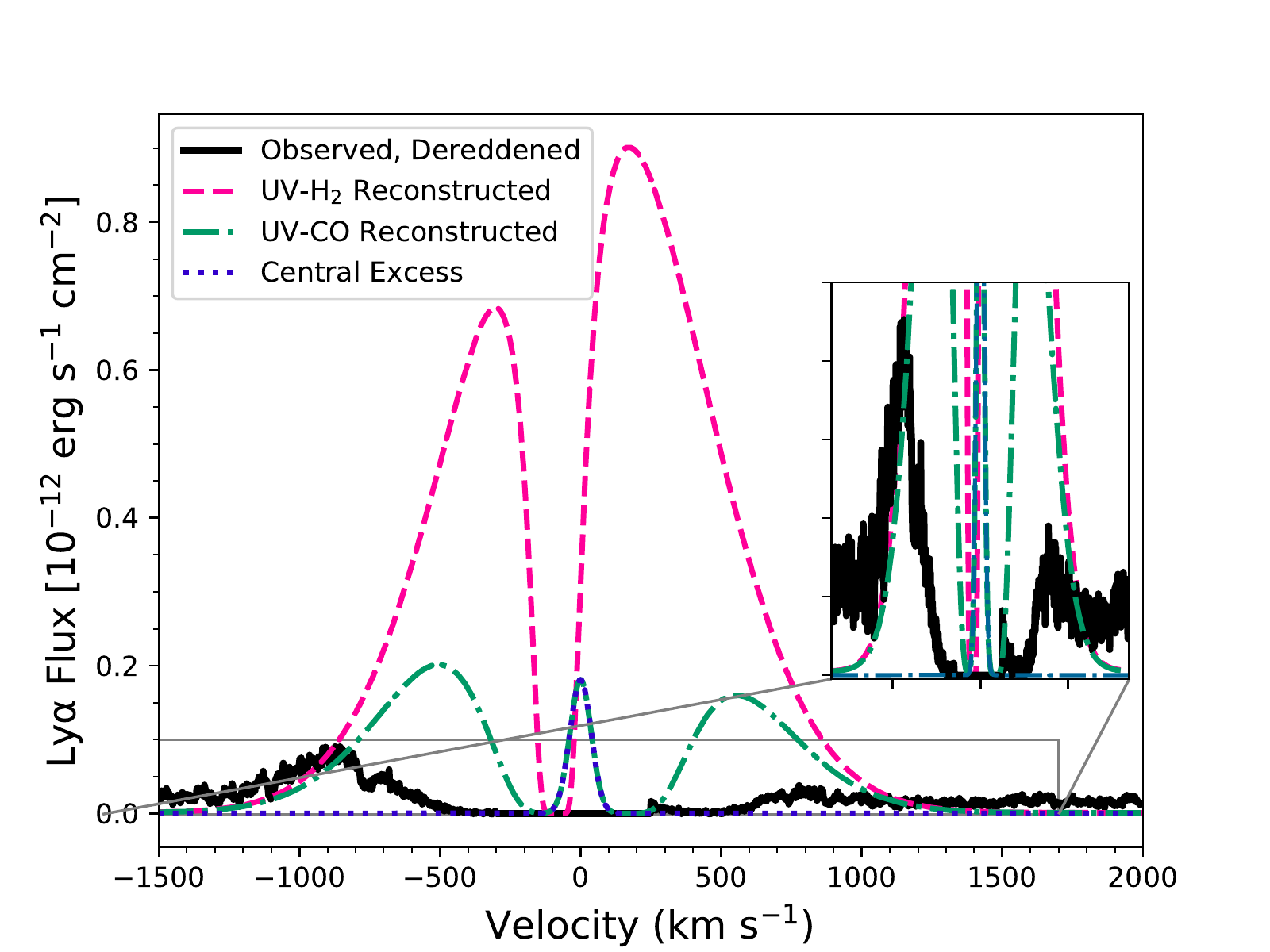} 
	\end{minipage}
	\begin{minipage}{0.5\textwidth}
	\centering
	\includegraphics[width=\linewidth]{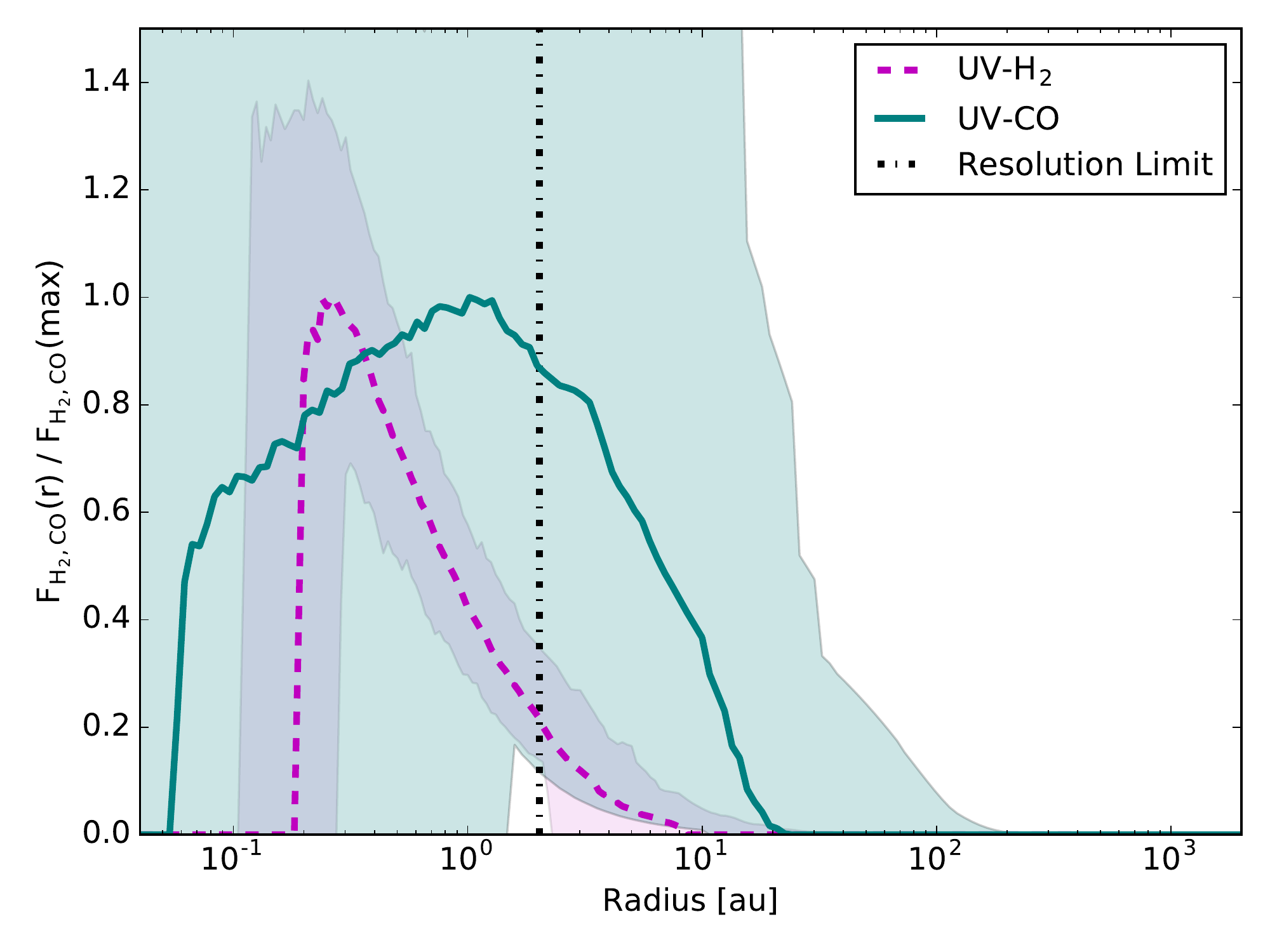} 
	\end{minipage}
\caption{Best-fit UV-CO 2-D radiative transfer (top left) and Gaussian models (top right), 100 UV-H$_2$ radiative transfer models with the smallest \textit{MSE}s (middle left), comparison of UV-H$_2$-/UV-CO-based and observed Ly$\alpha$ profiles (middle right) and radial distributions of flux for both species (bottom left) from the disk around DM Tau. Residuals on the best-fit UV-CO model are color-coded from blue to red, based on the Ly$\alpha$ pumping wavelength required to excite the upper level of the transition.}
\label{DMTau_H2CO}
\end{figure*}

\begin{figure*}[t!]
	\begin{minipage}{0.5\textwidth}
	\centering
	\includegraphics[width=\linewidth]{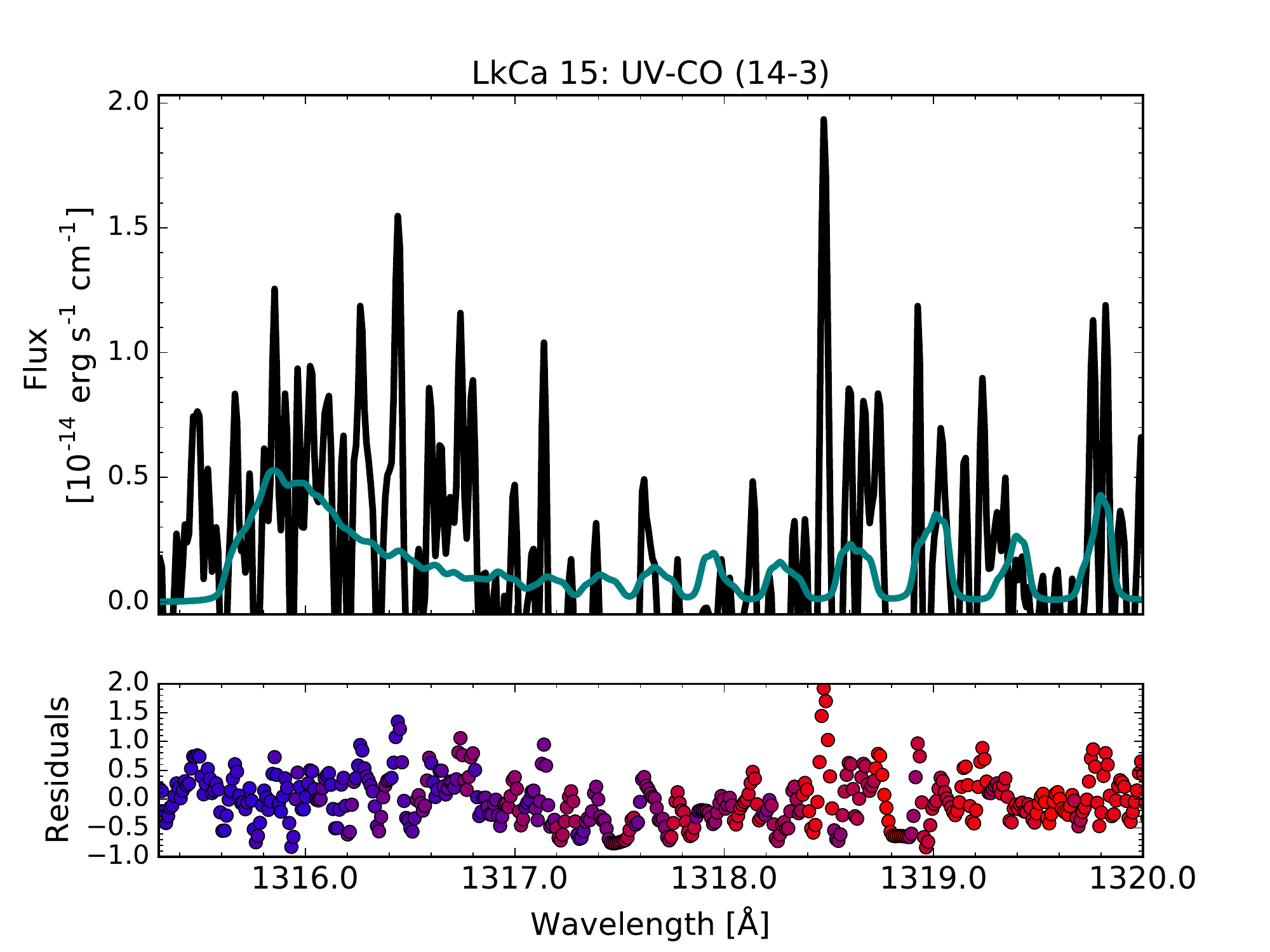}
	\end{minipage}
	\begin{minipage}{0.5\textwidth}
	\centering
	\includegraphics[width=\linewidth]{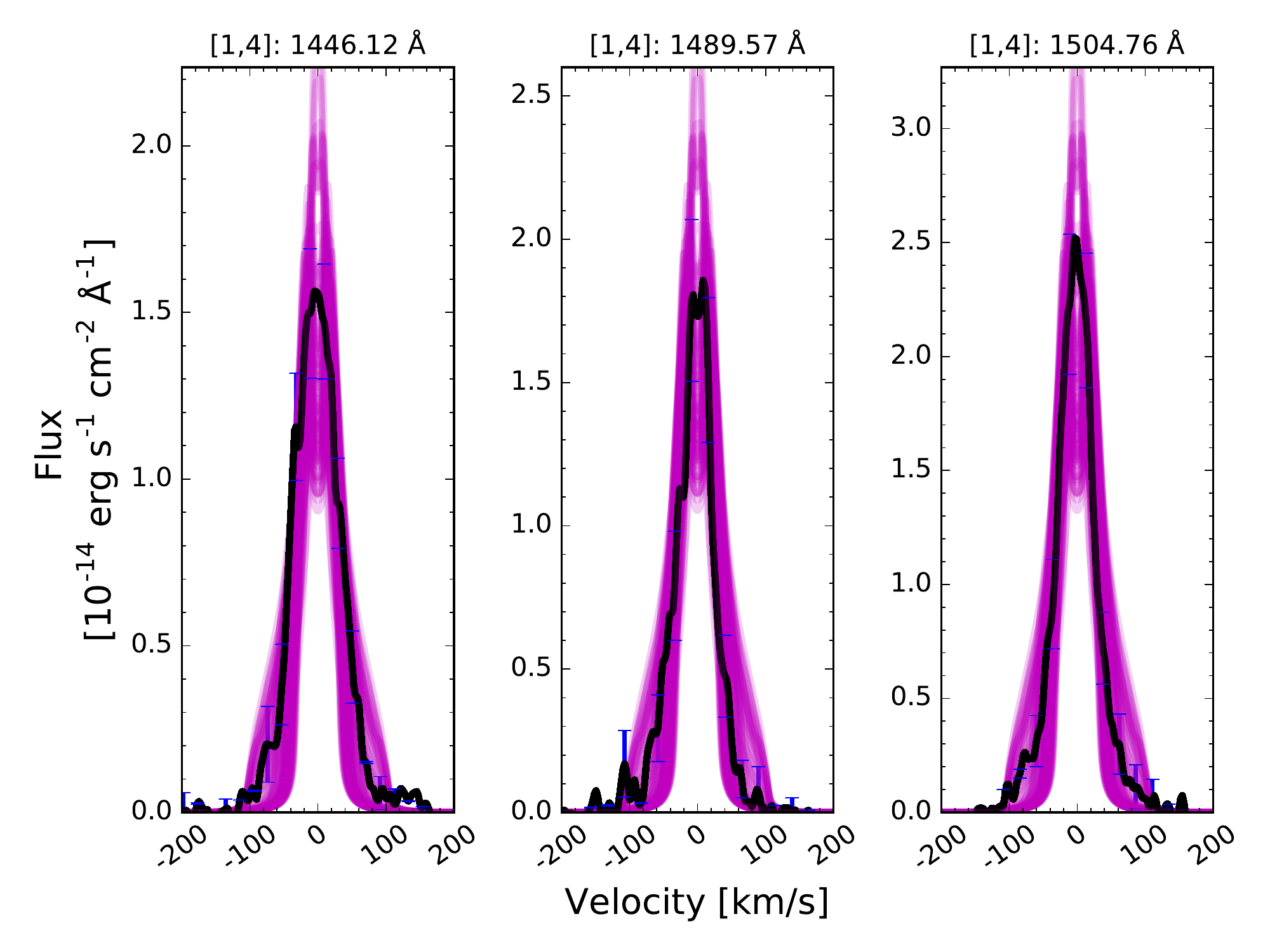} 
	\end{minipage}
	\begin{minipage}{0.5\textwidth}
	\centering
	\includegraphics[width=\linewidth]{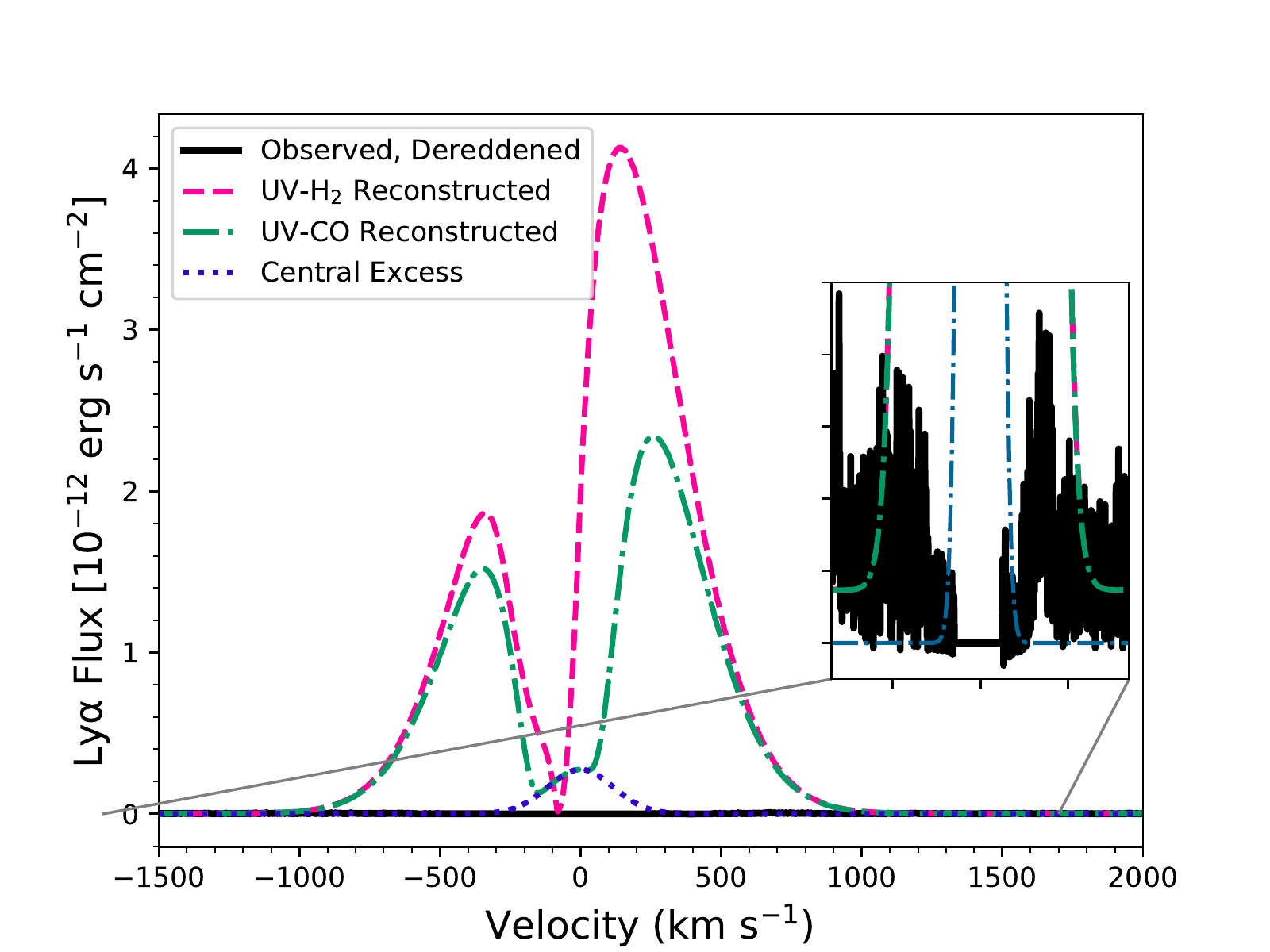} 
	\end{minipage}
	\begin{minipage}{0.5\textwidth}
	\centering
	\includegraphics[width=\linewidth]{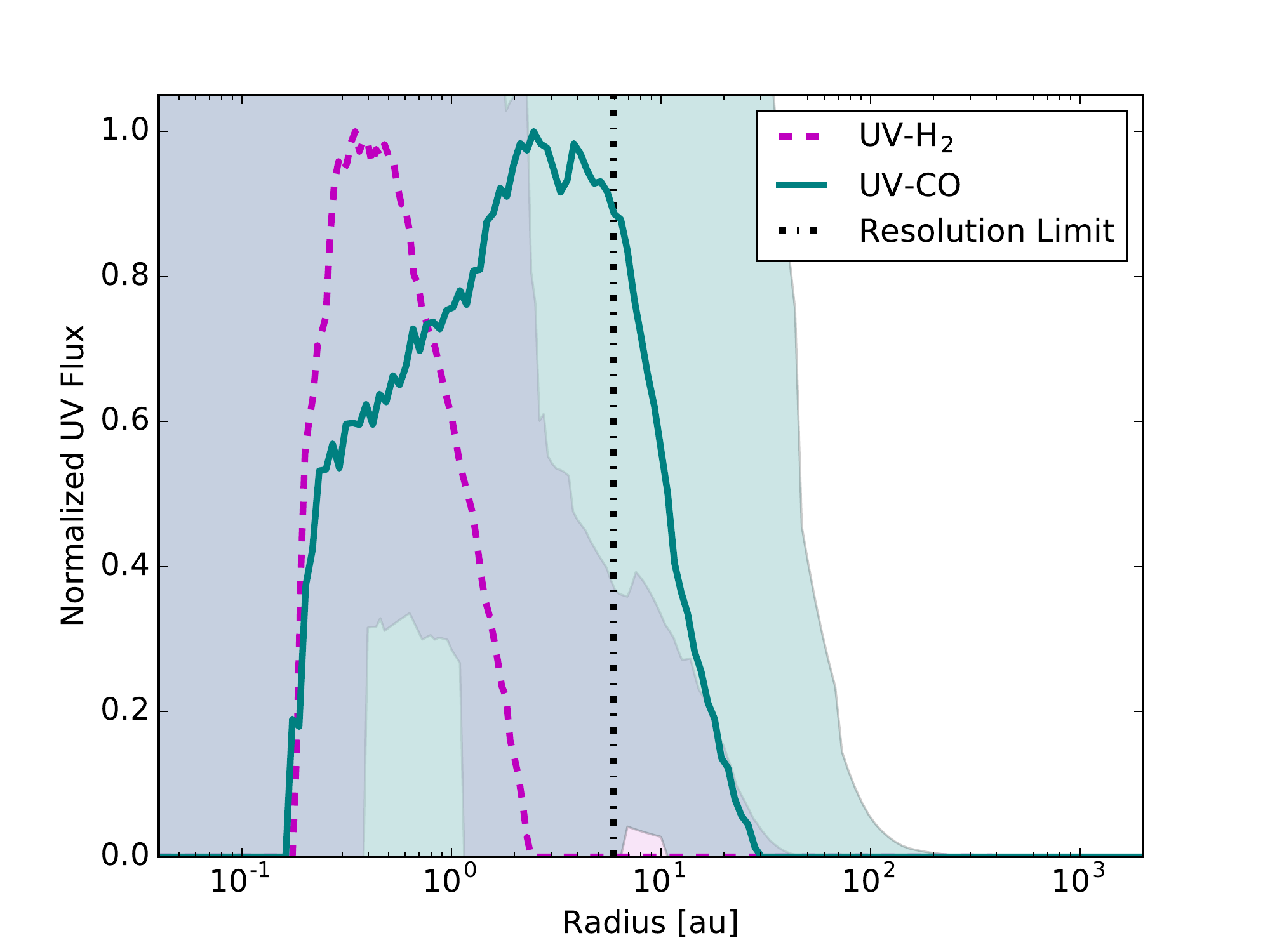} 
	\end{minipage}
\caption{Best-fit UV-CO model (top left), 100 UV-H$_2$ radiative transfer models with the smallest \textit{MSE}s (top right), comparison of UV-H$_2$-/UV-CO based and observed Ly$\alpha$ profiles (bottom left) and radial distributions of flux for both species (bottom right) from the disk around LkCa 15. Residuals on the best-fit UV-CO model are color-coded from blue to red, based on the Ly$\alpha$ pumping wavelength required to excite the upper level of the transition.}
\label{LkCa15_H2CO}
\end{figure*}

\begin{figure*}[t!]
	\begin{minipage}{0.5\textwidth}
	\centering
	\includegraphics[width=\linewidth]{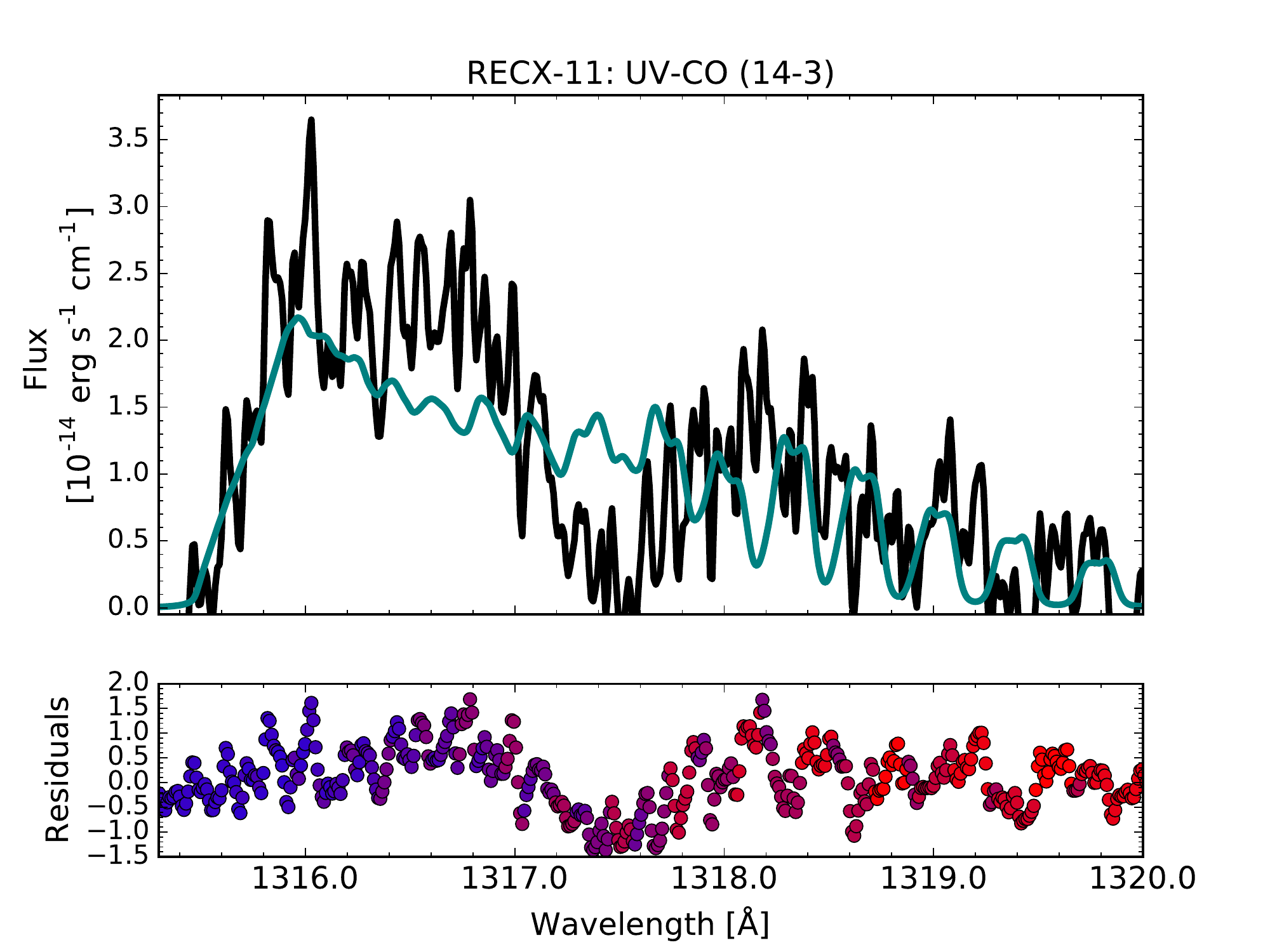}
	\end{minipage}
	\begin{minipage}{0.5\textwidth}
	\centering
	\includegraphics[width=\linewidth]{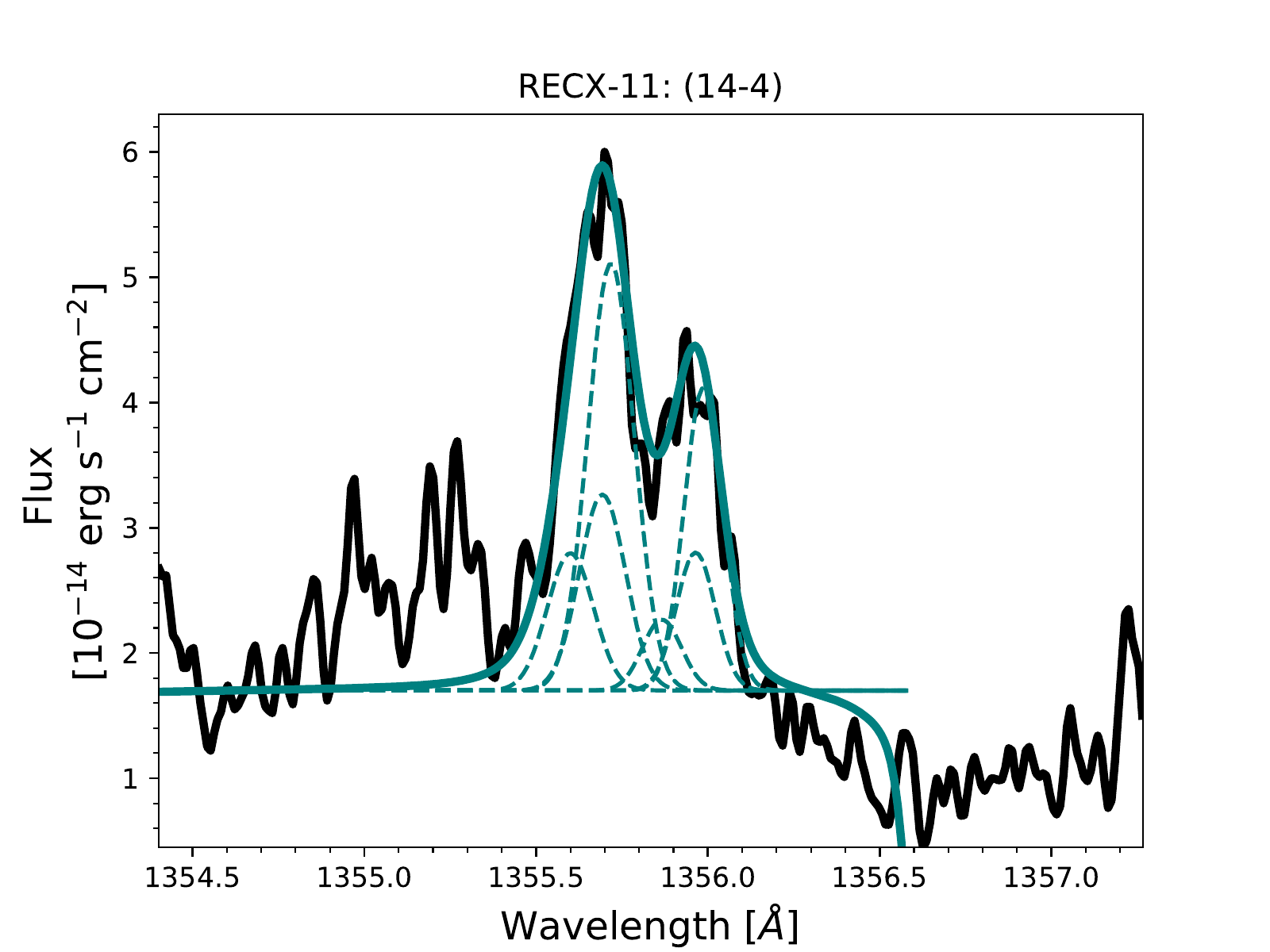}
	\end{minipage}
	\begin{minipage}{0.5\textwidth}
	\centering
	\includegraphics[width=\linewidth]{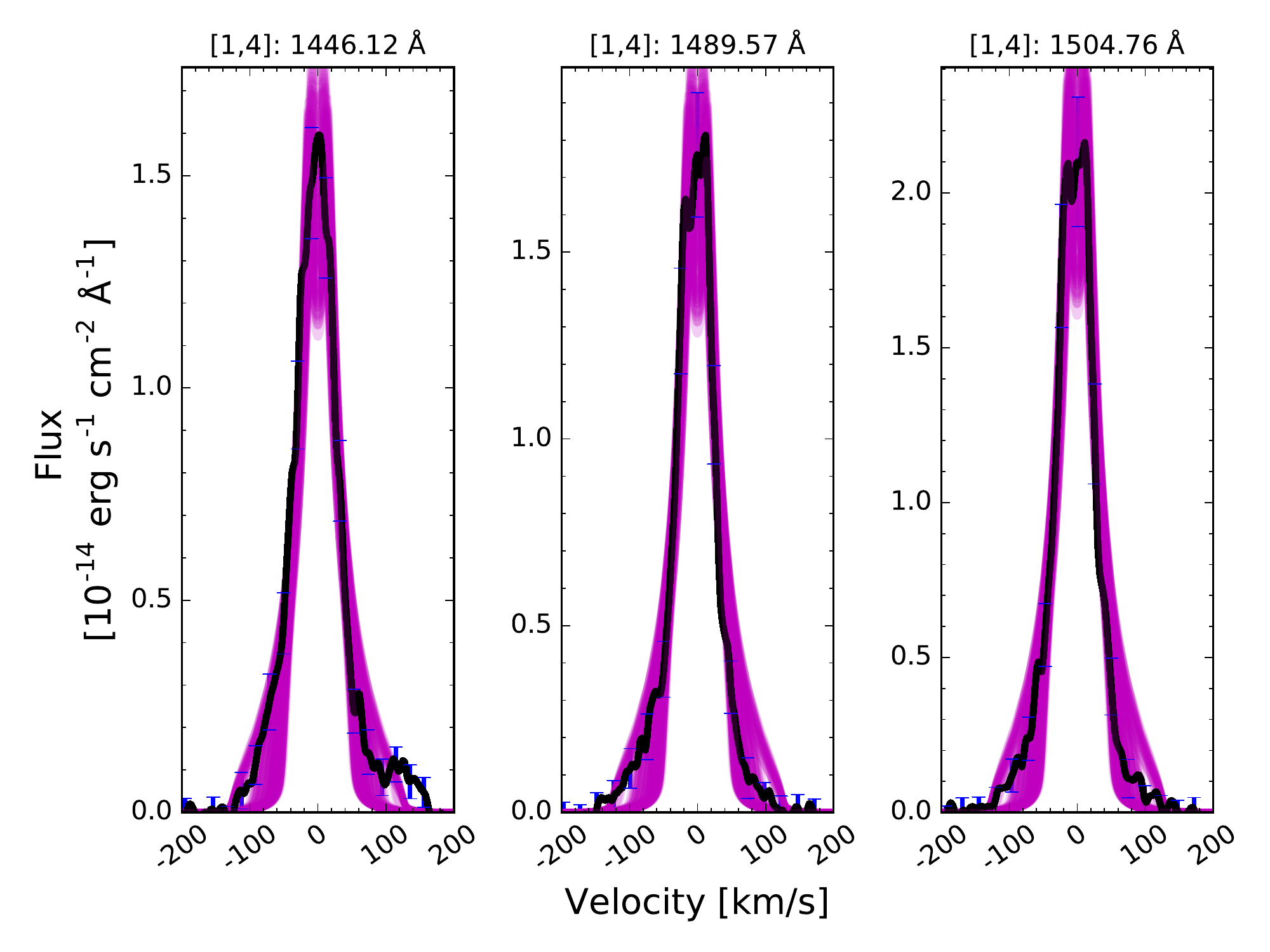} 
	\end{minipage}
	\begin{minipage}{0.5\textwidth}
	\centering
	\includegraphics[width=\linewidth]{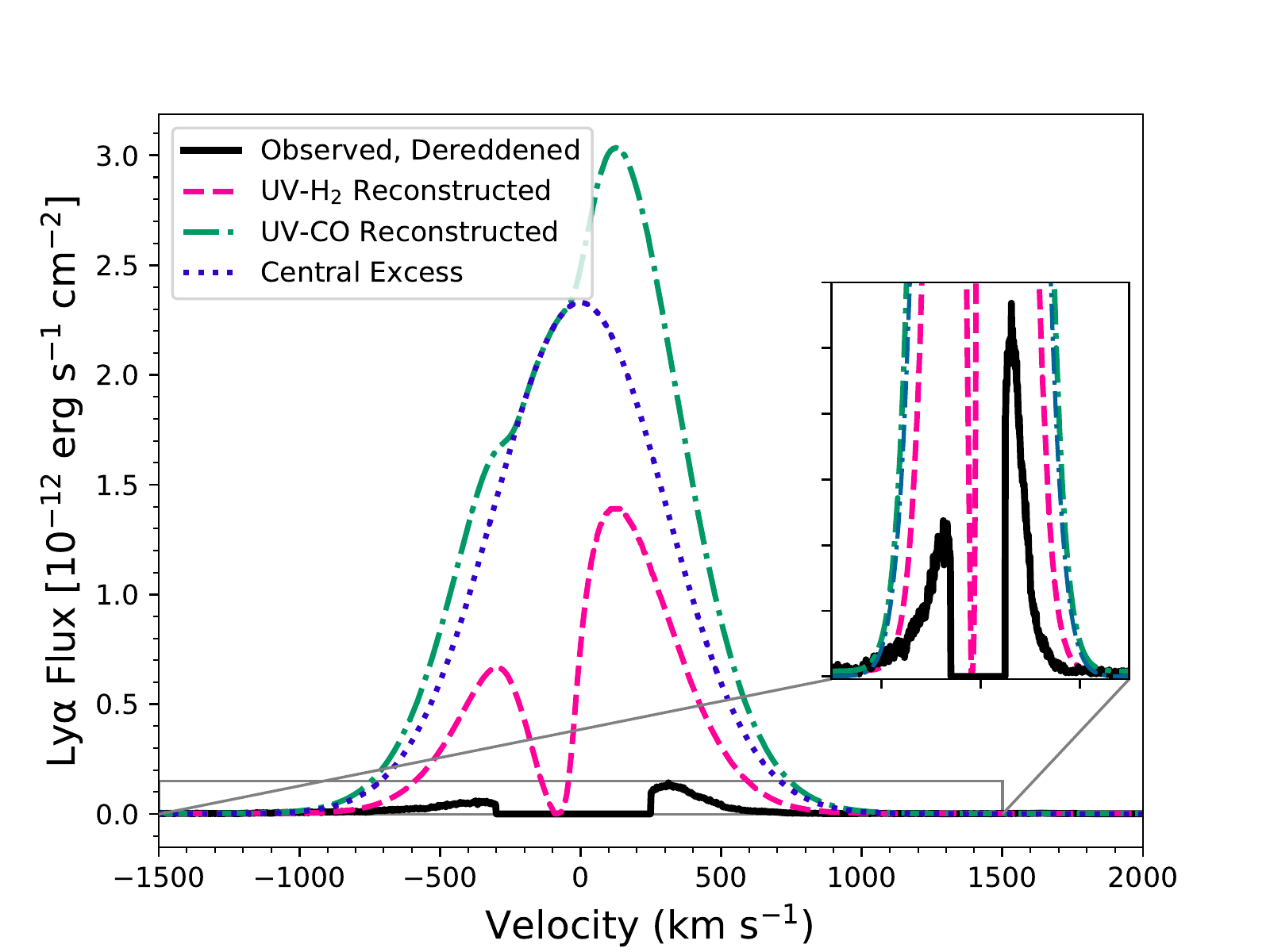} 
	\end{minipage}
	\begin{minipage}{0.5\textwidth}
	\centering
	\includegraphics[width=\linewidth]{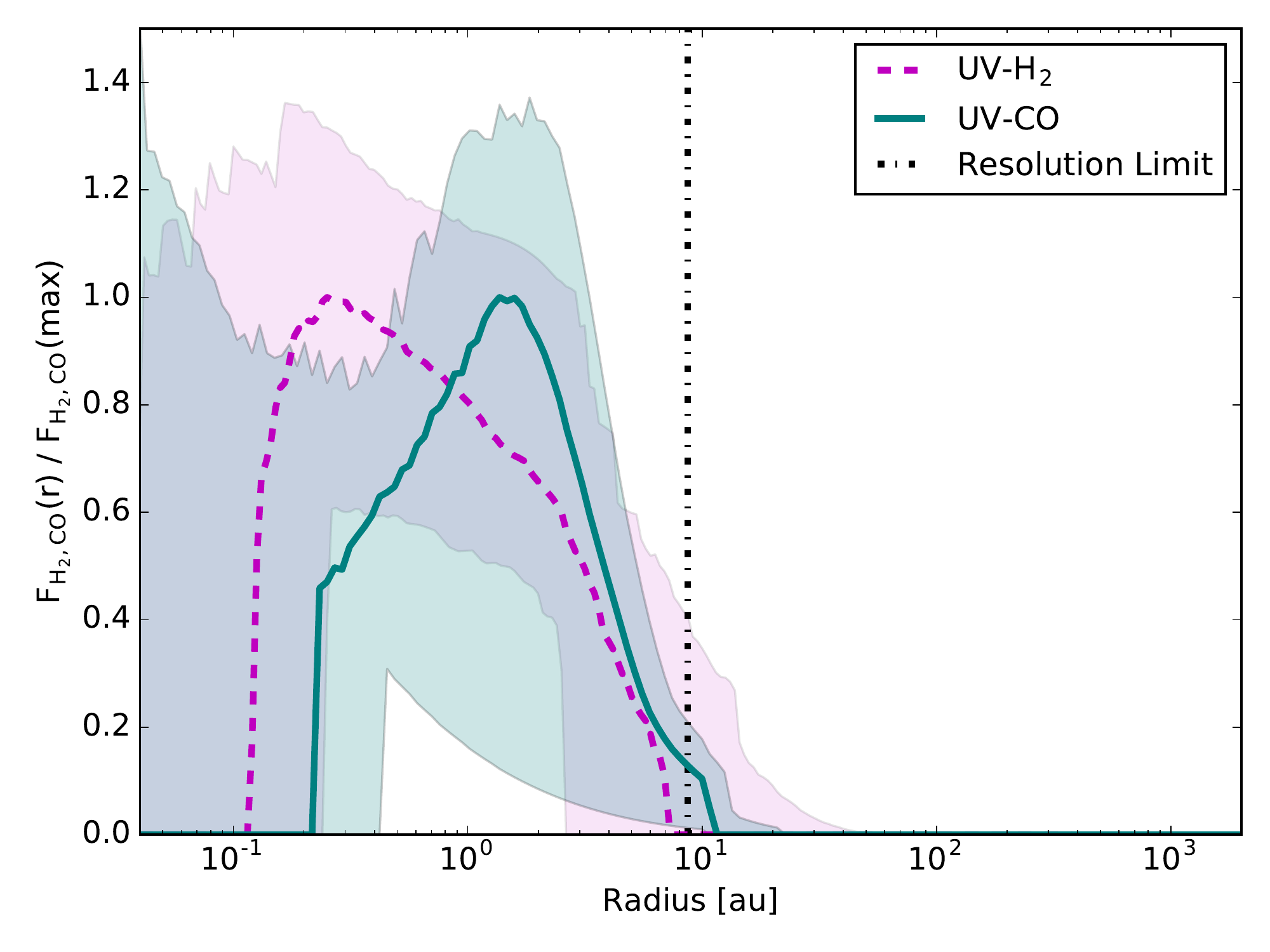} 
	\end{minipage}
\caption{Best-fit UV-CO 2-D radiative transfer (top left) and Gaussian models (top right), 100 UV-H$_2$ radiative transfer models with the smallest \textit{MSE}s (middle left), comparison of UV-H$_2$-/UV-CO-based and observed Ly$\alpha$ profiles (middle right) and radial distributions of flux for both species (bottom left) from the disk around RECX-11. Residuals on the best-fit UV-CO model are color-coded from blue to red, based on the Ly$\alpha$ pumping wavelength required to excite the upper level of the transition.}
\label{RECX11_H2CO}
\end{figure*}

\begin{figure*}[t!]
	\begin{minipage}{0.5\textwidth}
	\centering
	\includegraphics[width=\linewidth]{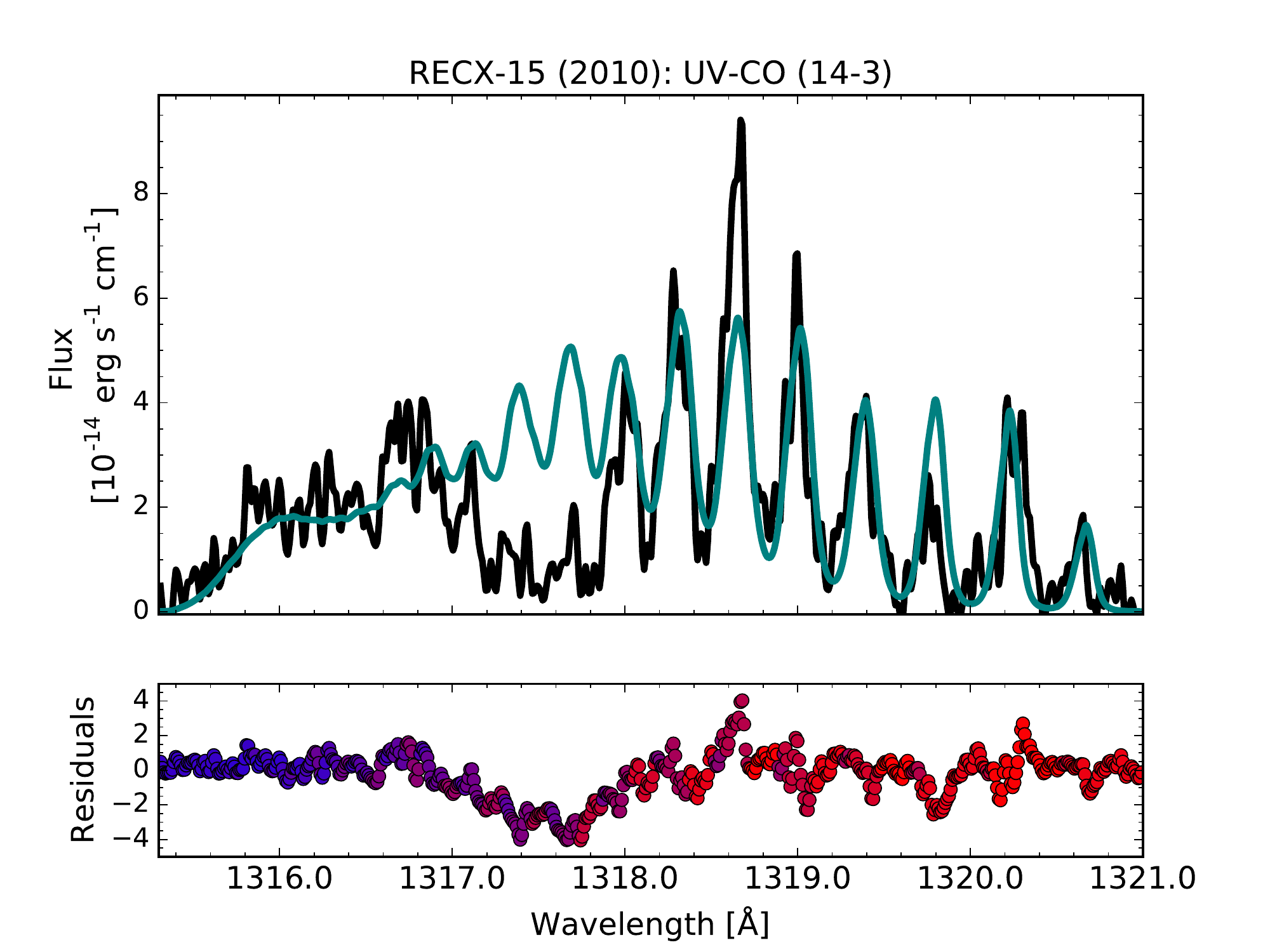}
	\end{minipage}
	\begin{minipage}{0.5\textwidth}
	\centering
	\includegraphics[width=\linewidth]{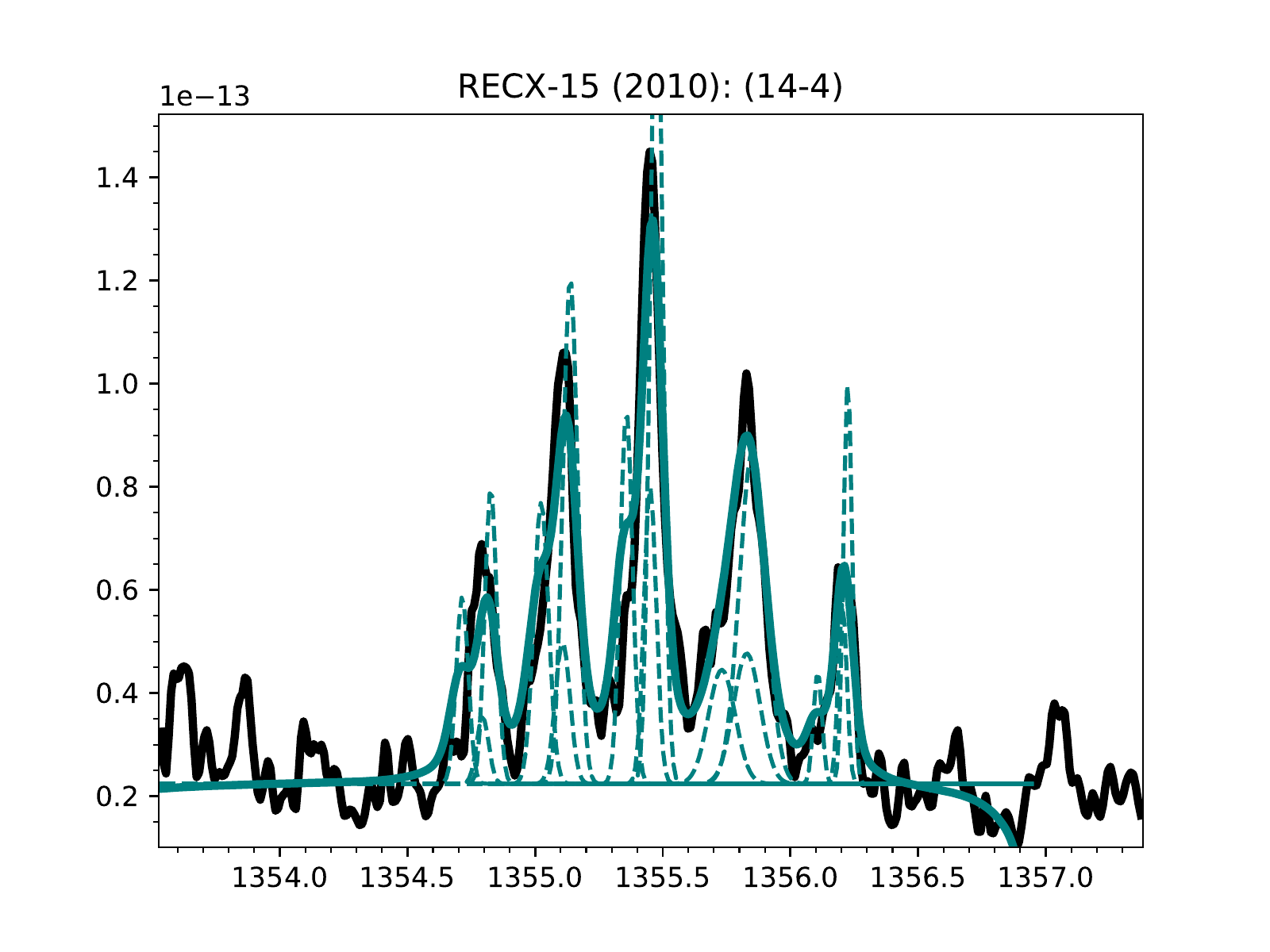}
	\end{minipage}
	\begin{minipage}{0.5\textwidth}
	\centering
	\includegraphics[width=\linewidth]{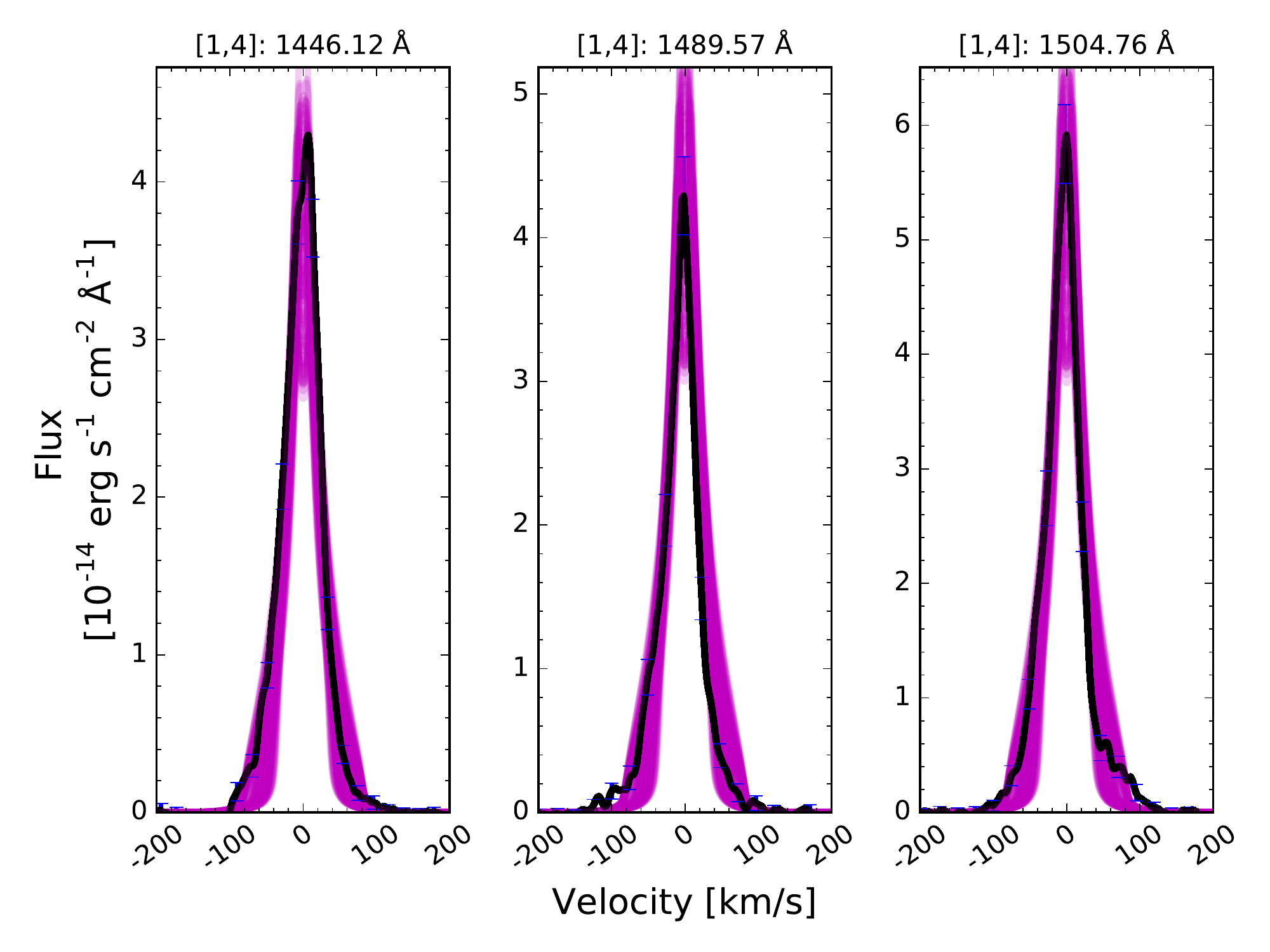} 
	\end{minipage}
	\begin{minipage}{0.5\textwidth}
	\centering
	\includegraphics[width=\linewidth]{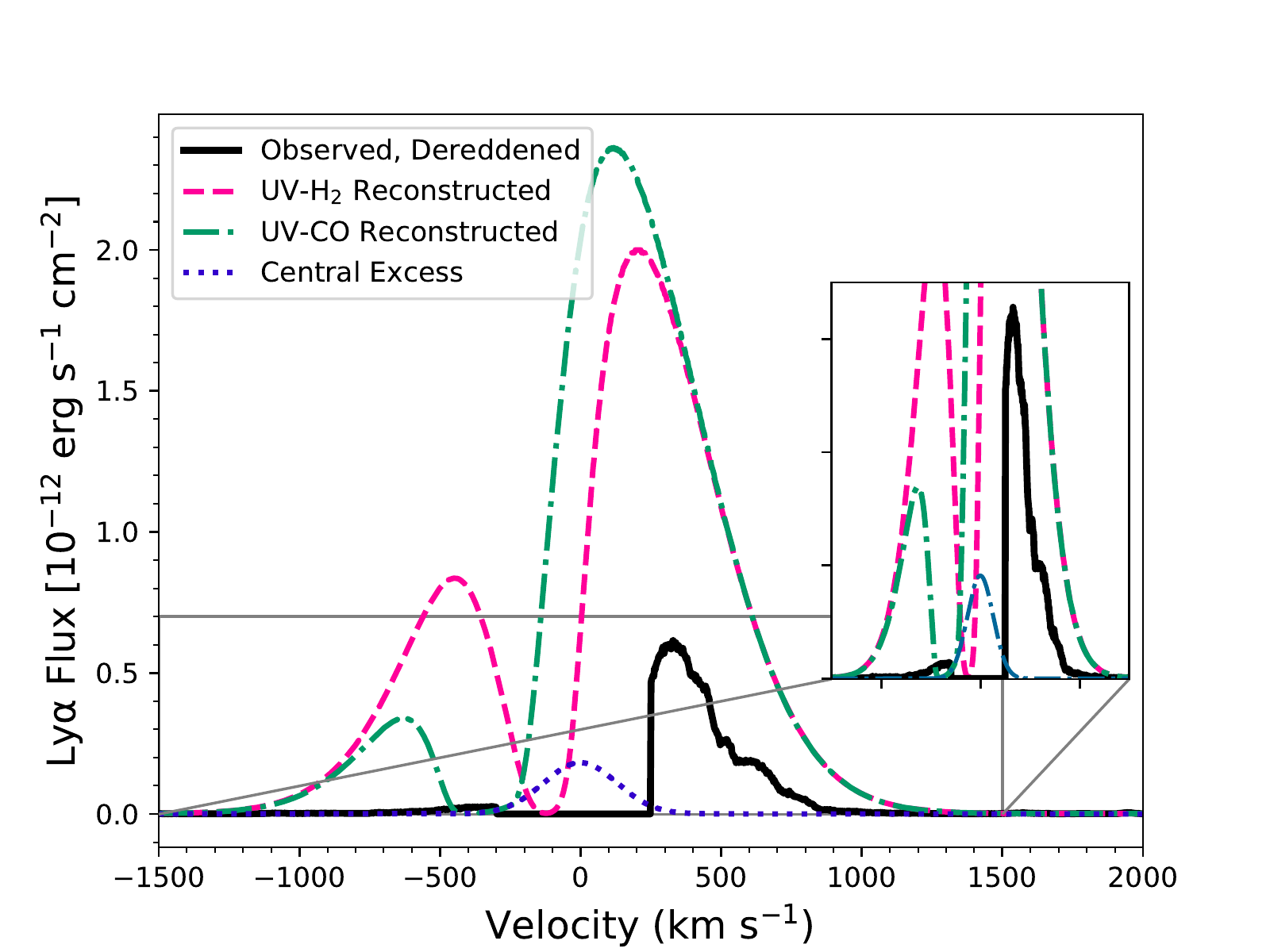} 
	\end{minipage}
	\begin{minipage}{0.5\textwidth}
	\centering
	\includegraphics[width=\linewidth]{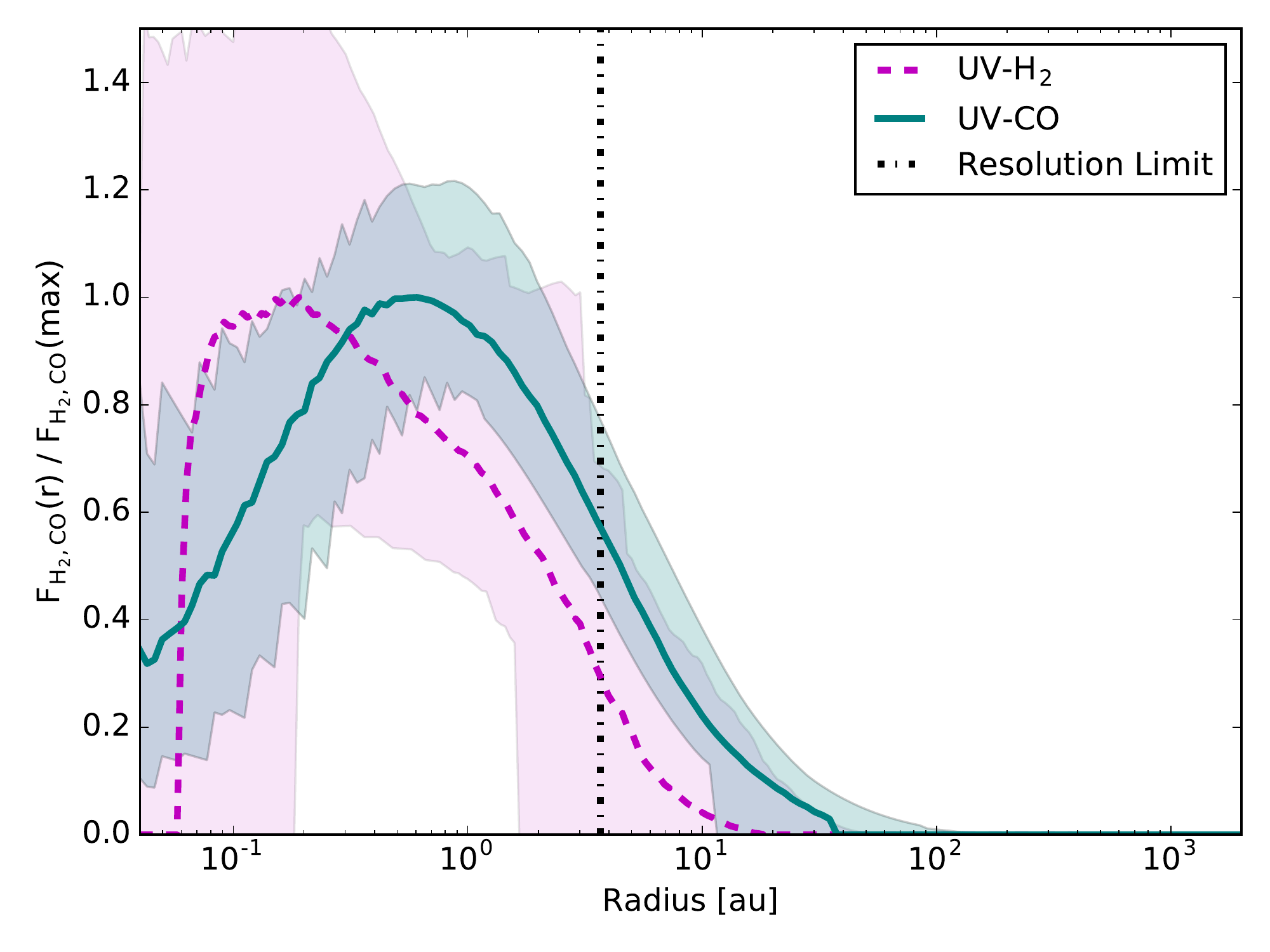} 
	\end{minipage}
\caption{Best-fit UV-CO 2-D radiative transfer (top left) and Gaussian models (top right), 100 UV-H$_2$ radiative transfer models with the smallest \textit{MSE}s (middle left), comparison of UV-H$_2$-/UV-CO-based and observed Ly$\alpha$ profiles (middle right) and radial distributions of flux for both species (bottom left) from the disk around RECX-15 (spectra from 2010). Residuals on the best-fit UV-CO model are color-coded from blue to red, based on the Ly$\alpha$ pumping wavelength required to excite the upper level of the transition.}
\label{RECX15_2010_H2CO}
\end{figure*}

\begin{figure*}[t!]
	\begin{minipage}{0.5\textwidth}
	\centering
	\includegraphics[width=\linewidth]{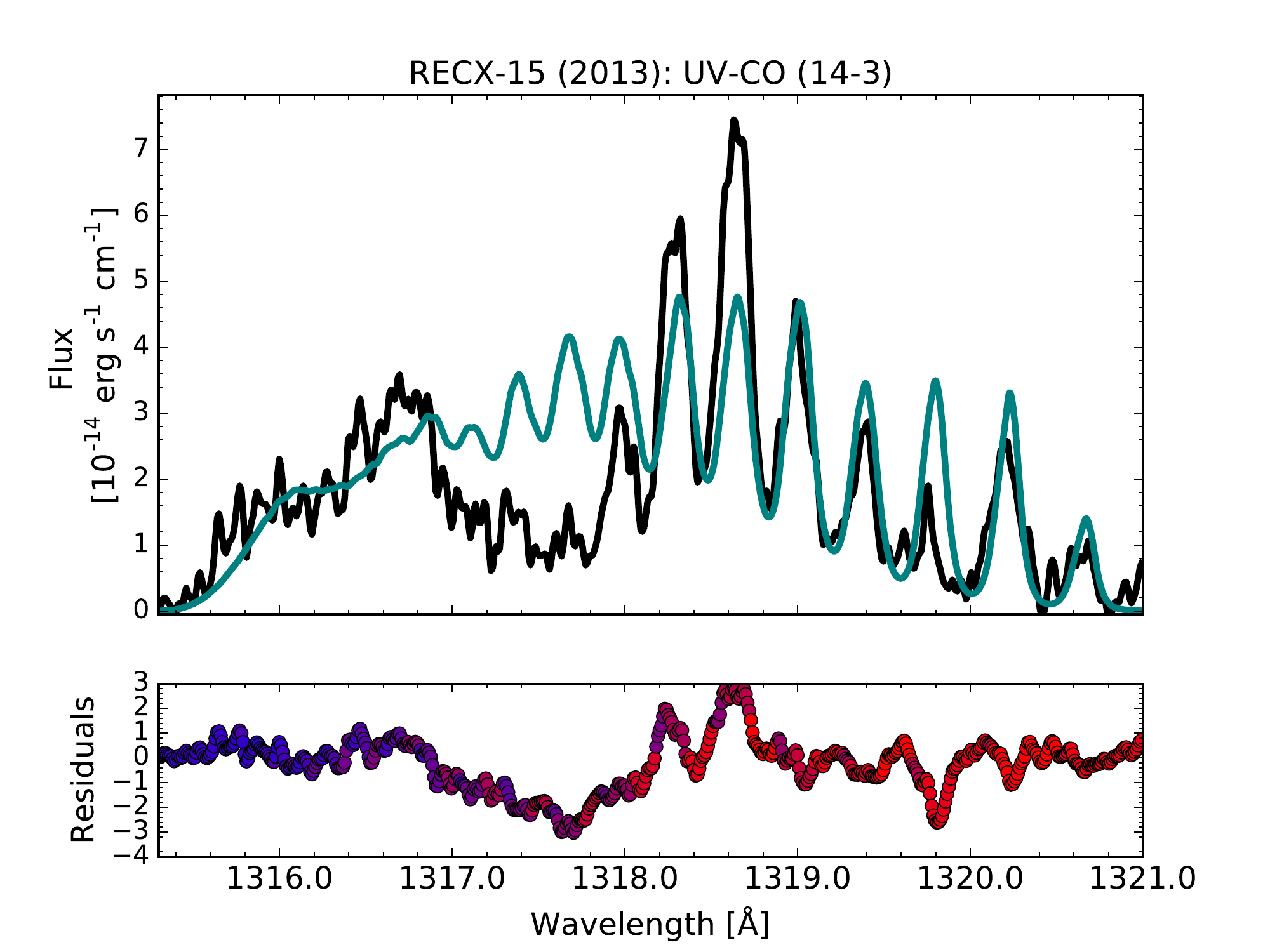}
	\end{minipage}
	\begin{minipage}{0.5\textwidth}
	\centering
	\includegraphics[width=\linewidth]{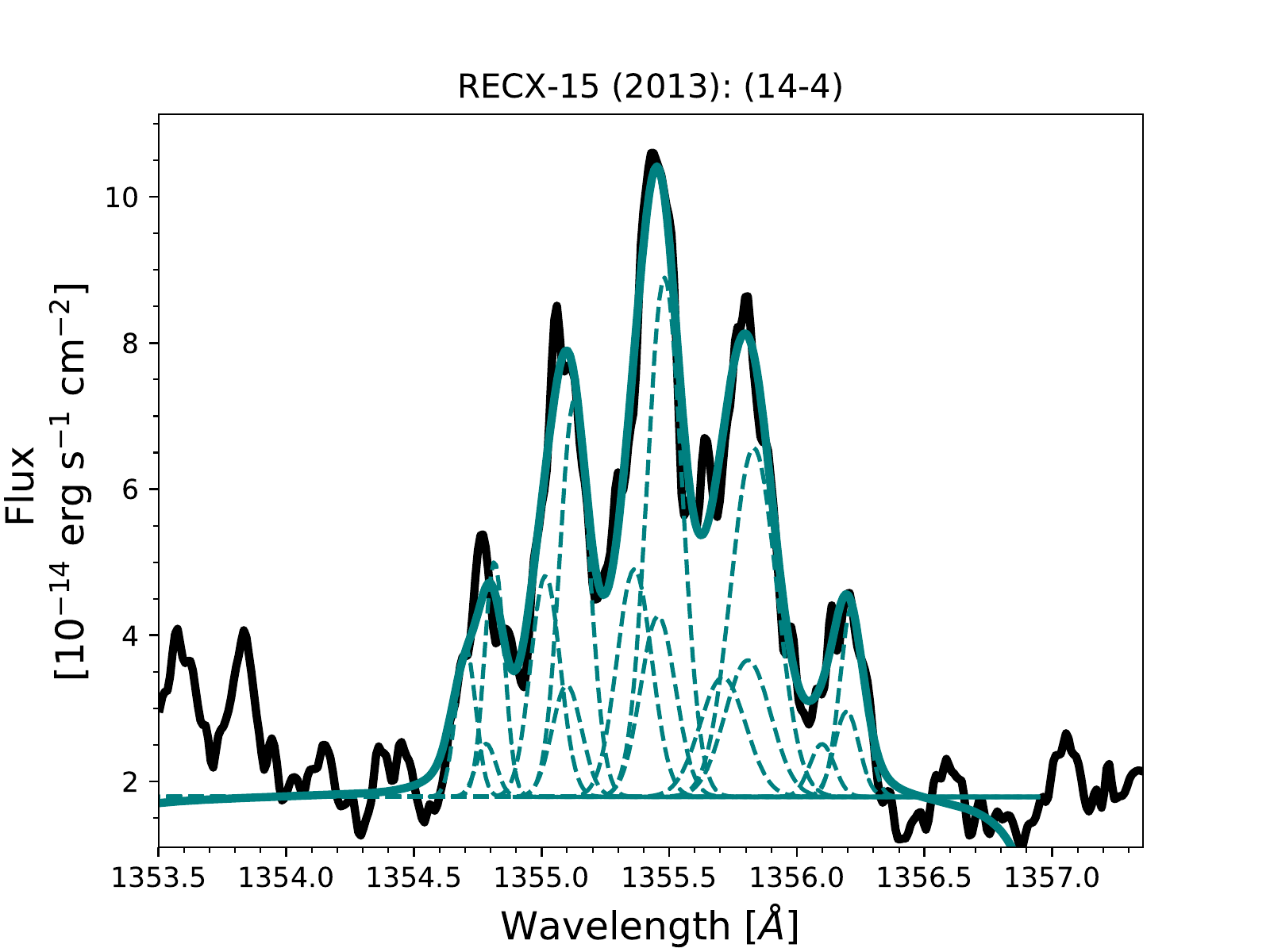}
	\end{minipage}
	\begin{minipage}{0.5\textwidth}
	\centering
	\includegraphics[width=\linewidth]{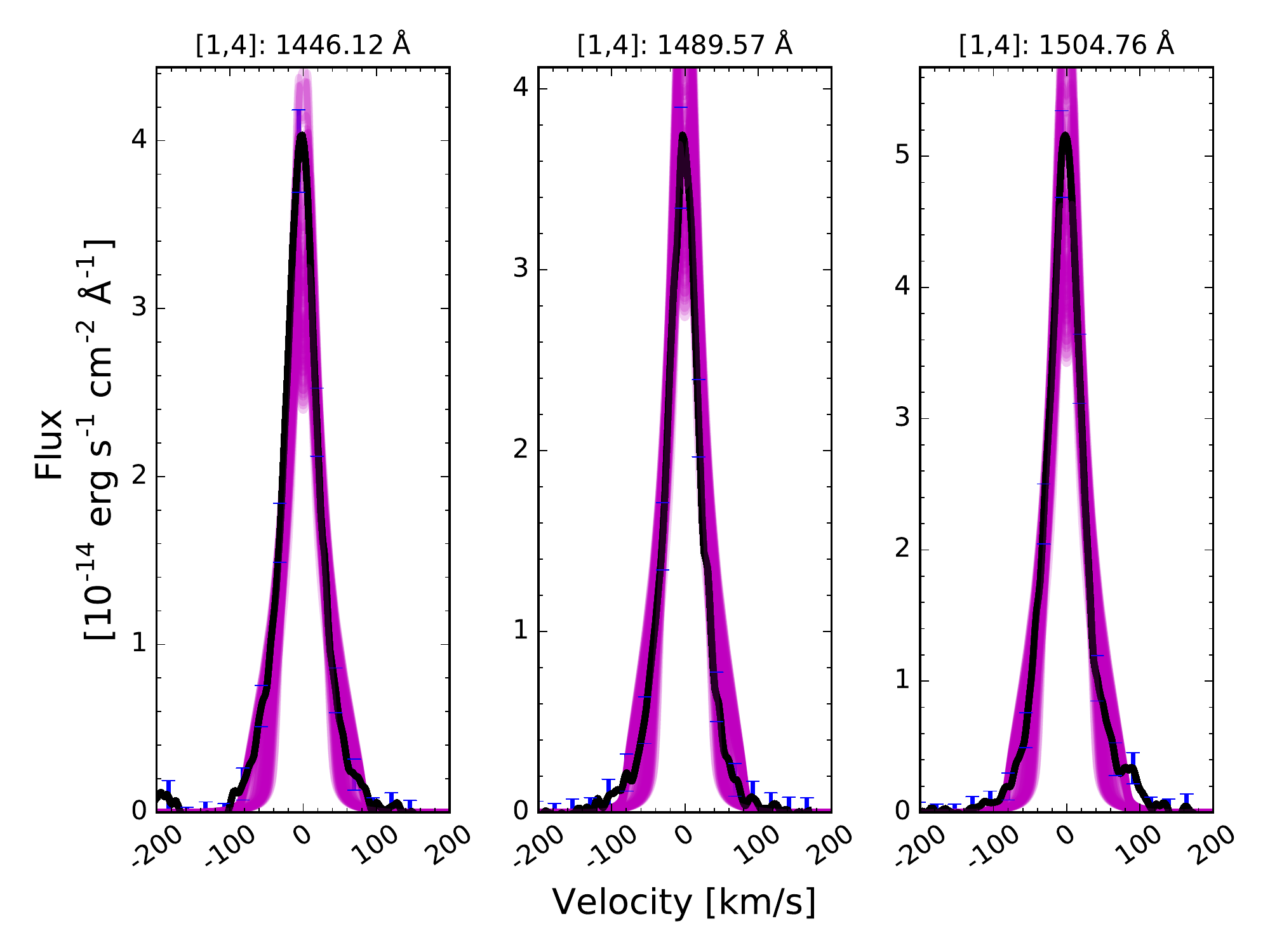} 
	\end{minipage}
	\begin{minipage}{0.5\textwidth}
	\centering
	\includegraphics[width=\linewidth]{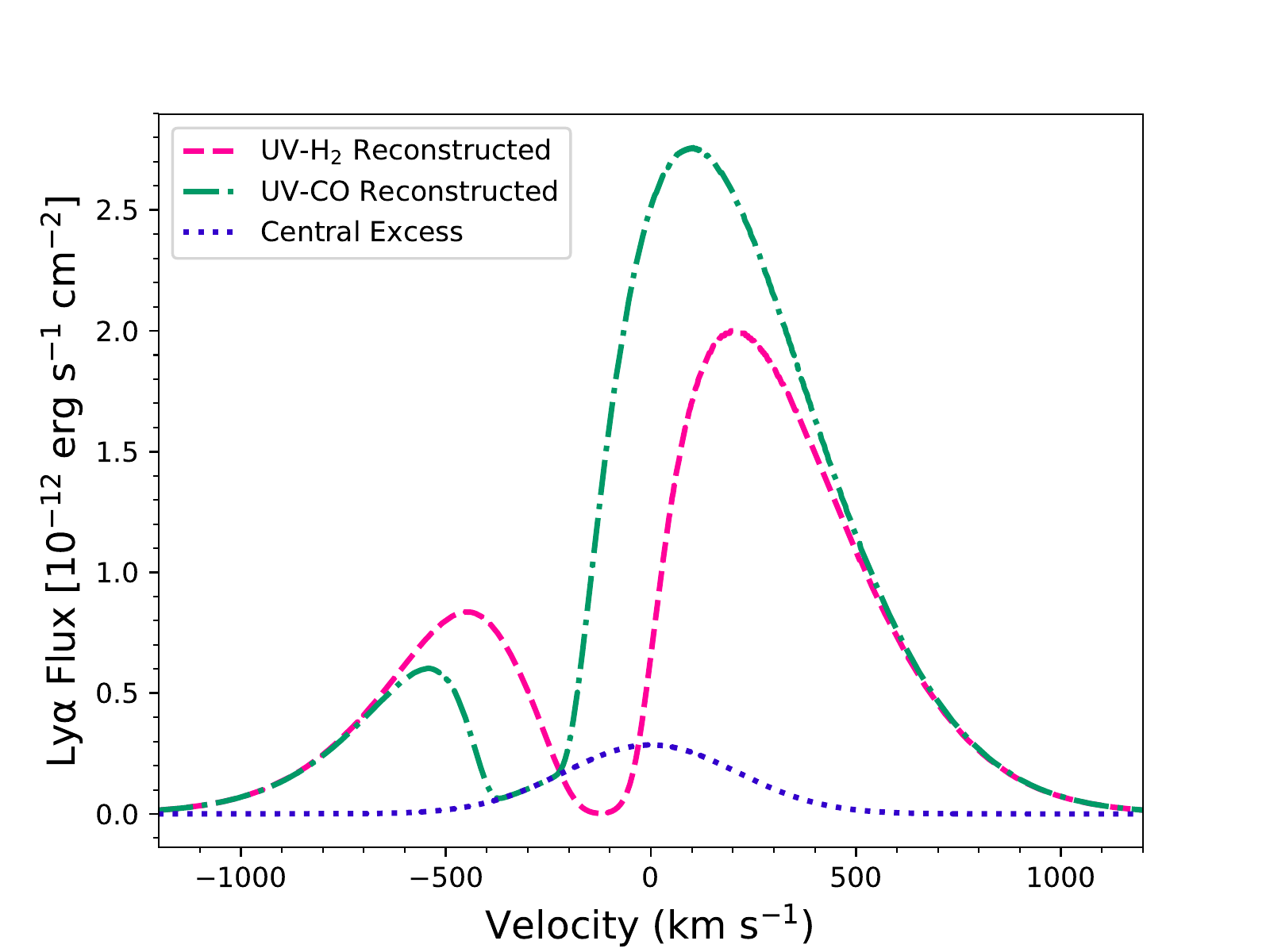} 
	\end{minipage}
	\begin{minipage}{0.5\textwidth}
	\centering
	\includegraphics[width=\linewidth]{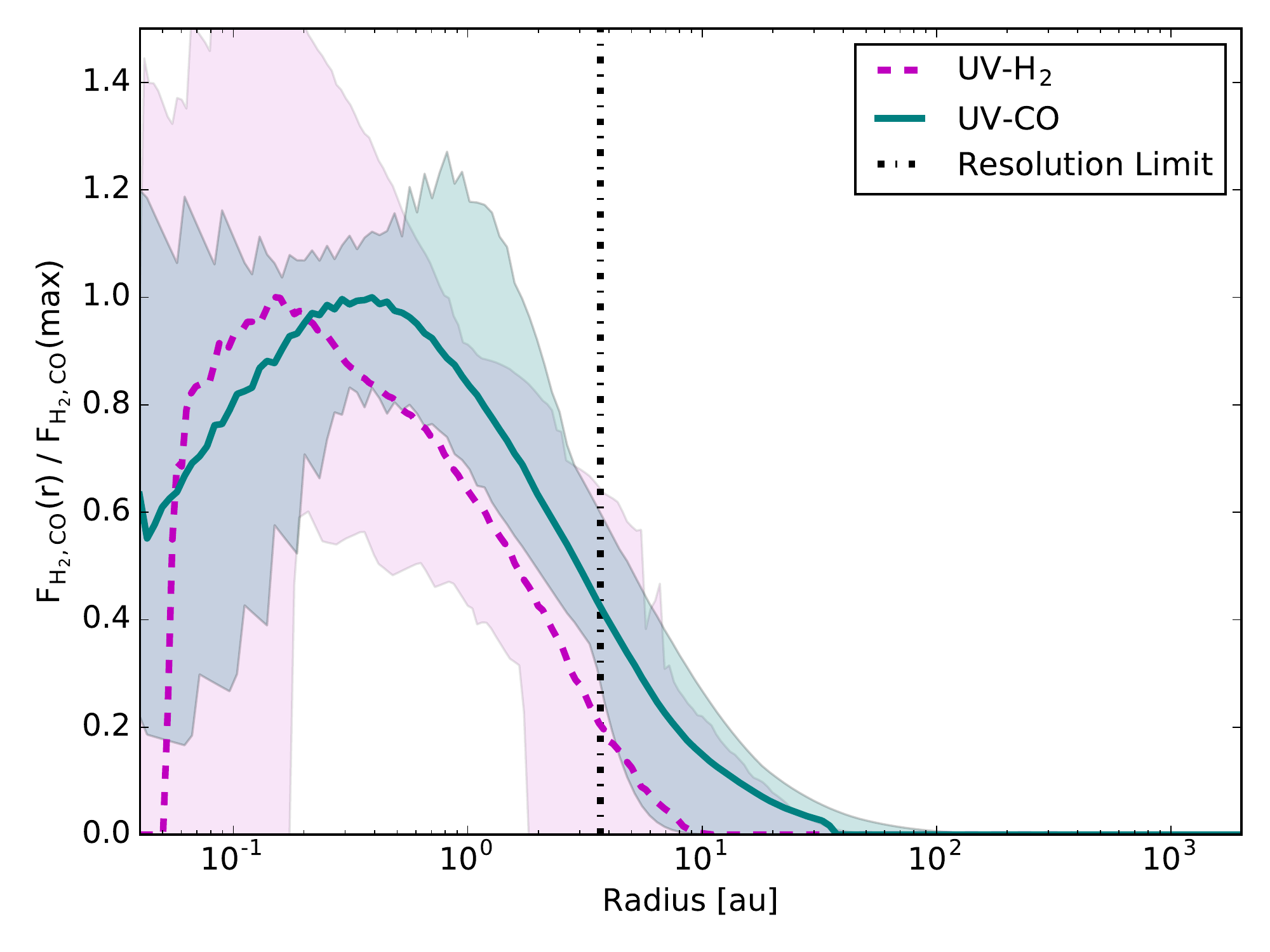} 
	\end{minipage}
\caption{Best-fit UV-CO 2-D radiative transfer (top left) and Gaussian models (top right), 100 UV-H$_2$ radiative transfer models with the smallest \textit{MSE}s (top right), comparison of UV-H$_2$-/UV-CO-based and observed Ly$\alpha$ profiles (bottom left) and radial distributions of flux for both species (bottom right) from the disk around RECX-15 (spectra from 2013). Residuals on the best-fit UV-CO model are color-coded from blue to red, based on the Ly$\alpha$ pumping wavelength required to excite the upper level of the transition.}
\label{RECX15_2013_H2CO}
\end{figure*}

\begin{figure*}[t!]
	\begin{minipage}{0.5\textwidth}
	\centering
	\includegraphics[width=\linewidth]{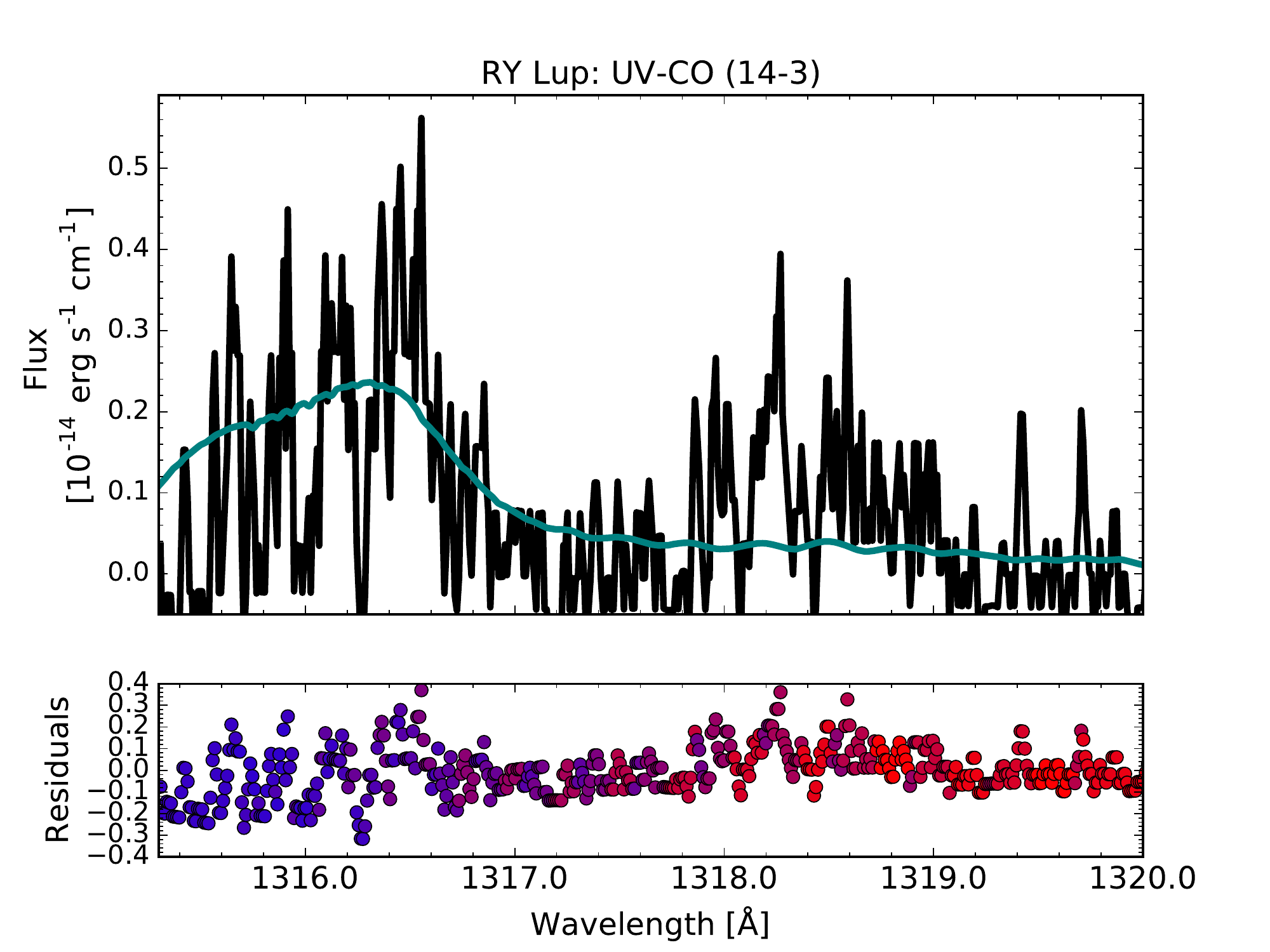}
	\end{minipage}
	\begin{minipage}{0.5\textwidth}
	\centering
	\includegraphics[width=\linewidth]{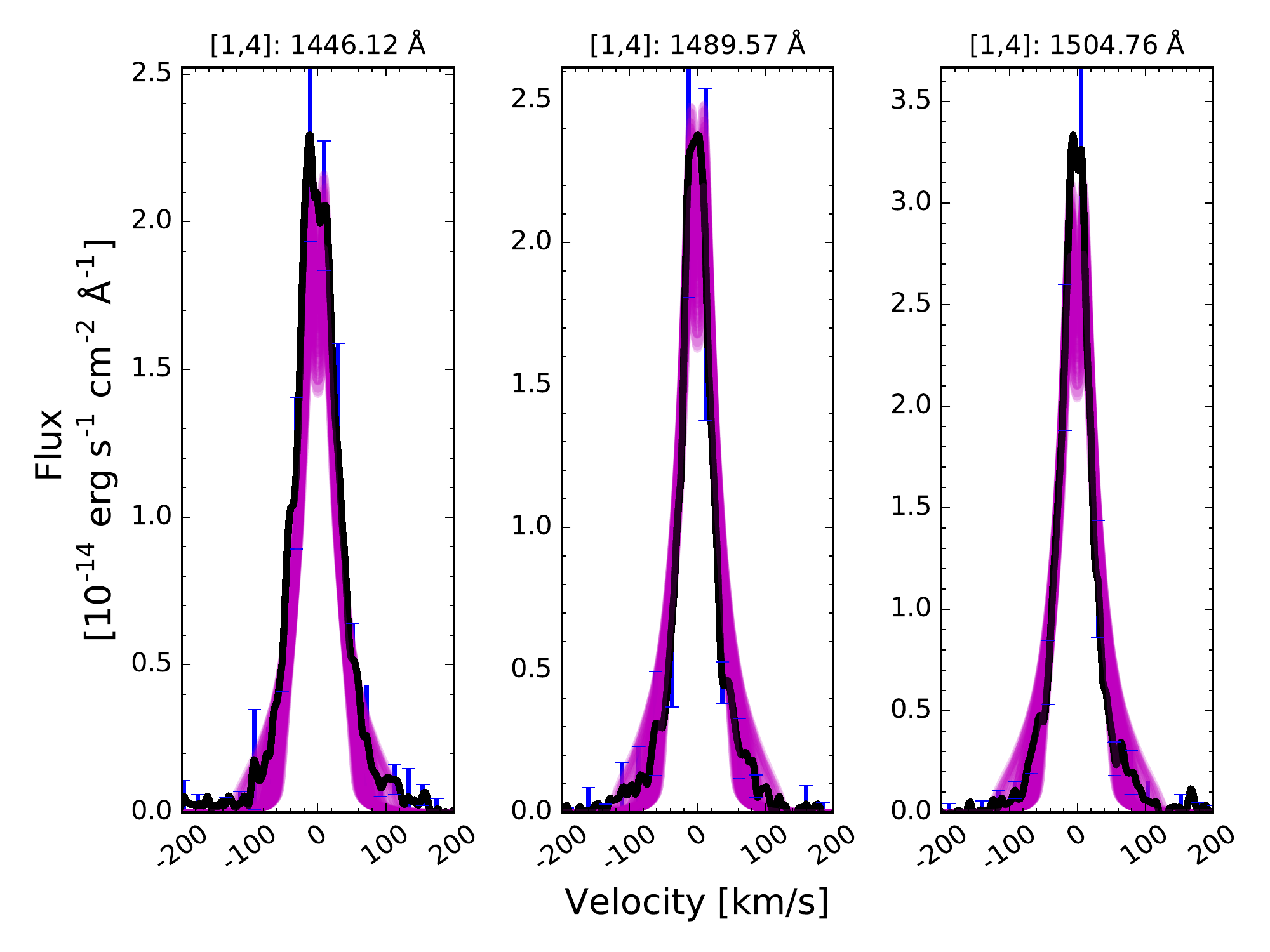} 
	\end{minipage}
	\begin{minipage}{0.5\textwidth}
	\centering
	\includegraphics[width=\linewidth]{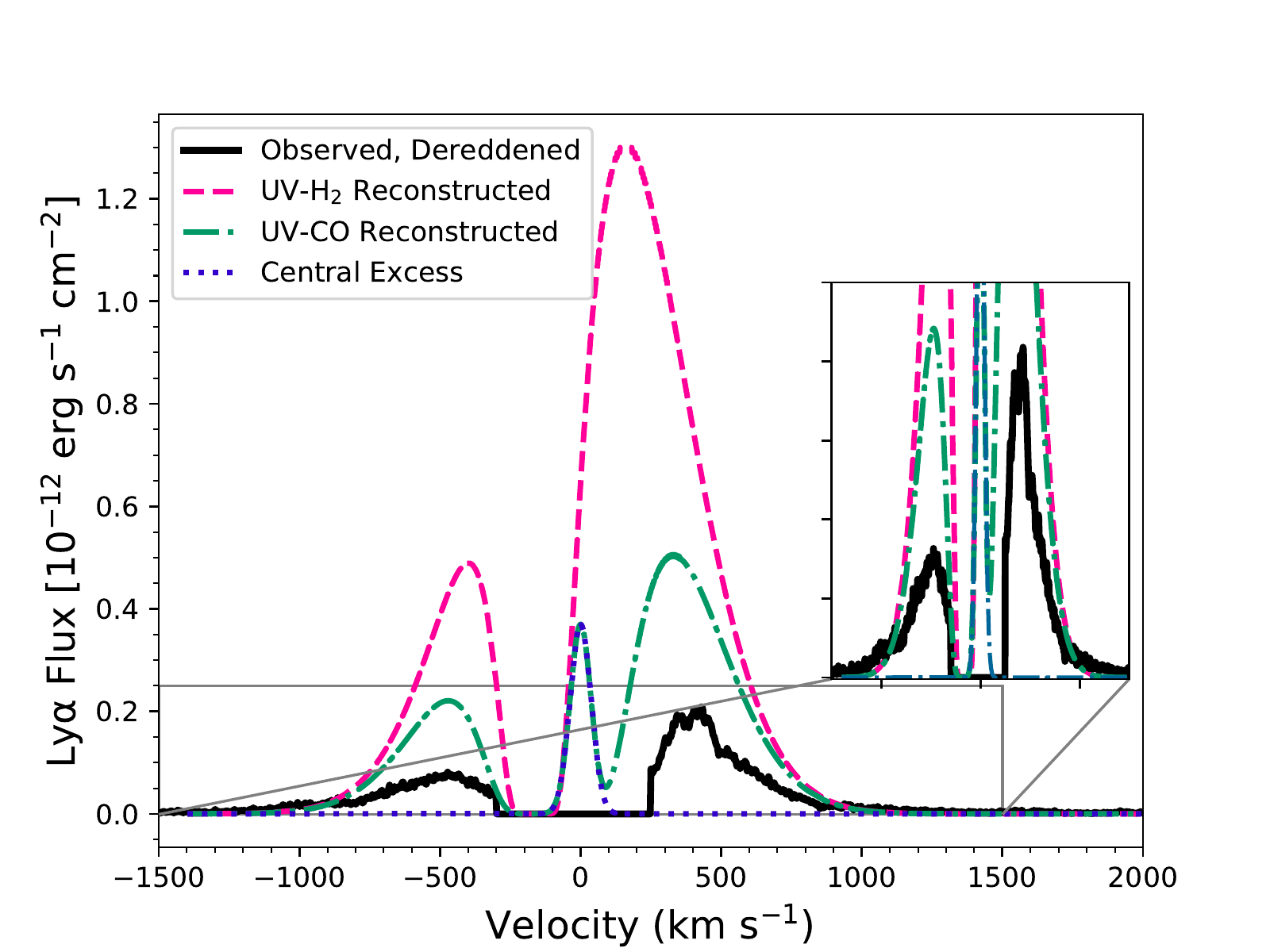} 
	\end{minipage}
	\begin{minipage}{0.5\textwidth}
	\centering
	\includegraphics[width=\linewidth]{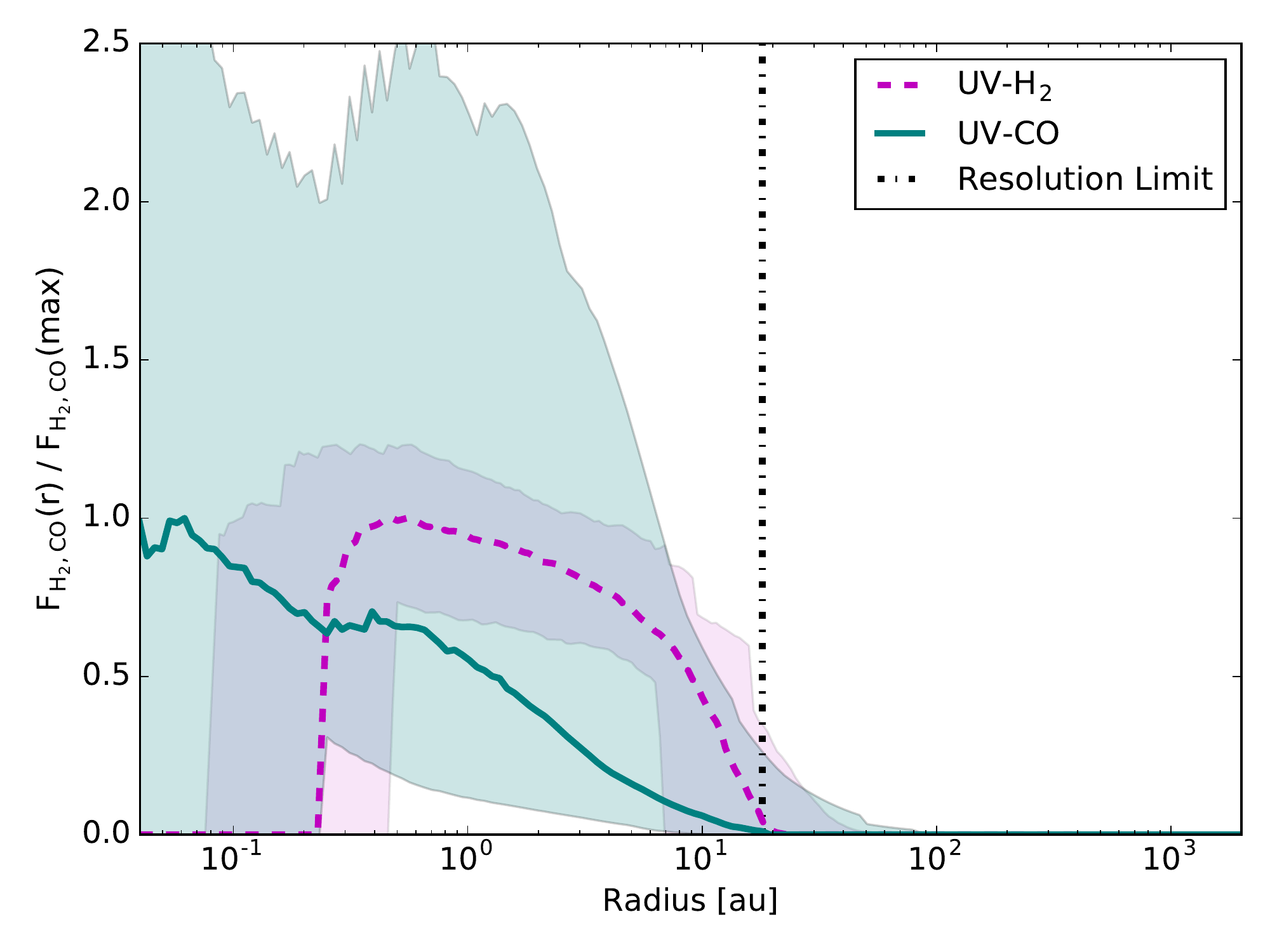} 
	\end{minipage}
\caption{Best-fit UV-CO model (top left), 100 UV-H$_2$ radiative transfer models with the smallest \textit{MSE}s (top right), comparison of UV-H$_2$-/UV-CO-based and observed Ly$\alpha$ profiles (bottom left) and radial distributions of flux for both species (bottom right) from the disk around RY Lupi. Residuals on the best-fit UV-CO model are color-coded from blue to red, based on the Ly$\alpha$ pumping wavelength required to excite the upper level of the transition.}
\label{RYLup_H2CO}
\end{figure*}

\begin{figure*}[t!]
	\begin{minipage}{0.5\textwidth}
	\centering
	\includegraphics[width=\linewidth]{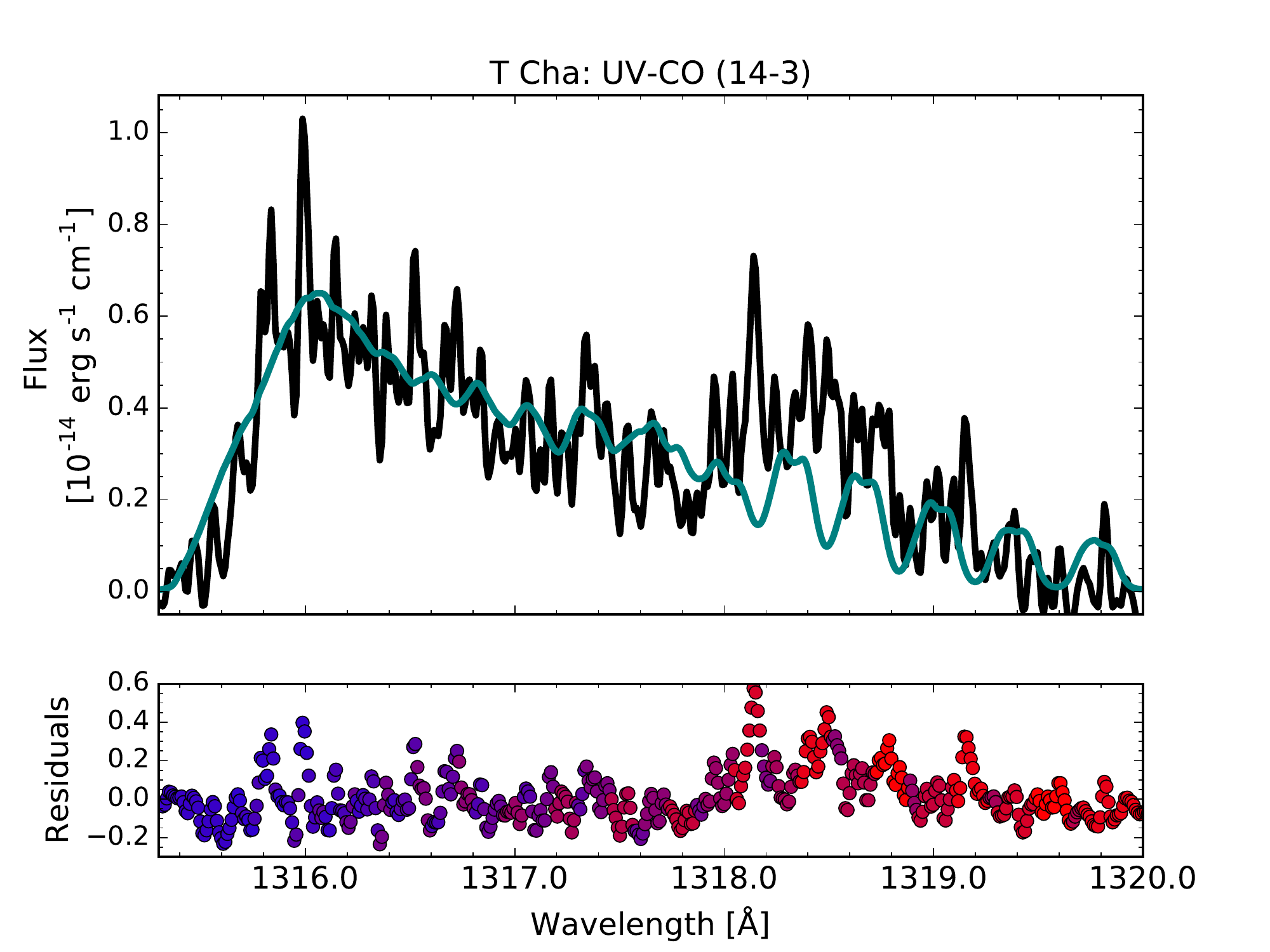}
	\end{minipage}
	\begin{minipage}{0.5\textwidth}
	\centering
	\includegraphics[width=\linewidth]{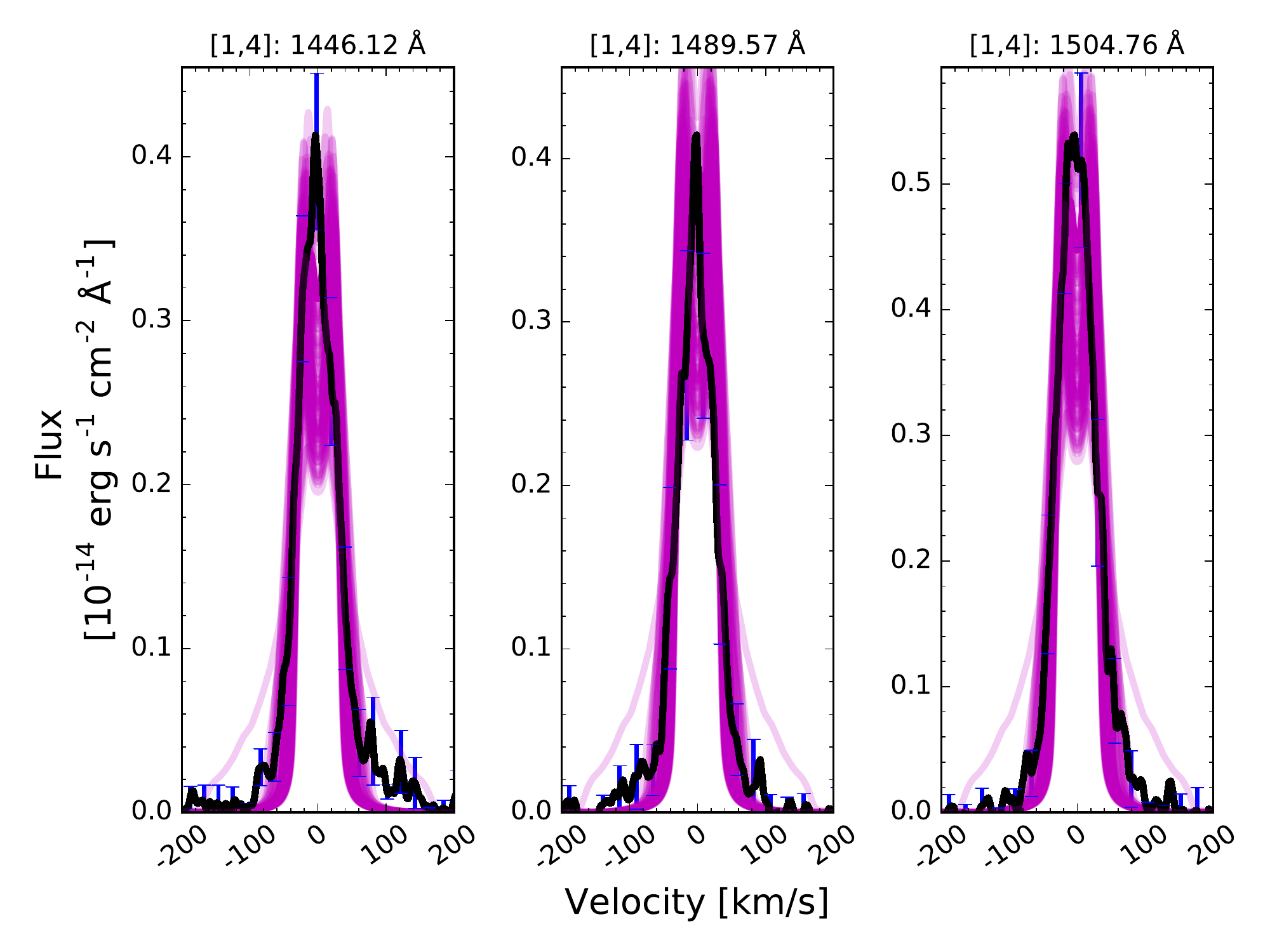} 
	\end{minipage}
	\begin{minipage}{0.5\textwidth}
	\centering
	\includegraphics[width=\linewidth]{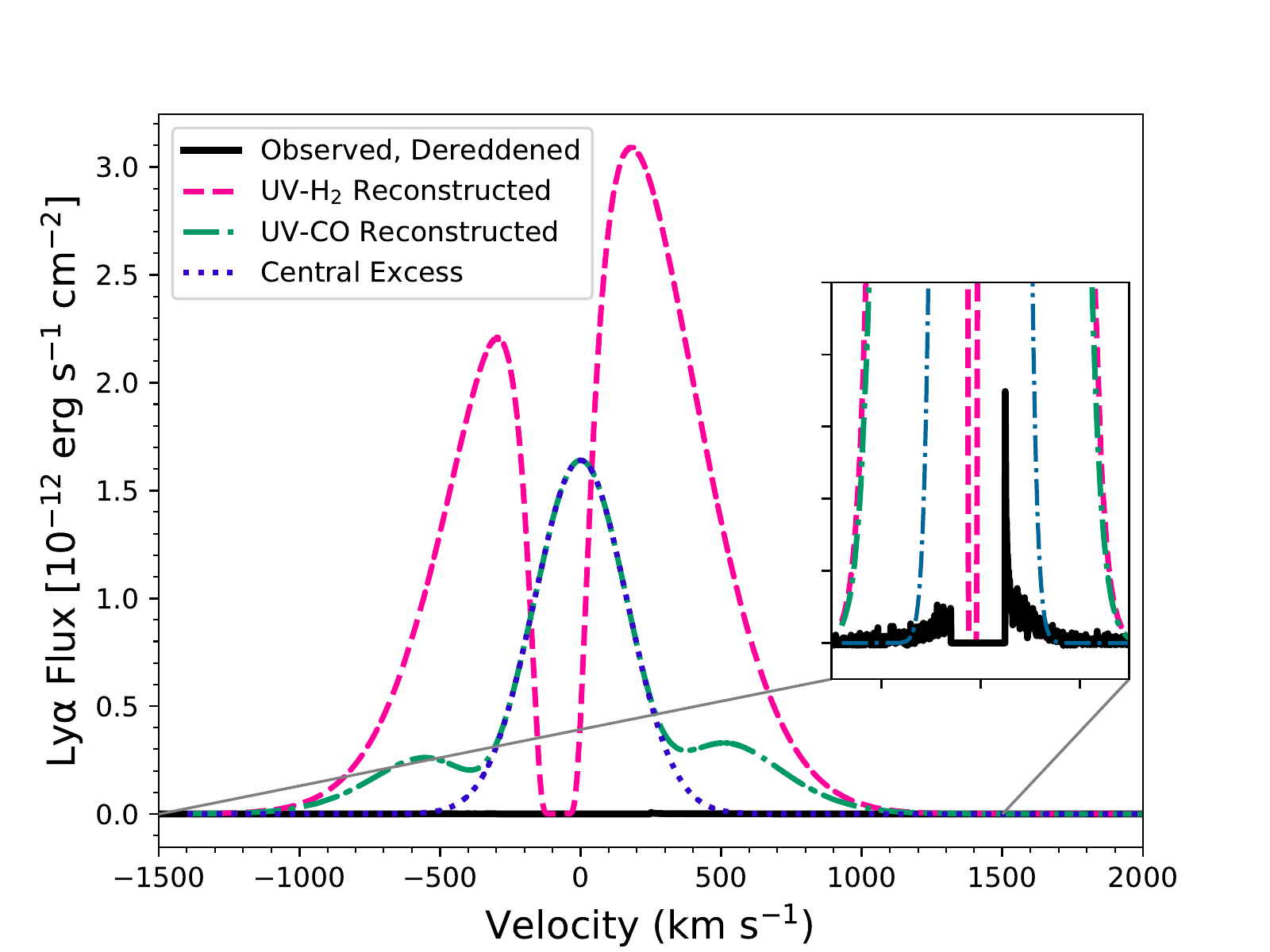} 
	\end{minipage}
	\begin{minipage}{0.5\textwidth}
	\centering
	\includegraphics[width=\linewidth]{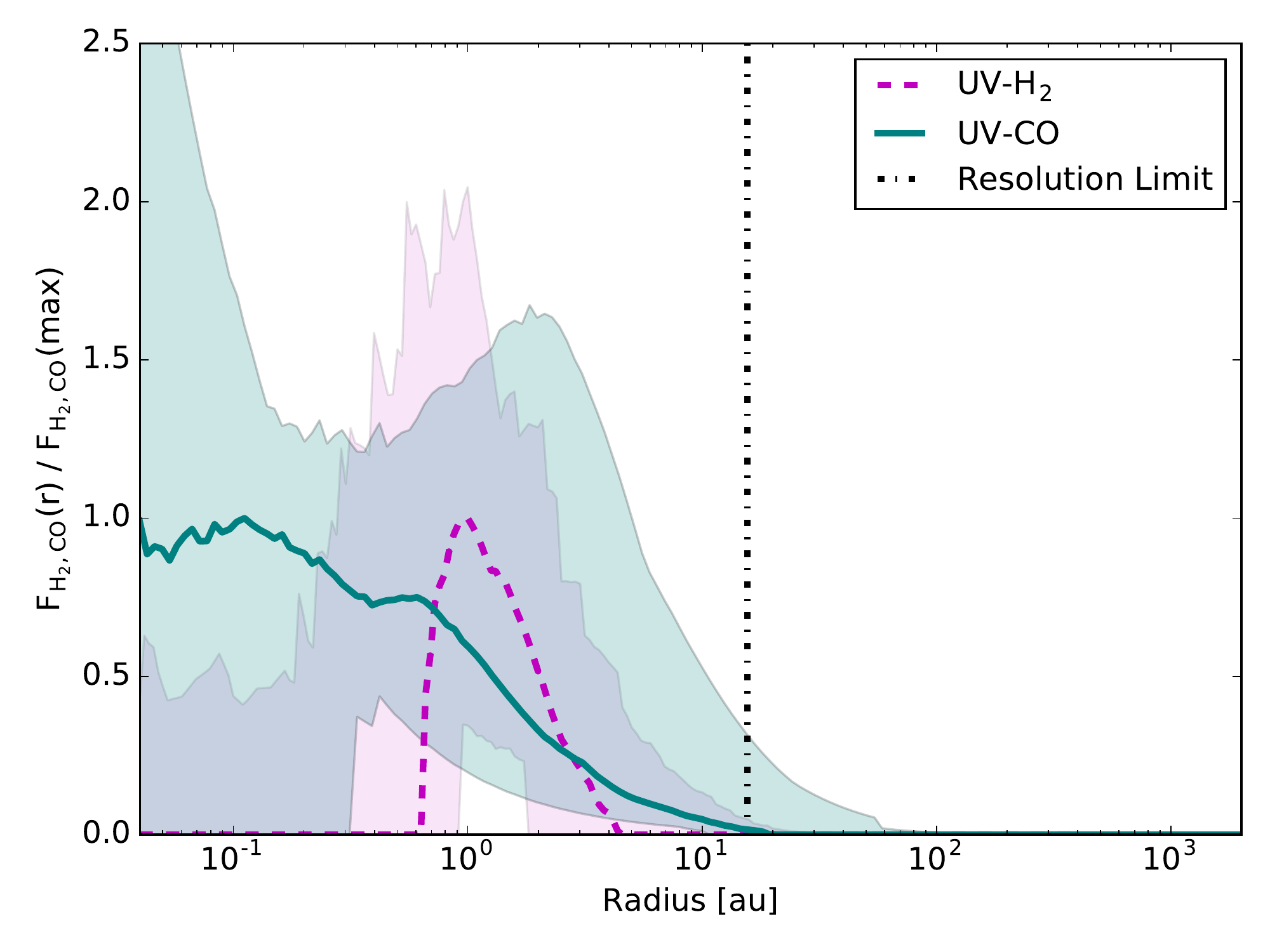} 
	\end{minipage}
\caption{Best-fit UV-CO model (top left), 100 UV-H$_2$ radiative transfer models with the smallest \textit{MSE}s (top right), comparison of UV-H$_2$-/UV-CO-based and observed Ly$\alpha$ profiles (bottom left) and radial distributions of flux for both species (bottom right) from the disk around T Cha. Residuals on the best-fit UV-CO model are color-coded from blue to red, based on the Ly$\alpha$ pumping wavelength required to excite the upper level of the transition.}
\label{TCha_H2CO}
\end{figure*}

\begin{figure*}[t!]
	\begin{minipage}{0.5\textwidth}
	\centering
	\includegraphics[width=\linewidth]{f15_57.pdf}
	\end{minipage}
	\begin{minipage}{0.5\textwidth}
	\centering
	\includegraphics[width=\linewidth]{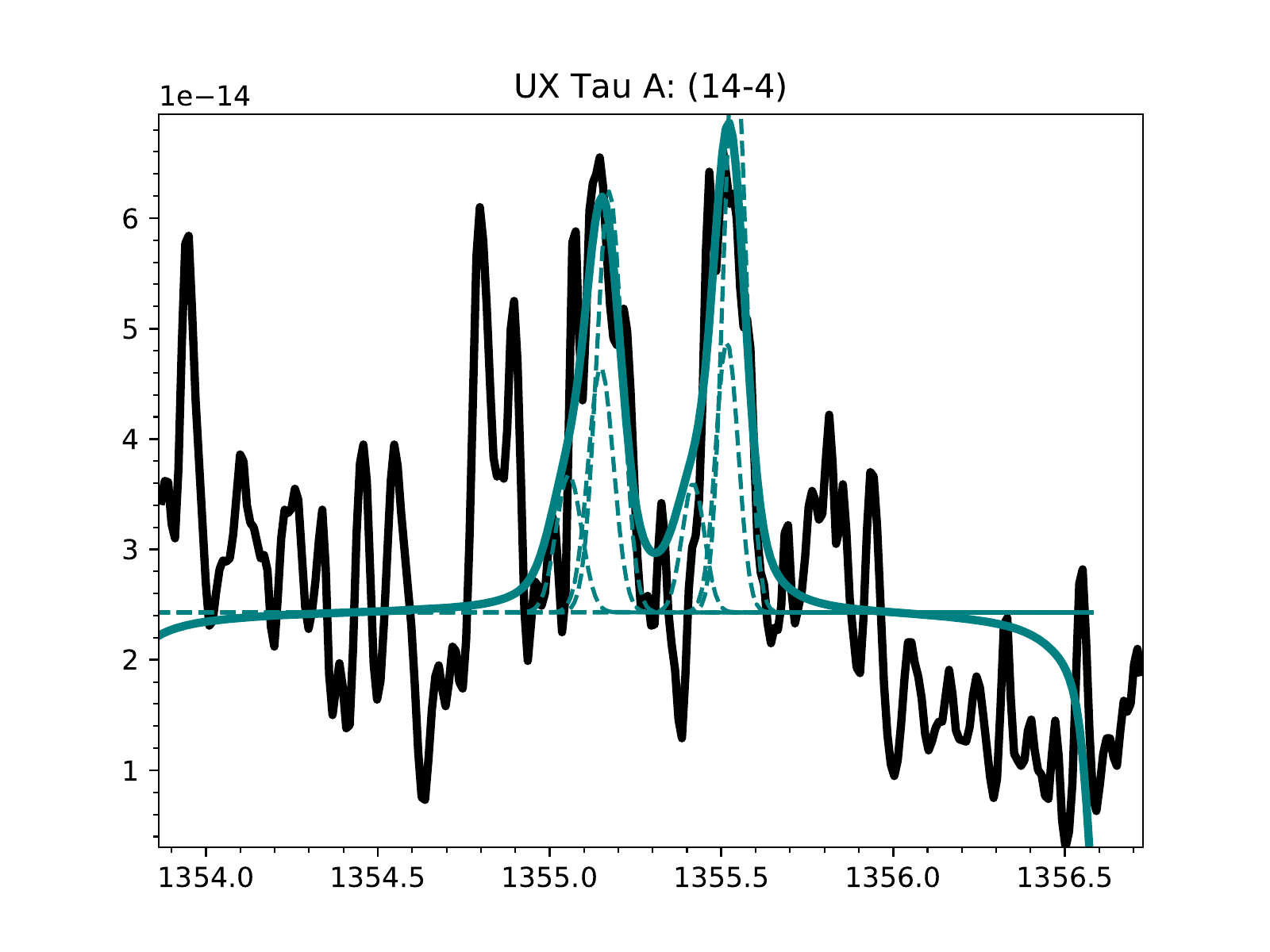}
	\end{minipage}
	\begin{minipage}{0.5\textwidth}
	\centering
	\includegraphics[width=\linewidth]{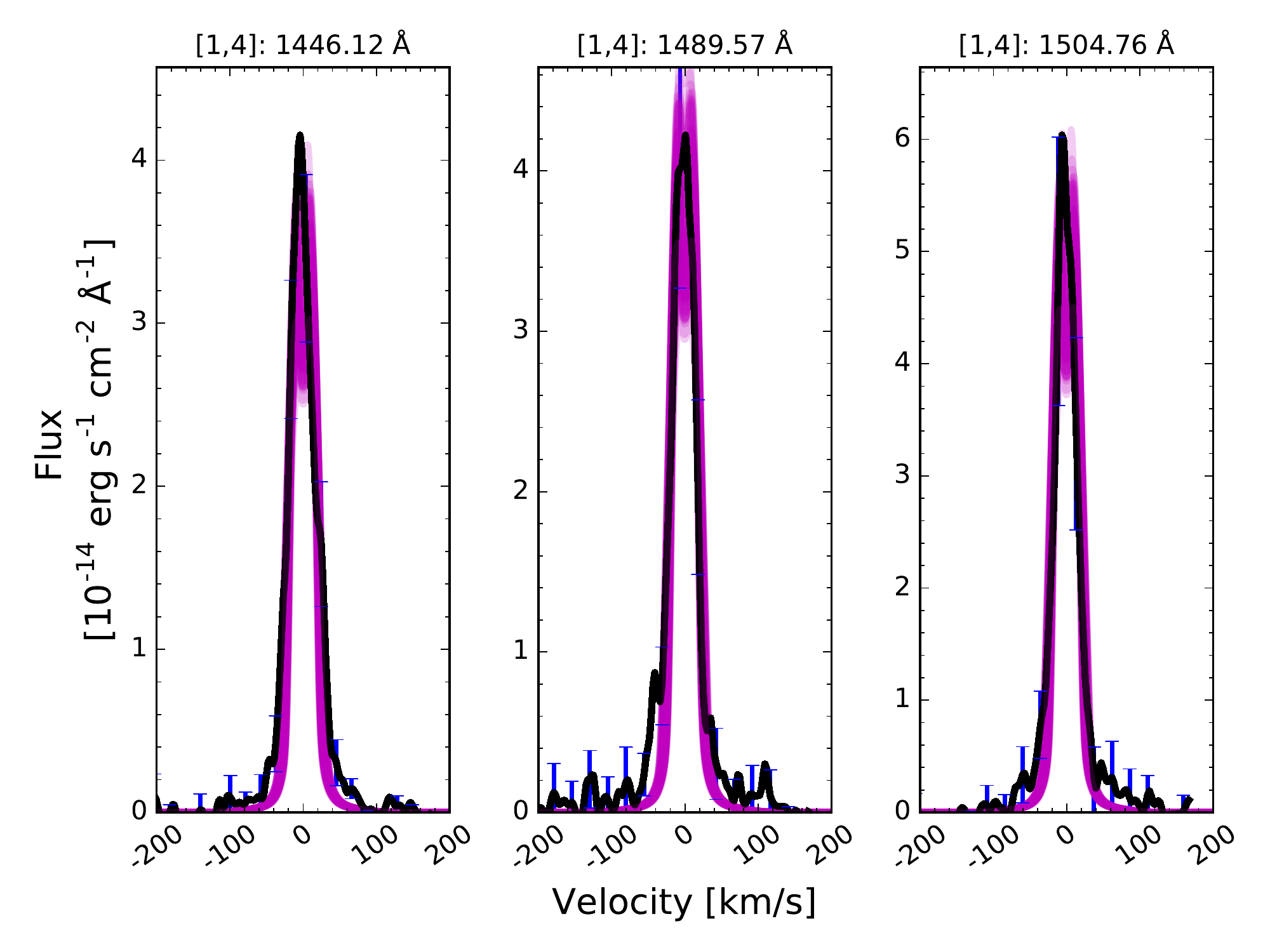} 
	\end{minipage}
	\begin{minipage}{0.5\textwidth}
	\centering
	\includegraphics[width=\linewidth]{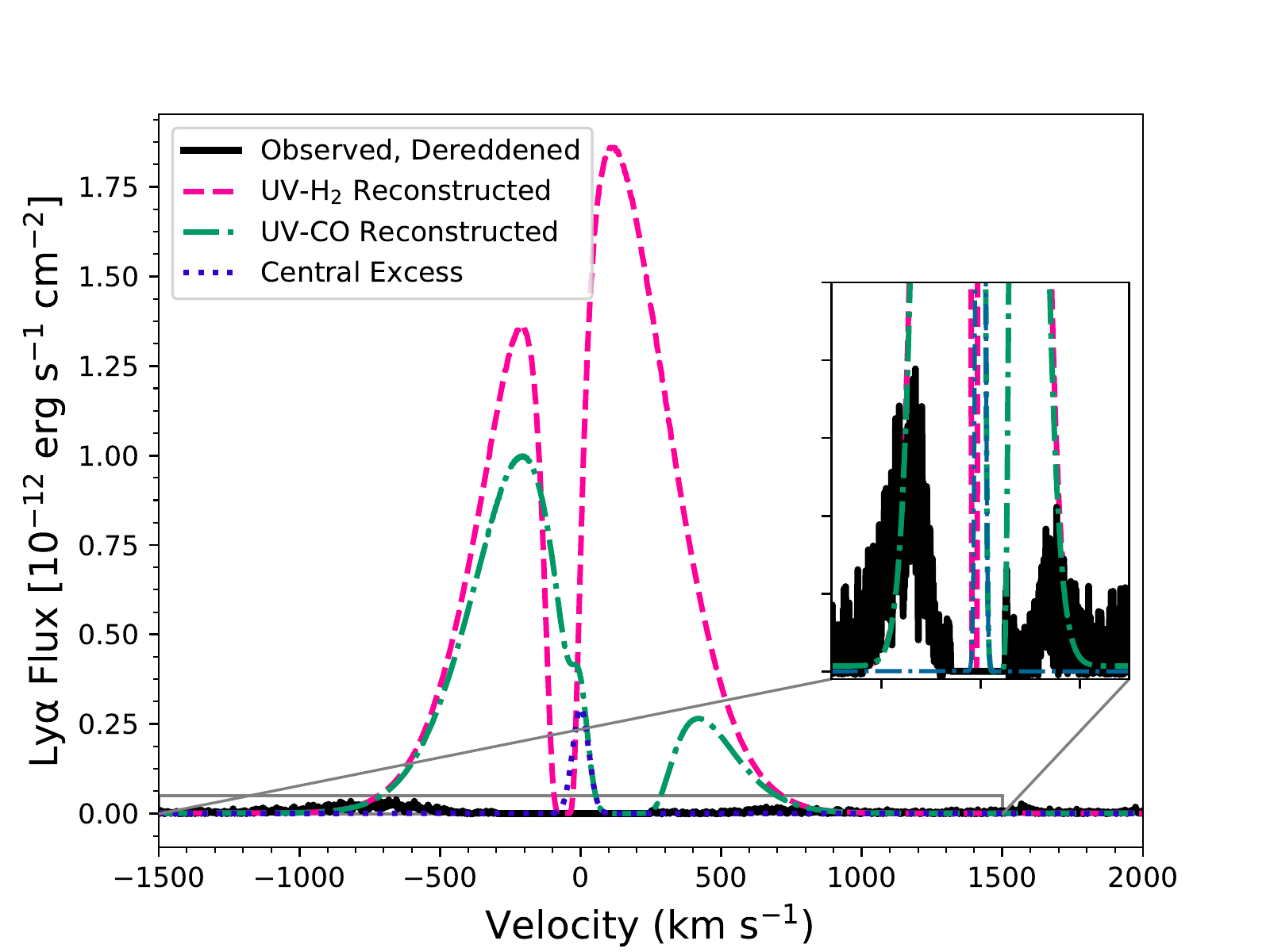} 
	\end{minipage}
	\begin{minipage}{0.5\textwidth}
	\centering
	\includegraphics[width=\linewidth]{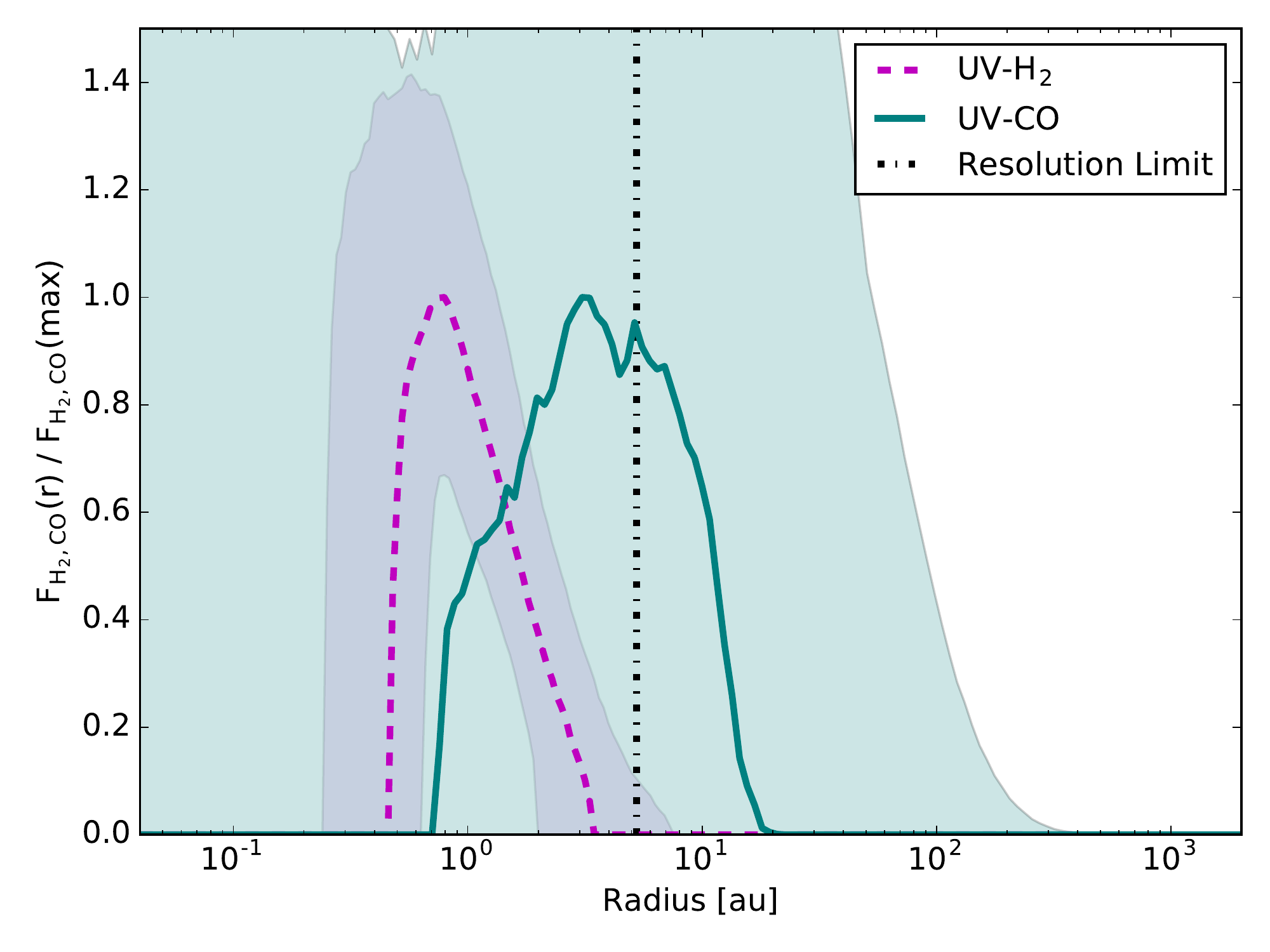} 
	\end{minipage}
\caption{Best-fit UV-CO 2-D radiative transfer (top left) and Gaussian models (top right), 100 UV-H$_2$ radiative transfer models with the smallest \textit{MSE}s (middle left), comparison of UV-H$_2$-/UV-CO-based and observed Ly$\alpha$ profiles (middle right) and radial distributions of flux for both species (bottom left) from the disk around UX Tau. The best-fit UV-H$_2$ model is unable to simultaneously capture emission from the high-velocity wings of the line profiles ($v > 75$ km s$^{-1}$) and the narrow line cores, so the true flux distribution likely extends closer to the central star than the inner radius shown here $\left( r < 0.4 \, \rm{AU} \right)$. Residuals on the best-fit UV-CO model are color-coded from blue to red, based on the Ly$\alpha$ pumping wavelength required to excite the upper level of the transition.}
\label{UXTau_H2CO}
\end{figure*}

\begin{figure*}[t!]
	\begin{minipage}{0.5\textwidth}
	\centering
	\includegraphics[width=\linewidth]{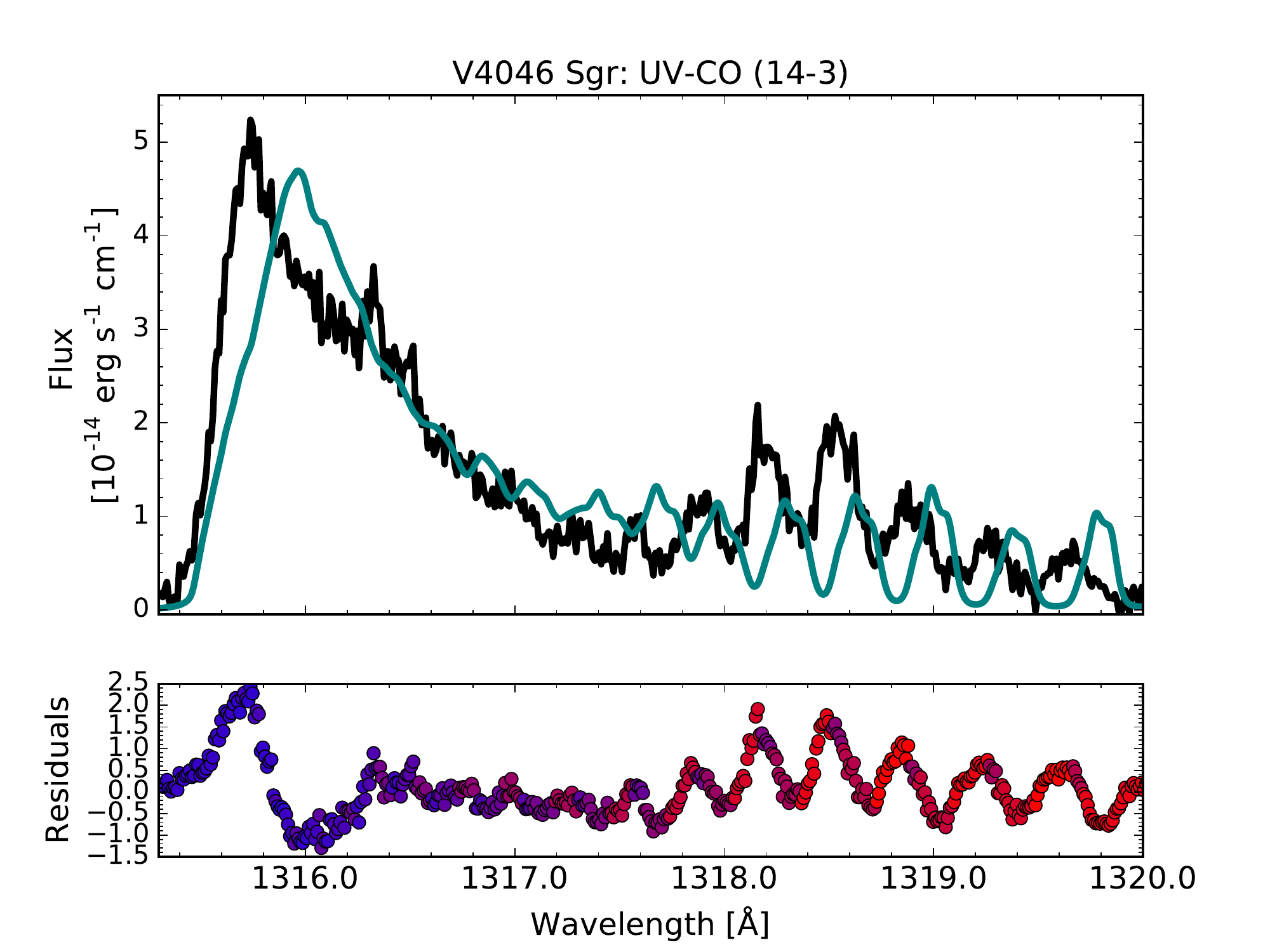}
	\end{minipage}
	\begin{minipage}{0.5\textwidth}
	\centering
	\includegraphics[width=\linewidth]{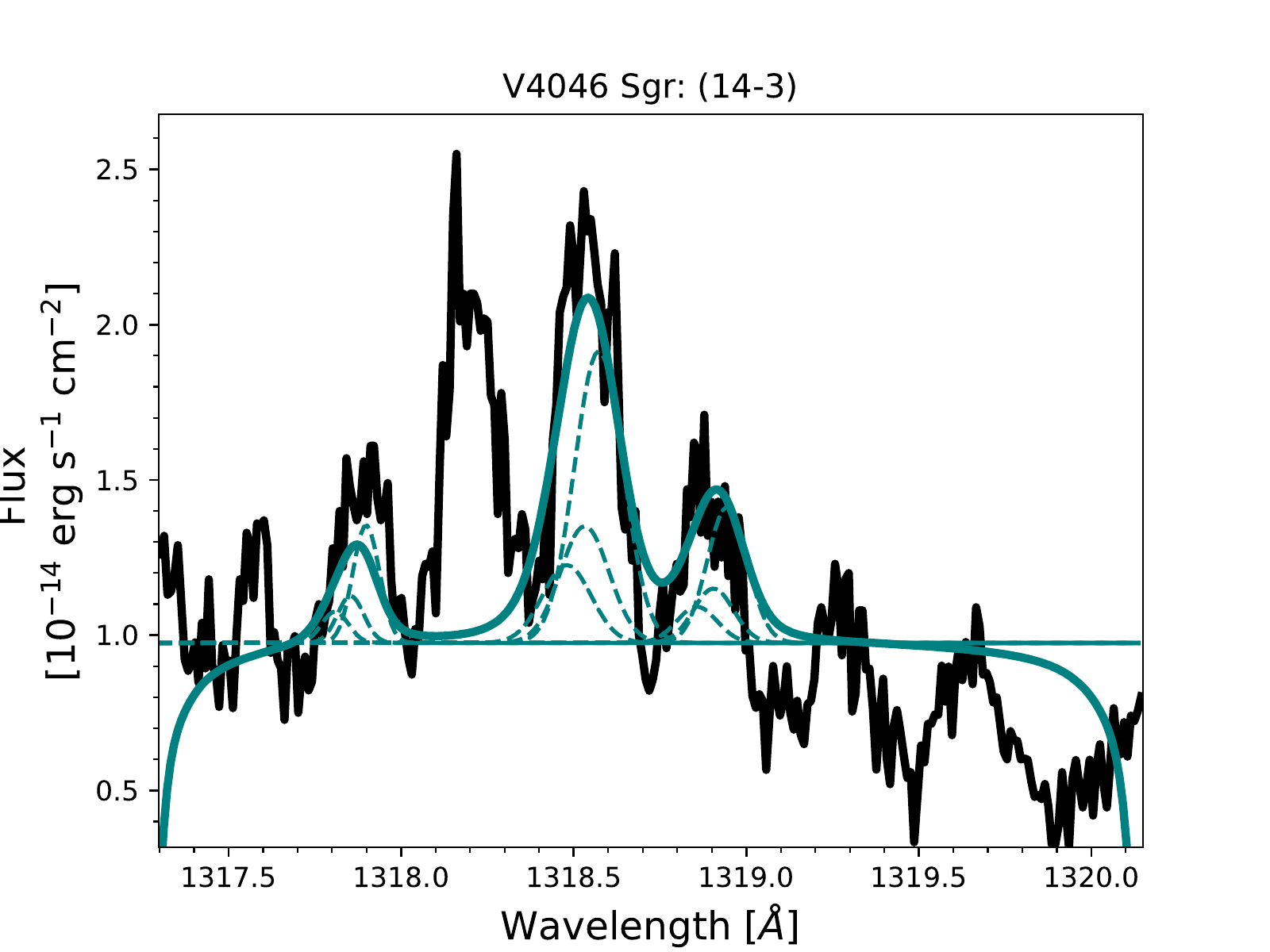}
	\end{minipage}
	\begin{minipage}{0.5\textwidth}
	\centering
	\includegraphics[width=\linewidth]{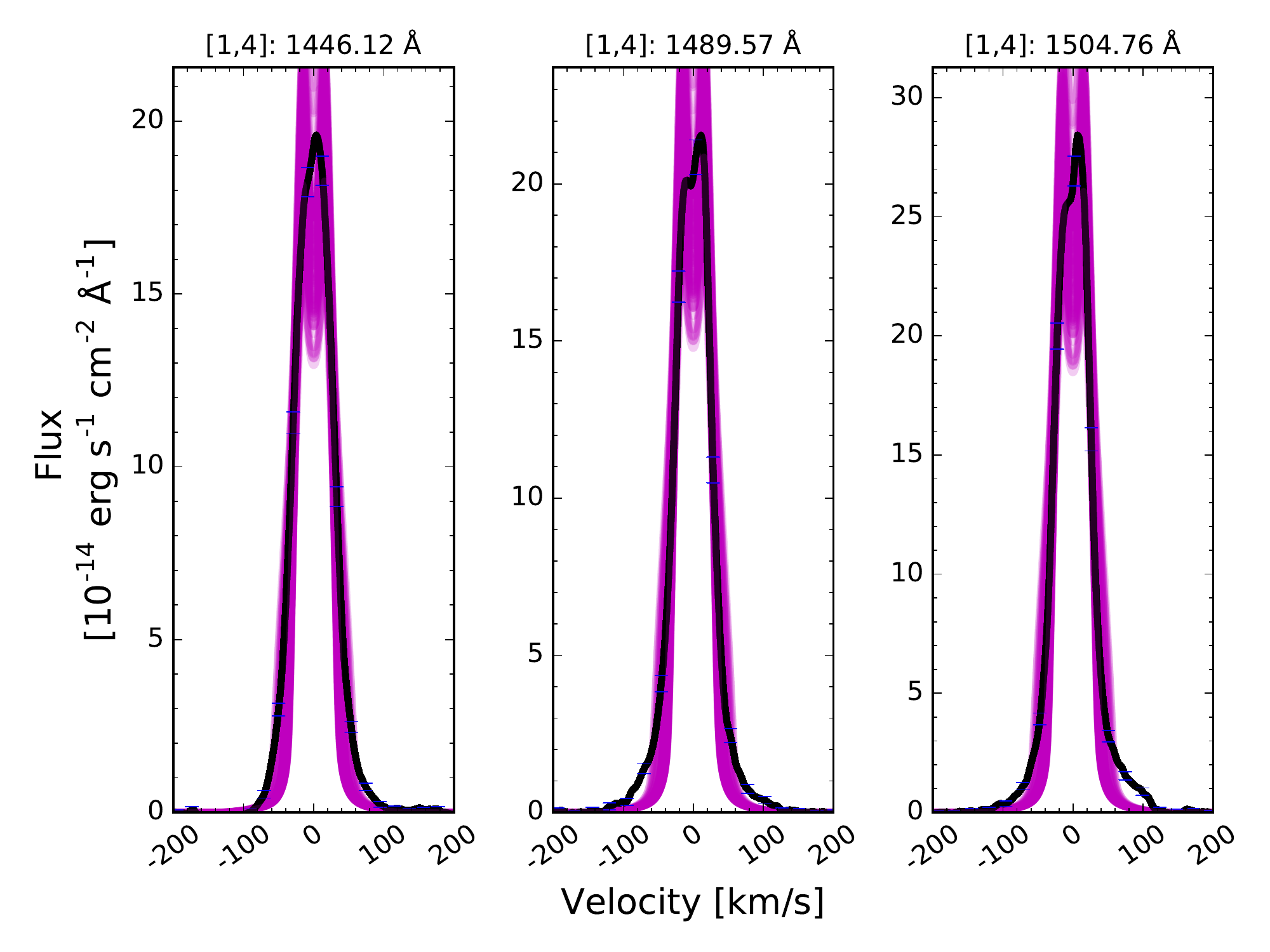} 
	\end{minipage}
	\begin{minipage}{0.5\textwidth}
	\centering
	\includegraphics[width=\linewidth]{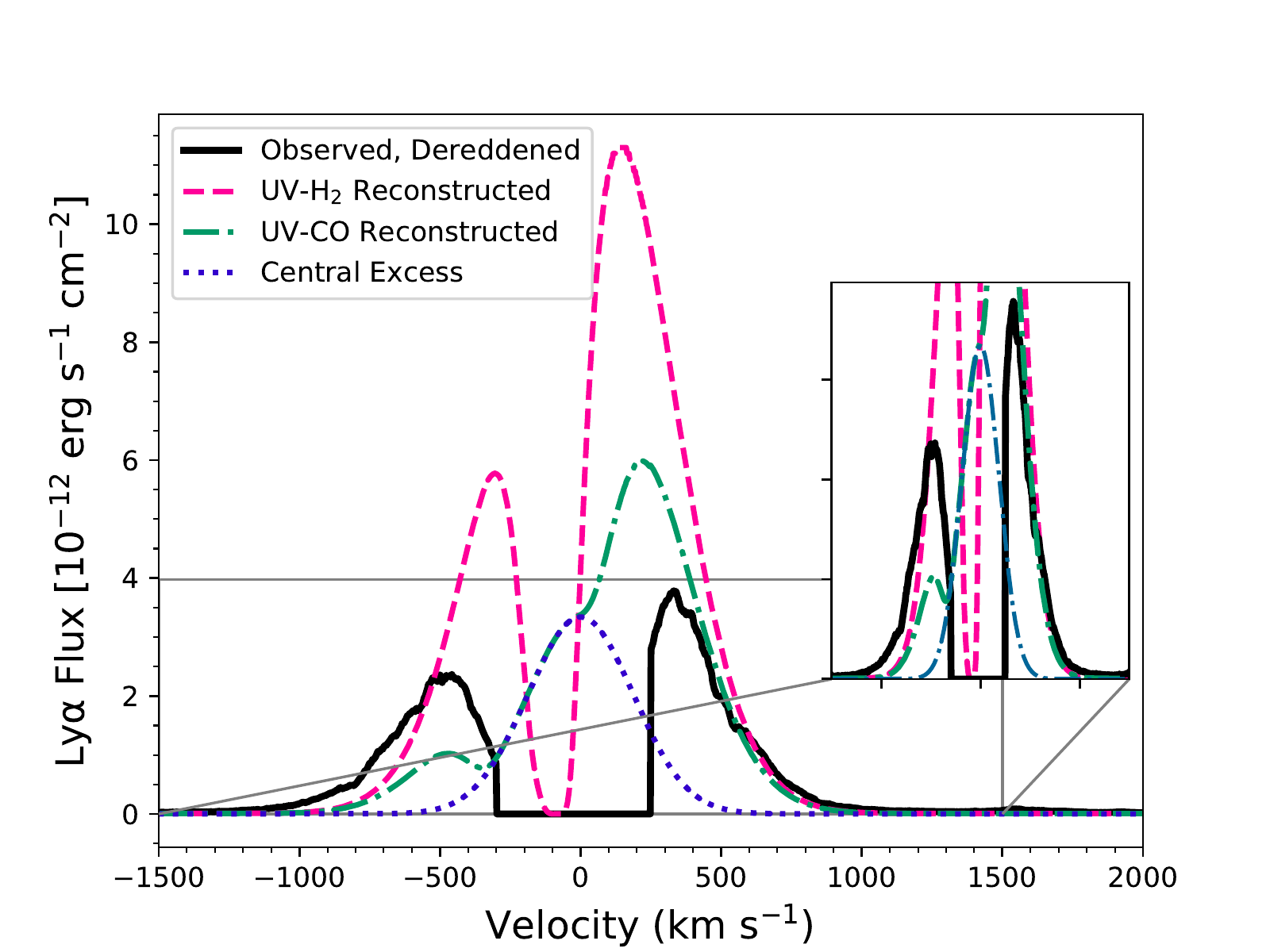} 
	\end{minipage}
	\begin{minipage}{0.5\textwidth}
	\centering
	\includegraphics[width=\linewidth]{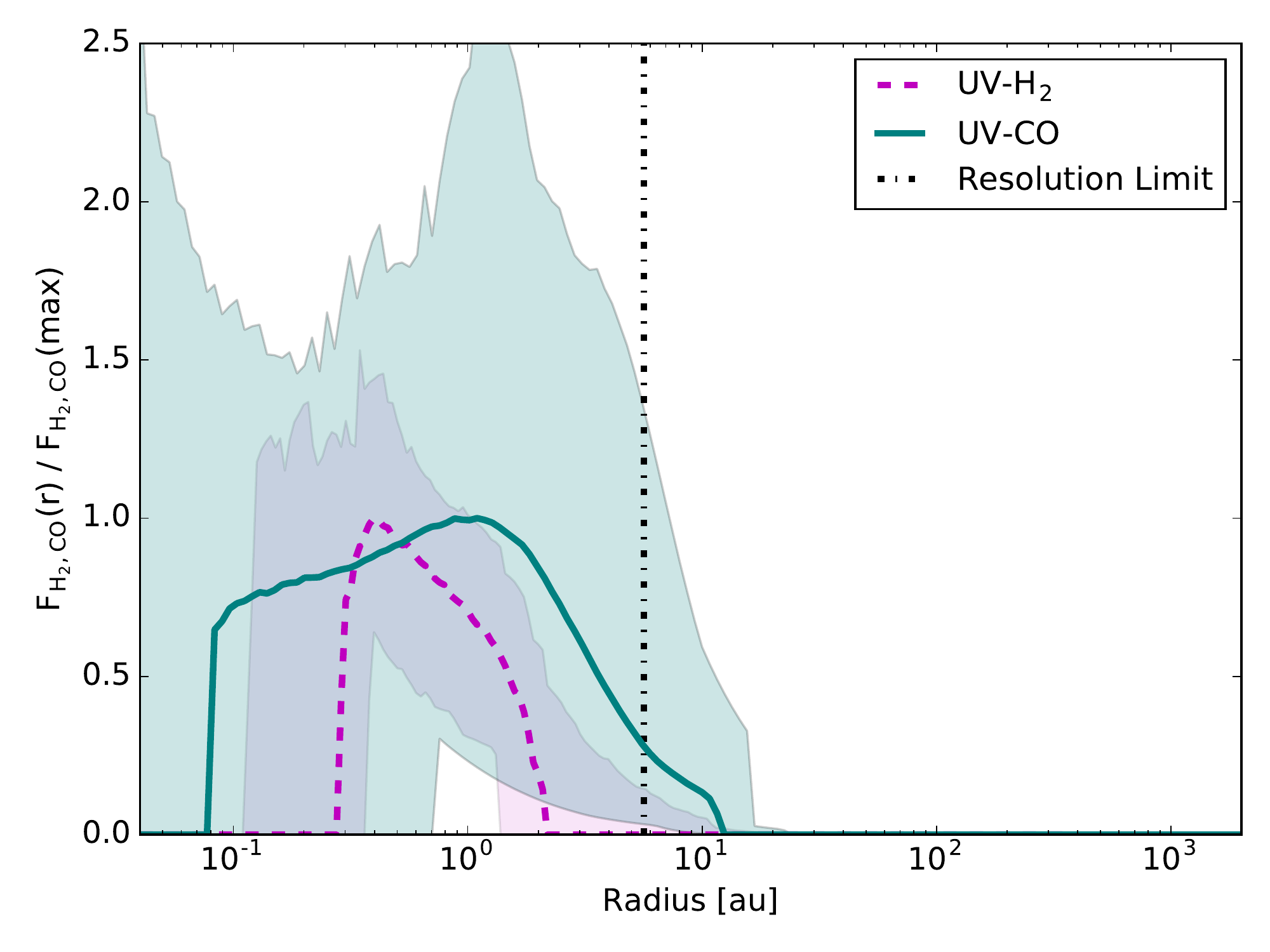} 
	\end{minipage}
\caption{Best-fit UV-CO 2-D radiative transfer (top left) and Gaussian models (top right), 100 UV-H$_2$ radiative transfer models with the smallest \textit{MSE}s (middle left), comparison of UV-H$_2$-/UV-CO-based and observed Ly$\alpha$ profiles (middle right) and radial distributions of flux for both species (bottom left) from the disk around V4046 Sgr. Residuals on the best-fit UV-CO model are color-coded from blue to red, based on the Ly$\alpha$ pumping wavelength required to excite the upper level of the transition.}
\label{V4046_H2CO}
\end{figure*}

\end{appendices}
 
\bibliographystyle{apj}
\bibliography{ms}

\end{document}